\documentclass[fleqn,usenatbib]{mnras}

\usepackage[T1]{fontenc}

\usepackage{newtxtext,newtxmath}
\usepackage{subfigure}

\usepackage{graphicx}	
\usepackage{amsmath}	

\usepackage{float}
\title[Constraints on the properties of local ULIRGs]
{Constraints on the active galactic nucleus and starburst activity of local ultraluminous infrared galaxies from a broad range of torus models}

\author[Charalambia Varnava]{\parbox{\linewidth}{Charalambia Varnava$^{1}$\thanks{E-mail: \href{mailto:varnava.haris@gmail.com}{varnava.haris@gmail.com} (CV)}, 
Andreas Efstathiou$^{1}$, Duncan Farrah$^{2,3}$ and Dimitra Rigopoulou$^{4}$
}
\\
\\
$^{1}$School of Sciences, European University Cyprus, Diogenes street, Engomi, 1516 Nicosia, Cyprus\\
$^{2}$Department of Physics and Astronomy, University of Hawaii, 2505 Correa Road, Honolulu, HI 96822, USA\\
$^{3}$Institute for Astronomy, University of Hawaii, 2680 Woodlawn Drive, Honolulu, HI 96822, USA\\
$^{4}$Oxford Astrophysics, Denys Wilkinson Building, University of Oxford, Keble Road, Oxford OX1 3RH, UK
}

\date{Accepted 2025 February 19. Received 2025 2025 February 12; in original form 2024 2024 November 22}

\pubyear{2024}

\begin{document}
\label{firstpage}
\pagerange{\pageref{firstpage}--\pageref{lastpage}} 
\maketitle

\begin{abstract}
{In this paper we further explore the properties of the HERschel Ultra Luminous Infrared Galaxy Survey (HERUS) sample of 42 local ultraluminous infrared galaxies (ULIRGs) with our recently developed Bayesian spectral energy distribution (SED) fitting code SMART (Spectral energy distributions Markov chain Analysis with Radiative Transfer models). SMART fits SEDs exclusively with multicomponent radiative transfer models. Mid-infrared spectroscopy can be included in the fitting at a spectral resolution matched to that of the radiative transfer models. We fit the SEDs of the HERUS ULIRGs with four different models for the active galactic nucleus (AGN) torus, a starburst and a spheroidal galaxy model, to put constraints on the AGN fraction of the galaxies and their star formation rate (SFR). Two of the AGN torus models we explored are smooth and two are two-phase. We find that, in most cases, a smooth tapered AGN torus provides the best fit to the data. We also find that solutions with other torus models may predict AGN and total luminosities up to an order of magnitude or more lower, but very rarely higher than the best-fitting model. In contrast, we find that, with minor exceptions, the predicted SFR and stellar mass of the ULIRGs are generally robustly estimated irrespective of the assumed torus model.  This is despite the fact that one of the AGN torus models we use assumes fluffy grains with high emissivity in the far-infrared and submillimetre, which could potentially reduce the contribution of a starburst at those wavelengths and reduce the SFR. 
}
\end{abstract}


\section{Introduction}\label{sec:intro}

Local ultraluminous infrared galaxies (ULIRGs) have been studied extensively in the last few decades since their discovery by IRAS in the 1980s \citep{houck85,soif86}. They are an important population in their own right for studies in the local Universe, as their intense star formation and active galactic nucleus (AGN) activity make them the most luminous galaxies \citep{genzel98,rigop99,far03,far22,efs14,efs22}. Perhaps more  significantly, though, local ULIRGs are important laboratories for understanding the role of mergers in triggering extreme star formation and AGN activity that is deeply obscured by dust. Studying these processes aids the interpretation of observations of similar events that took place in the most luminous galaxies in the history of the Universe, such as submillimetre galaxies \citep{barg98,hugh98,efs03,casey14,rowan18}, hyperluminous infrared galaxies \citep{rowan93,rowan00,far02,ver02,efstathiou06}, quasars and hot dust obscured galaxies \citep{eisen12,bridge13,efs21,varn24}.

The popular evolutionary scenario of ULIRGs proposes that these systems are the result of the merger of gas-rich spiral galaxies. According to the scenario, towards the end of the merger we see the emergence of an accreting supermassive black hole (SMBH) that clears dust and gas through outflows and leads to an unobscured quasar and an elliptical galaxy \citep{sanders88,far22}. This evolutionary scenario has been supported by some authors \citep{surace98,canal01,hou11}, but questioned by others \citep{genzel01,tacc02,dasyra06,rodr10}. More recent observations \citep{aalto15,iman19} have cast further doubts on the evolutionary scenario, by presenting evidence for the presence of compact nuclei, hinting at the presence of powerful deeply embedded AGN in the centres of interacting cool ULIRGs. To assess the validity of this scenario, we need good estimates of the AGN torus luminosity in local ULIRGs and quasars, which presumably lie at different phases of this evolutionary sequence. This is complicated by the presence of dust in the form of a toroidal structure around the nucleus, according to the AGN unification models \citep{anton93,ramos17}. 

Another important issue is how robust the estimates of star formation rate (SFR) in local ULIRGs and other extreme starbursts that existed in the history of the Universe are. Current galaxy formation models (e.g. \citealt{lacey16}) have difficulty producing these extreme starbursts at high redshift. It is generally acknowledged that fitting the spectral energy distributions (SEDs) with radiative transfer models for starbursts and AGN can give the best estimates of these key physical quantities, as they encapsulate the physics of ULIRGs much better than other methods like energy balance, such as Bayesian Analysis of Galaxies for Physical Inference and Parameter EStimation (BAGPIPES, \citealt{carnall18}), BayEsian Analysis of GaLaxy sEds (BEAGLE, \citealt{cheval16}), Code Investigating GALaxy Emission (CIGALE, \citealt{noll09,boq19}), Multi-wavelength Analysis of Galaxy Physical Properties (MAGPHYS, \citealt{dac08}) and Prospector \citep{johnson21}.

\cite{smart} recently developed the new Bayesian SED fitting code SMART (Spectral energy distributions Markov chain Analysis with Radiative Transfer models; \citealt{ascl}), which fits SEDs exclusively with radiative transfer models and allows the best possible decomposition of the SEDs into the starburst, AGN torus and host galaxy components that is possible with current data. They applied this new code to study the HERschel Ultraluminous Infrared Galaxy Survey (HERUS) sample \citep{far13} of 42 local ULIRGs at $z<0.27$,  with excellent multiwavelength photometry and mid-infrared spectrophotometry obtained with the Infrared Spectrograph (IRS) onboard the Spitzer Space Telescope. \cite{smart} but also \cite{efs22} found that the galaxies in this sample have high rates of star formation but also a significant contribution from an AGN. 

SMART allows the use of four different AGN torus models \citep{efstathiou95,fritz06,sie15,stal16}. In this paper we study the effect of the assumed AGN torus model on key physical quantities, such as SFR, stellar mass, AGN torus luminosity and AGN fraction. Our method allows us to determine the most reliable estimates of the AGN fraction and SFR for the HERUS sample that can be obtained with the currently available data. The Spitzer spectroscopy data of the HERUS ULIRGs are included in the fitting at a spectral resolution, which is matched to that of the radiative transfer models. 

Our analysis builds on the results of \cite{efs22}, who fitted the HERUS sample with the Bayesian SED fitting code SATMC (Spectral energy distribution Analysis Through Markov Chains; \citealt{john13}). In \cite{smart} we carried out an analysis of this sample with only one of the AGN torus models discussed in this paper, namely the smooth tapered discs of \cite{efstathiou95}. In this paper we improve on the analysis of \cite{efs22} in a number of ways: \cite{efs22} do look at different torus models, but only state which is preferred. In this paper go a step further and also examine how the choice of the AGN torus model affects the derived starburst and host parameters. More specifically, we compare the physical quantities derived by the fits with the four different model combinations and this analysis allows us to determine the preferred geometry for the AGN obscurer. At the same time, we are able to assess how the choice of torus model affects the estimates of key physical quantities like SFR and stellar mass. It is important to note that one of the torus models assumes fluffy grains, which have higher emissivity in the far-infrared and submillimetre, and this can potentially affect the estimates of SFR. Furthermore, here we show the SED fit plots with all the different model combinations, in order to gain a better understanding of the differences between the fits and assessing the contribution of each model component to the SED. We also present a short discussion of all the objects in the sample and list all of the extracted physical quantities, using the four model combinations. These results were only partly presented in \cite{efs22}. Lastly, unlike SATMC, which uses the Metropolis-Hastings algorithm \citep{metropolis53,hastings70}, SMART employs the \textit{emcee} code \citep{foreman13} for the MCMC sampling, which performs much faster.

This paper complements the study of \cite{pap25}, who studied a sample of 200 galaxies at $z=2$, selected in the ELAIS-N1 field of the Herschel Extragalactic Legacy Project (HELP; \citealt{shi19,shi21}) at 250 $\mu m$, and also fitted with four different AGN torus models. The sample of \cite{pap25} has only photometry data from the ultraviolet to the submillimetre.

This paper is organized as follows: In Section \ref{sec:description} we describe the sample, as well as the SED fitting method and the models in Section \ref{sec:results} we present our results and in Section \ref{sec:discussion} we discuss them further. In Section \ref{sec:conclusion} we present our conclusions. Throughout this work we assume $H_0=70$\,km\,s$^{-1}$\,Mpc$^{-1}$, $\Omega=1$ and $\Omega_{\Lambda}=0.7$.

\section{Description of the data, the SED fitting method and the radiative transfer models}\label{sec:description}

\subsection{Sample details}

The HERUS sample was described in more detail in \cite{far13,pea16,cle18,efs22,smart}. The sample comprises all 40 ULIRGs from the IRAS PSC-z survey \citep{sau00} with 60 $\mu m$ fluxes greater than 2 Jy, together with three randomly selected ULIRGs with lower 60 $\mu m$ fluxes, IRAS 00397-1312 (1.8 Jy), IRAS 07598+6508 (1.7 Jy) and IRAS 13451+1232 (1.9 Jy). We have excluded the quasar 3C~273 as it is a Blazar, to give a sample of 42 objects. The sample is not complete, but it includes nearly all known ULIRGs at $z<0.27$ and therefore gives an almost unbiased benchmark of local ULIRGs. All 42 objects were observed with the IRS instrument \citep{houck04} on board Spitzer and by Herschel, as part of both the HERUS and SHINING surveys \citep{fisch10,sturm11,hail12,gonz13}. The IRS data included in the fit with SMART have a wavelength grid, which is separated in steps of 0.05 in the log of the rest wavelength. We also use additional points around the 9.7 $\mu m$ silicate feature and the polycyclic aromatic hydrocarbon molecule (PAH) features to the equally spaced wavelength grid. More multiwavelength photometry was added for all galaxies in the sample, as described in \cite{efs22}. 

\subsection{Description of the SED fitting method}

Our method allows us to explore the impact of four different AGN torus models and therefore constrain the properties of the obscuring torus, but also quantify the uncertainties in the AGN fraction and SFR of the fitted galaxies. Each torus model is fitted in combination with the starburst model of \cite{efstathiou00}, as revised by \cite{efstathiou09}, and the spheroidal model of \cite{efs21}:

\begin{enumerate}

\item The smooth AGN torus model originally developed by \cite{efstathiou95} is part of the CYGNUS (CYprus models for Galaxies and their NUclear Spectra) collection of radiative transfer models. More details of the implementation of this combination of models within SATMC code are given in \cite{efs21} and \cite{efs22}. This model assumes a tapered disc geometry (the thickness of the disc increases linearly with distance from the black hole in the inner part of the torus, but assumes a constant thickness in the outer part).

\item The smooth AGN torus model of \cite{fritz06} assumes a flared disc geometry (the thickness of the disc increases linearly with distance from the black hole).

\item The two-phase AGN torus model SKIRTOR of \cite{stal16} also assumes a flared disc geometry.

\item The two-phase AGN torus model of \cite{sie15}, unlike the other three torus models listed above, assumes that dust grains are fluffy and therefore have a higher emissivity in the far-infrared and submillimetre. This model is therefore expected to give the stronger contribution from the AGN in that part of the spectrum. 
\end{enumerate}

The anisotropy of the emission from the torus is a feature of all torus models considered in this study and requires a correction to the observed AGN luminosity, in order to obtain their intrinsic luminosity. \cite{efstathiou06}, \cite{efs22}, \cite{smart} and \cite{varn24} defined the anisotropy correction factor $A$, which represents the factor by which we need to multiply the observed luminosity to derive the true luminosity:
\begin{equation}
A(\theta_i) = {{\int_0^{\pi/2} ~~S(\theta_i') ~~sin \theta_i' ~~d \theta_i' } \over {S(\theta_i)}}~~,
\end{equation}
where $\theta_i$ is the torus inclination and $S(\theta_i)$ represents the bolometric emission over the relevant wavelength range. $A(\theta_i)$ is generally different for the infrared and bolometric luminosities and is significant for all the AGN torus models incorporated in this study. A higher A factor would apply for bolometric luminosities. The high anisotropy factors are due to the very strong dependence of the spectra emitted by the torus as a function of inclination (see, for example, Fig. 5 in \citealt{efstathiou95} and \citealt{fritz06}). In the mid-infrared the emission can vary by more than an order of magnitude. In Tables \ref{tab:extractedB_CYGNUS}, \ref{tab:extractedB_Fritz}, \ref{tab:extractedB_SKIRTOR} and \ref{tab:extractedB_Sie} we present the anisotropy correction factor $A$ predicted by the four different torus models.

SMART can extract the fitted model parameters but also a number of other physical quantities, which are listed in Table \ref{tab:derived}. The SFR of the starburst and the spheroidal component are computed self-consistently by the radiative transfer models, which incorporate the stellar population synthesis models of \cite{bru93,bru03}. We assume a Salpeter initial mass function (IMF) with a metallicity 40 per cent of solar for the spheroidal component and solar for the starburst component. A detailed description of all the radiative transfer models used in this work is given in Section \ref{sec:models}.

\subsection{Description of the radiative transfer models}\label{sec:models}

In this study we use libraries of radiative transfer models for the emission of starburst episodes, spheroidal host galaxies, AGN tori and polar dust. These are part of the CYGNUS collection.

\subsubsection{Starburst model}
CYGNUS includes a starburst model, which is described in \citet{efstathiou00} and \citet{efstathiou09}. This model has three parameters, which are the giant molecular clouds' initial optical depth ($\tau_v=50-250$), the time constant of the exponentially decaying SFR ($\tau_{*}=15-35$ Myr) and the age of the starburst ($t_{*}=5-35$ Myr).

\subsubsection{Spheroidal host model}
The CYGNUS spheroidal host model was described in more detail in \cite{efs21}. The models of \cite{bru93,bru03} are first used in combination with an assumed star formation history (SFH) to compute the spectrum of starlight, which is illuminating the dust throughout the model galaxy. Although the spectrum of starlight is assumed to be constant throughout the galaxy, its intensity is assumed to vary according to a S\'ersic profile with $n=4$, which is equivalent to de Vaucouleurs's law. For the SFH we assume a delayed exponential ($\dot{M}_{\ast} \propto t \times e^{-t/\tau^s}$), where $\tau^s$ is the e-folding time of the exponential.

The spheroidal model has three parameters, which are the e-folding time of the assumed delayed exponential SFH ($\tau^s=0.125-8$ Gyr), the optical depth of the galaxy from its centre to its surface ($\tau_v^s=0.1-15$) and a parameter that controls the bolometric intensity of stellar emission relative to that of the bolometric intensity of starlight in the solar neighborhood ($\psi^s=1-17$). The library used in this paper was computed assuming all the galaxies have an age equal to the age of the Universe at a redshift of $z=0.1$. We assumed that all the stars in the galaxy formed with a Salpeter IMF out of gas with a metallicity of 40\% of solar.

\subsubsection{AGN torus models}
\begin{figure*}
\centering
\includegraphics[width=.9\linewidth]{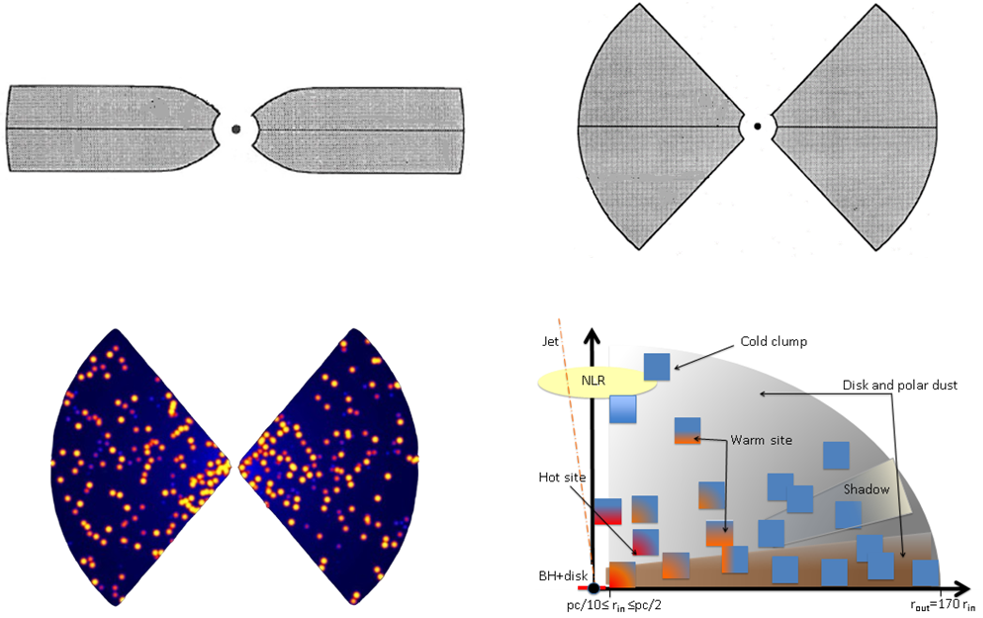}
\vspace{10pt}
\caption{Schematic diagrams that show cross-sections of the four different AGN torus models that are currently incorporated in SMART.
            {\itshape Top left:} CYGNUS tapered torus model of Efstathiou \& Rowan-Robinson (1995).
            {\itshape Top right:} flared torus model of Fritz et al. (2006).
            {\itshape Bottom left:} SKIRTOR two-phase flared torus model of Stalevski et al. (2016).
		  {\itshape Bottom right:} two-phase fluffy dust torus model of Siebenmorgen et al. (2015) (only one quadrant is shown).}
\label{fig:4AGN}
\end{figure*}

In Fig. \ref{fig:4AGN} we show schematic diagrams of the cross-sections of the four different AGN torus models that are currently incorporated in SMART. We describe these models in the following paragraphs.

The CYGNUS AGN torus model has four parameters, which are the equatorial optical depth at 1000\AA ~($\tau_{uv}=250-1450$), the ratio of outer to inner disc radius ($r_2/r_1=20-100$), the half-opening angle of the torus ($\theta_o=30\degr-75\degr$) and the inclination of the torus ($\theta_i=0\degr - 90\degr$). We assume that the density distribution in the tapered disc falls off with distance from the SMBH $r$ as $r^{-1}$.

The AGN torus model of \cite{fritz06} also assumes a smooth torus. Its density distribution is defined as $ \rho (r, \theta) \propto r^{\beta} e^{-\gamma|cos \theta|}$, so it has two more parameters than the CYGNUS model, $\beta$ and $\gamma$. In this work we fix $\beta=0$ and $\gamma=4$. This gives the same number of free parameters as the CYGNUS AGN torus model. The four free parameters of the model of \cite{fritz06} are the equatorial optical depth at 9.7$\mu m$ ($\tau_{9.7\mu m}=1-10$), the ratio of outer to inner radius ($r_2/r_1=10-150$), the half-opening angle of the torus ($\theta_o=20\degr-70\degr$) and the inclination of the torus ($\theta_i=0\degr-90\degr$). For these parameters we use the full range available in the library. It is notable that the model of \cite{fritz06} has a lower resolution in inclination compared to the CYGNUS model, with 10 points covering the range $0\degr-90\degr$. The model is linearly interpolated between these values to obtain the same resolution we have in CYGNUS.

The SKIRTOR model assumes a two-phase geometry. More specifically, it is assumed that the torus dust lies in discrete clouds that are embedded in a smooth distribution. The dust density distribution in the torus follows $ \rho (r, \theta) \propto r^{-p} e^{-q|cos \theta|}$.  We fix the parameters $p$ and $q$ to 1.  As argued in \cite{smart} and \cite{varn24}, fixing the parameter $p$ of the SKIRTOR model to the value of 1, gives the best agreement for the silicate absorption features of obscured quasars, so we adopted the same approach in this study. We have four remaining free parameters, which are the same as those in the model of \cite{fritz06}. For these parameters we also explore the full range available in the library ($\tau_{9.7\mu m}=3-11$, $r_2/r_1=10-30$, $\theta_o=20\degr-70\degr$ and $\theta_i=0\degr-90\degr$). The SKIRTOR model also has 10 inclinations that cover the range $0\degr-90\degr$, so again the templates are linearly interpolated to have the same resolution as in the CYGNUS fits.

As in the case of the SKIRTOR model, the model of \cite{sie15} assumes a two-phase geometry. The dust density distribution in this model follows an isothermal disc profile, which is defined by equations (3) and (4) in \cite{sie15}. There are no parameters to fix, in order to specify the density profile in this case. A particular feature of this model is that it assumes that dust covers the whole sphere around the black hole, i.e. the half-opening angle of the torus is assumed to be zero. Also, unlike the other three torus models listed above, this model assumes that the dust grains are fluffy and have higher emissivity in the far-infrared and submillimeter compared to normal interstellar grains. The parameters of the model of \cite{sie15} are the cloud volume filling factor ($V_c=1.5-77$ per cent), the optical depth (in the V band) of the individual clouds ($A_c=0-45$), the optical depth (in the V band) of the disc mid-plane ($A_d=50-500$) and the inclination, which takes 9 values corresponding to bins at 86\degr, 80\degr, 73\degr, 67\degr, 60\degr, 52\degr, 43\degr, 33\degr and 19\degr measured from the pole. Similar to the case of the SKIRTOR and \cite{fritz06} models, the model is interpolated to have the same resolution in inclination as the CYGNUS model. The model also has as a parameter the inner radius of the dusty torus. We select the value of $1000 \times 10^{15} cm$, which for an AGN luminosity of $10^{11} L_\odot$ gives a temperature at the inner torus radius of about 1000K.

\subsubsection{Polar dust model}
For the galaxies IRAS~05189-2524, IRAS~07598+6508 and IRAS~13451+1232 we explore the impact of adding a component of polar dust in the fit, as we see evidence that this addition can improve the fit. The model assumes that polar dust is concentrated in discrete spherical optically thick clouds, all of which are assumed to have constant temperature for all dust grains \citep{efstathiou06}. The model assumes the same multi-grain dust mixture as in the starburst radiative transfer model \citep{efstathiou09}, but the small grains and PAHs are assumed to be destroyed by the strong radiation field of the AGN to which these clouds are directly exposed. Unlike \cite{efs22}, in SMART the temperature of the polar dust clouds $T_p$  is a free parameter in the fit and is assumed to vary in the range  800K$-$1200K. We assume that all clouds have an optical depth from the center to the surface in the V band of 100.

\begin{figure}
\centering
\includegraphics[width=1.\linewidth]{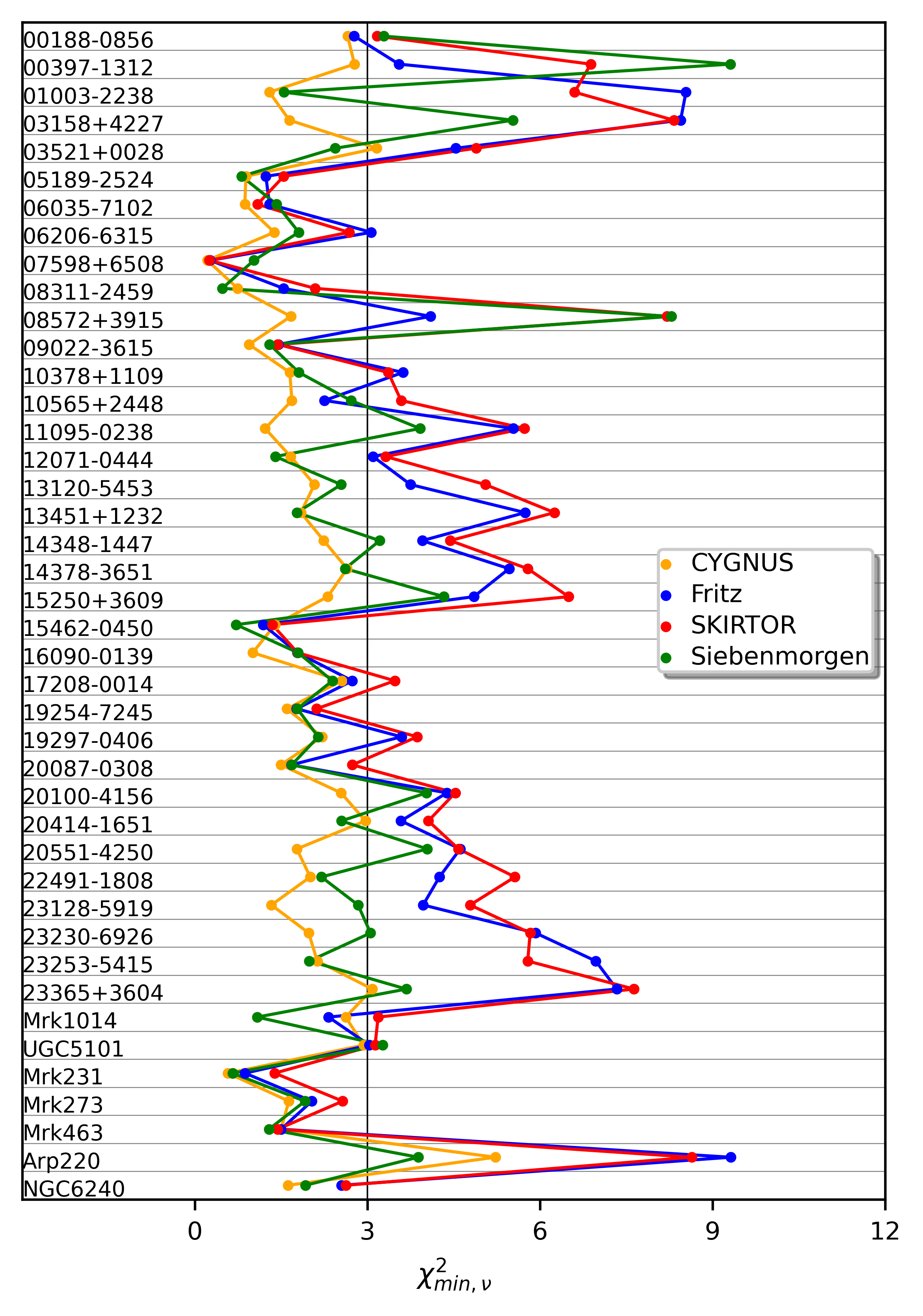}
\caption{Comparison of the minimum reduced $\chi^2$ ($\chi^{2}_{min, \nu}$) of all objects fitted with the CYGNUS tapered torus, Fritz et al. (2006) flared torus, SKIRTOR two-phase flared torus and Siebenmorgen et al. (2015) two-phase fluffy dust torus models. The CYGNUS and Siebenmorgen et al. (2015) models give systematically better fits than the Fritz et al. (2006) and SKIRTOR models. CYGNUS is usually better overall.}
\label{fig:chi_2}
\end{figure}

\begin{table}
	\centering
\caption{Derived physical quantities for the starburst, AGN torus and spheroidal host combination, as well as the symbol used. The luminosities are integrated over $1-1000~\mu m$.}
	\label{tab:derived}
\begin{tabular}{llllll} 
		\hline
		Physical quantity &  Symbol\\
		\hline  
Observed AGN torus luminosity          & $L_{AGN}^{o}$\\  
Corrected AGN torus luminosity         & $L_{AGN}^{c}$\\
Polar dust AGN luminosity              & $L_{p}$ \\  
Starburst luminosity                   & $L_{SB}$\\   
Spheroidal host luminosity             & $L_{sph}$\\
Total corrected luminosity             & $L_{tot}^{c}$\\   
Starburst SFR (averaged over its age)   & $\dot{M}_*^{age}$\\
Spheroidal SFR                         & $\dot{M}_{sph}$\\
Total SFR                              & $\dot{M}_{tot}$\\ 
Starburst stellar mass                 & $M^{*}_{SB}$\\
Spheroidal stellar mass                & $M^{*}_{sph}$\\ 
Total stellar mass                     & $M^{*}_{tot}$\\     
AGN fraction                           & $F_{AGN}$\\ 
Anisotropy correction factor           & $A$\\
		\hline
	\end{tabular}
\end{table}

\section{Results}\label{sec:results}

In \cite{smart} we compared the results of the fits with the CYGNUS AGN torus model with those obtained by \cite{efs22} and demonstrated that the results obtained with the two methods are in very good agreement. The novelty of SMART is that it is specifically designed for fitting with radiative transfer models. In this paper we present the fits with all the combinations of models, as this analysis was not presented in \cite{efs22}. Furthermore, here we compare the physical quantities derived by the fits with the four model combinations and list the values of the extracted physical quantities. These results lead to a direct comparison of the fits and a better exploration of the impact of the four different AGN torus models, in order to constrain the properties of the obscuring torus. Our approach also allows us to quantify the uncertainties in the AGN fraction and SFR of the fitted galaxies.

Our analysis allows us to assess which AGN torus model best fits the observational data for the HERUS sample. In Figs \ref{fig:HERUS-resultsA}, \ref{fig:HERUS-resultsB}, \ref{fig:HERUS-resultsC}, \ref{fig:HERUS-resultsD}, \ref{fig:HERUS-resultsE}, \ref{fig:HERUS-resultsF} and \ref{fig:HERUS-resultsG} we present the ultraviolet to millimetre SED fits of the HERUS sample with all the model combinations.

A good first approach to see which model gives the best fit to the data is to calculate the minimum reduced $\chi^2$ of each fit. A comparison plot of the minimum reduced $\chi^2$ provided by the fits with all four combinations of models is given in Fig. \ref{fig:chi_2}. It is clear that the CYGNUS AGN torus model nearly always gives better fits than the \cite{fritz06} and SKIRTOR models and usually fits better than the \cite{sie15} model. A discussion of the effect of this on the predicted AGN torus luminosities follows. A similar conclusion was reached in \cite{efs22}, using the SATMC code.

The extracted physical quantities of the fits with the CYGNUS combination of models are given in Tables \ref{tab:extractedA_CYGNUS} and \ref{tab:extractedB_CYGNUS}. In Tables \ref{tab:extractedA_Fritz} and \ref{tab:extractedB_Fritz} we present the same quantities, with the CYGNUS AGN torus model being replaced by the model of \cite{fritz06}. The corresponding quantities of the fits with the SKIRTOR AGN torus model are given in Tables \ref{tab:extractedA_SKIRTOR} and \ref{tab:extractedB_SKIRTOR}, while in Tables \ref{tab:extractedA_Sie} and \ref{tab:extractedB_Sie} we show the extracted physical quantities of the fits with the AGN torus model of \cite{sie15}. To explore these results better, we plot some of the quantities derived by the four different combinations of models for the HERUS sample against the value derived by the best-fitting model. This allows us to see how well the physical quantities can be constrained. 

\begin{figure*}
\centering
\includegraphics[width=70mm]{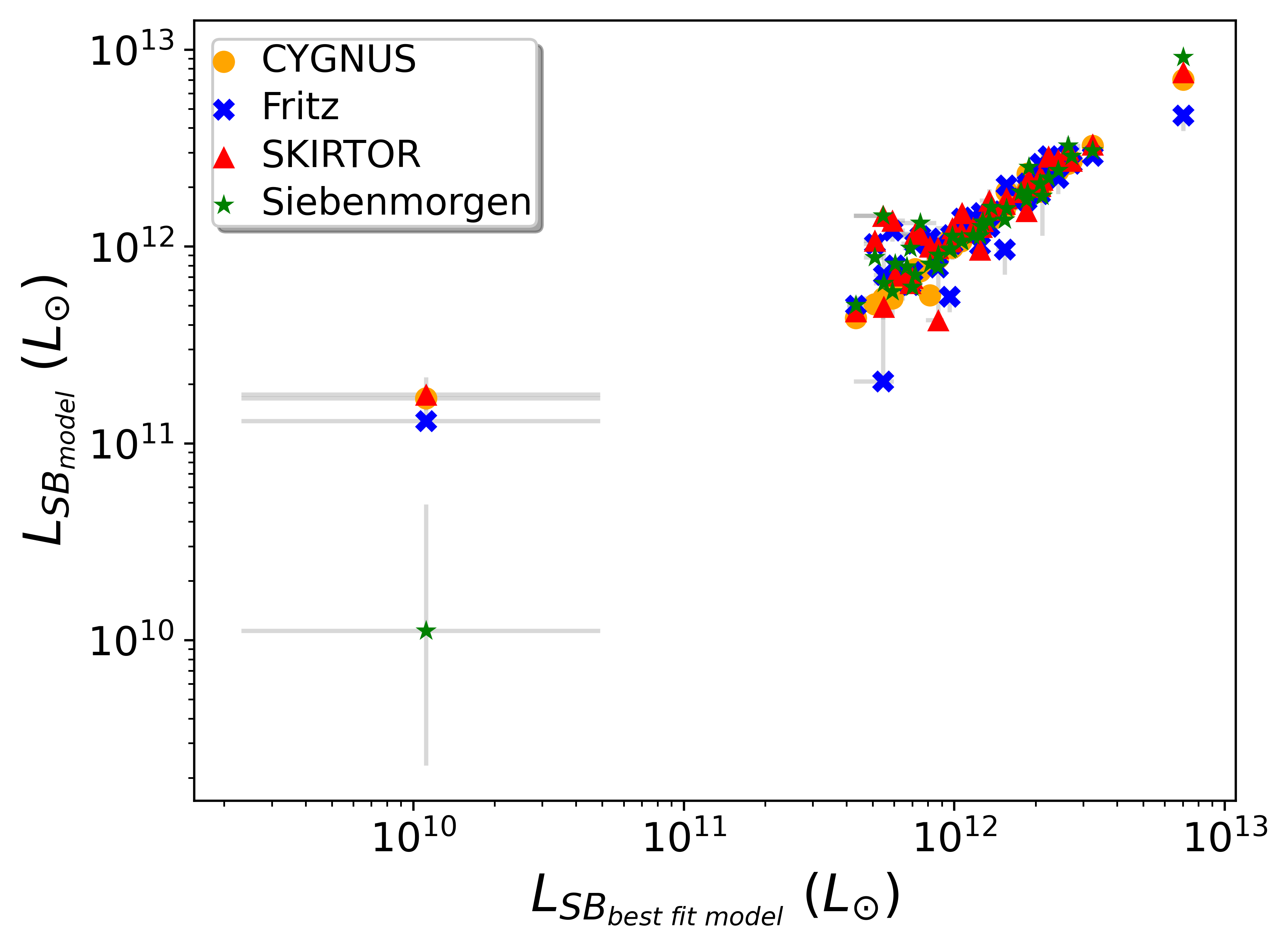} \hspace{30pt}
\includegraphics[width=70mm]{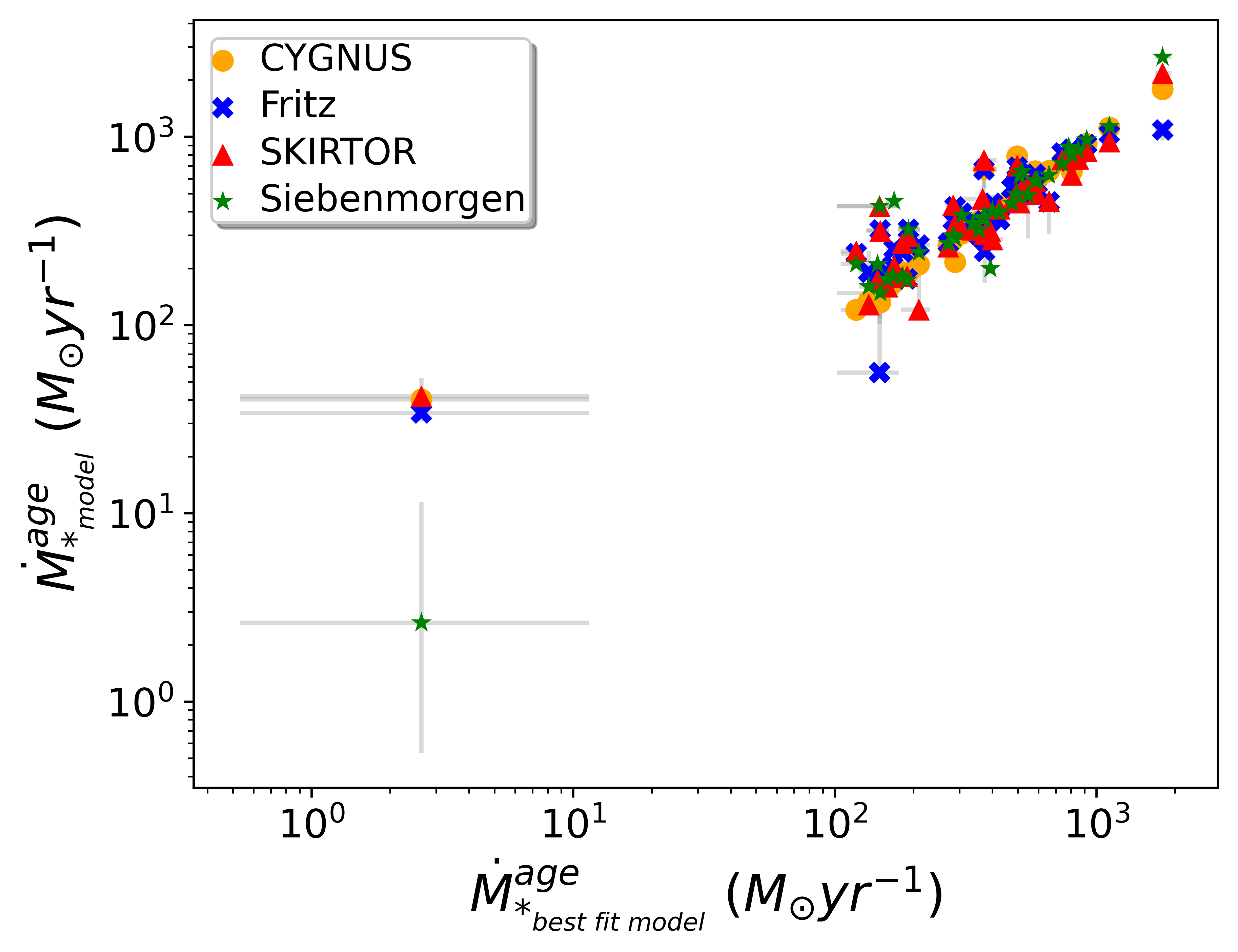}
$$
$$
\includegraphics[width=70mm]{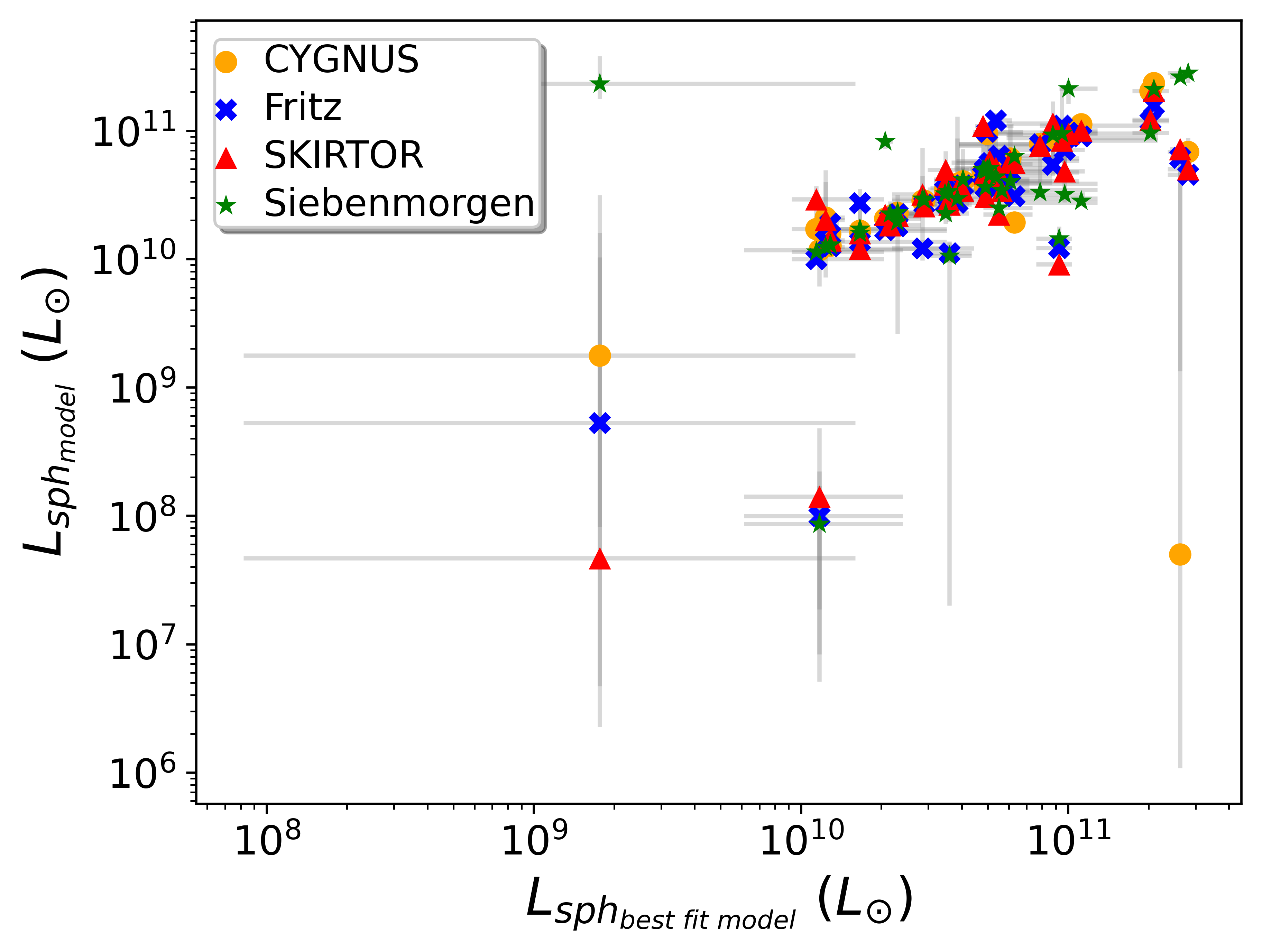} \hspace{30pt}
\includegraphics[width=70mm]{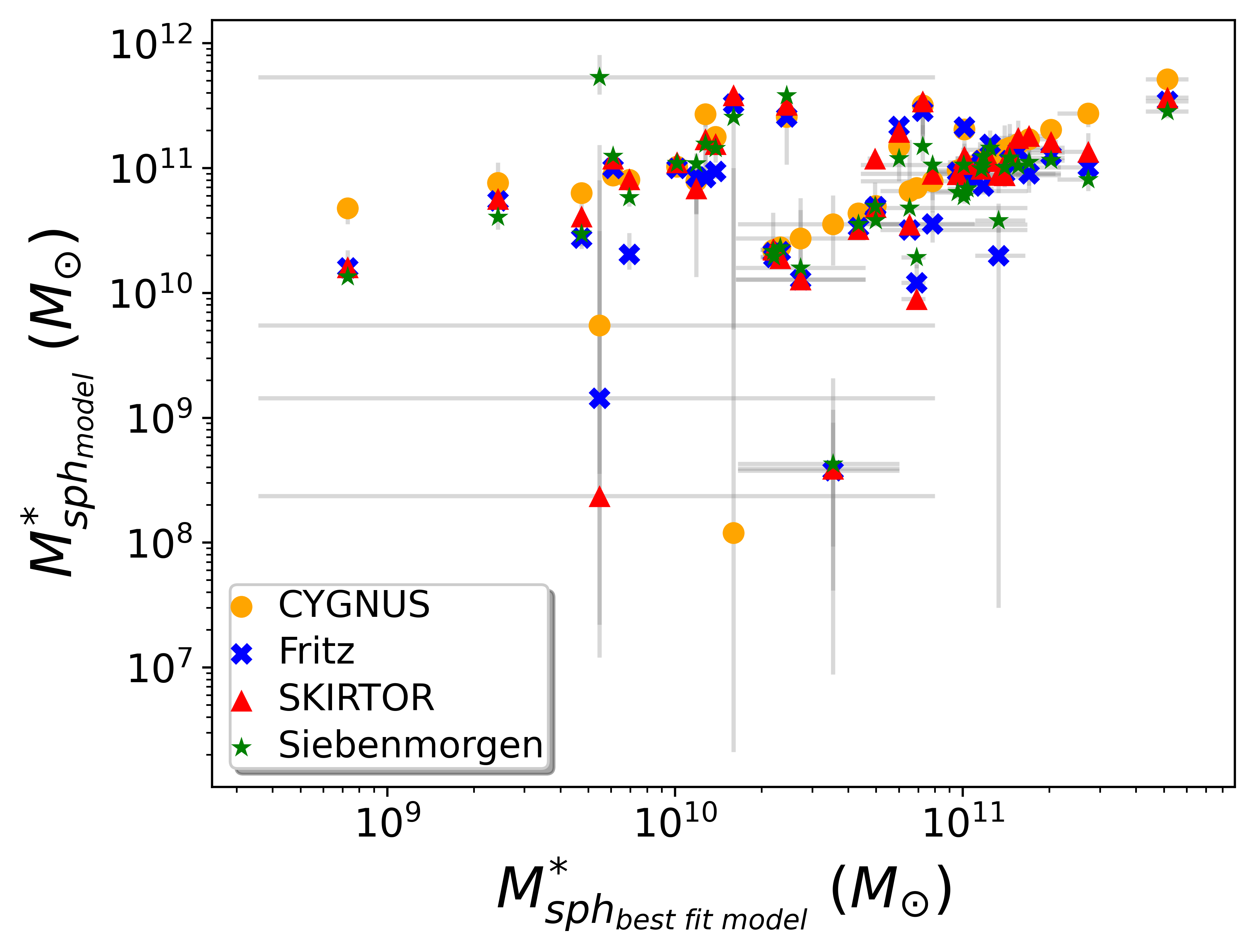}
\vspace{10pt}
\caption{Plot of physical quantities extracted by the four different combinations of models against the value derived by the best-fitting model: CYGNUS tapered torus model (orange), flared torus model of \citealt{fritz06} (blue), SKIRTOR two-phase flared torus model (red), two-phase fluffy dust torus model of \citealt{sie15} (green).
            {\itshape Top left:} Starburst luminosity.
            {\itshape Top right:} Starburst SFR.
            {\itshape Bottom left:} Spheroidal host luminosity.
		  {\itshape Bottom right:} Stellar mass.}
\label{fig:quant1}
\end{figure*}

\begin{figure*}
\centering
\includegraphics[width=70mm]{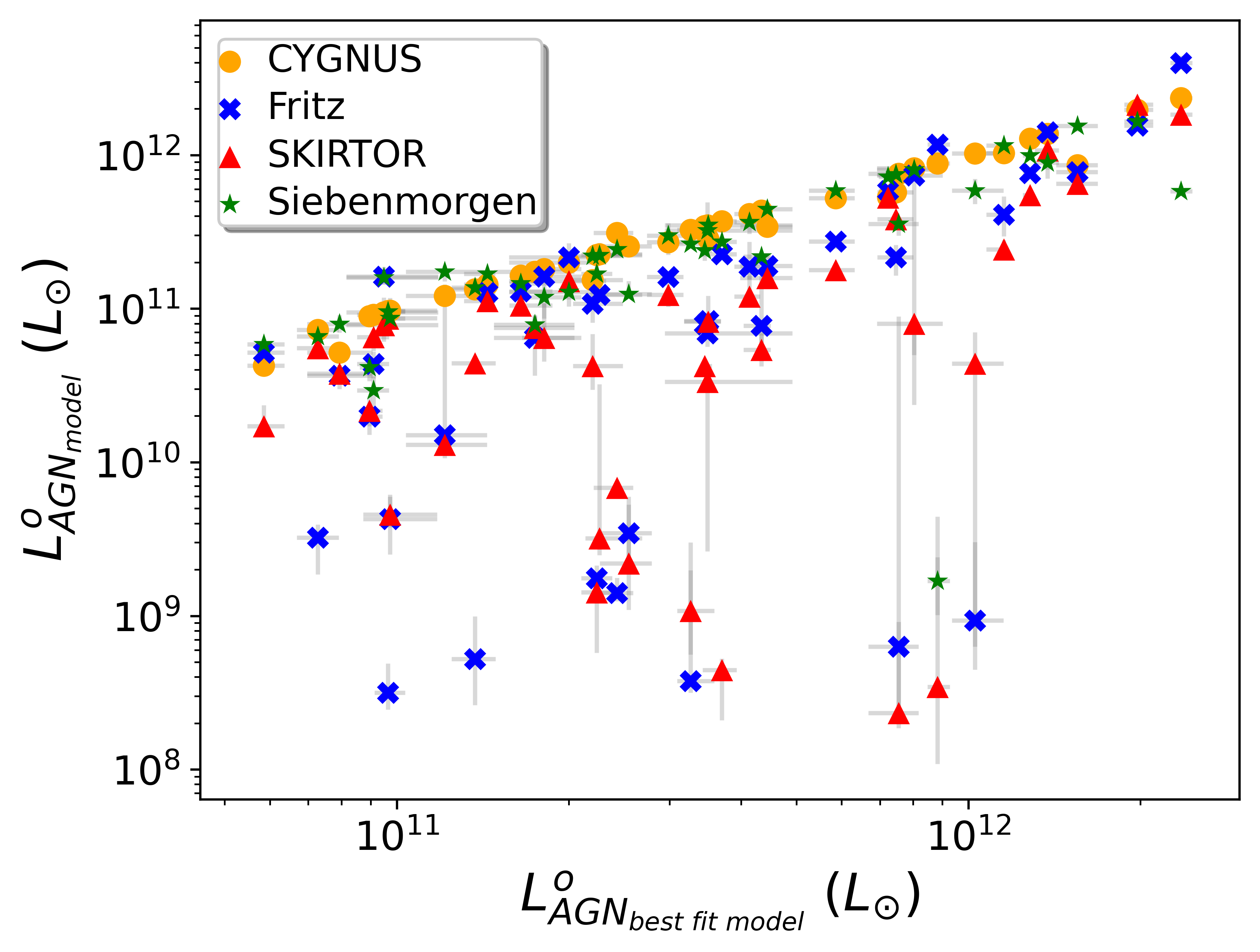} \hspace{30pt}
\includegraphics[width=70mm]{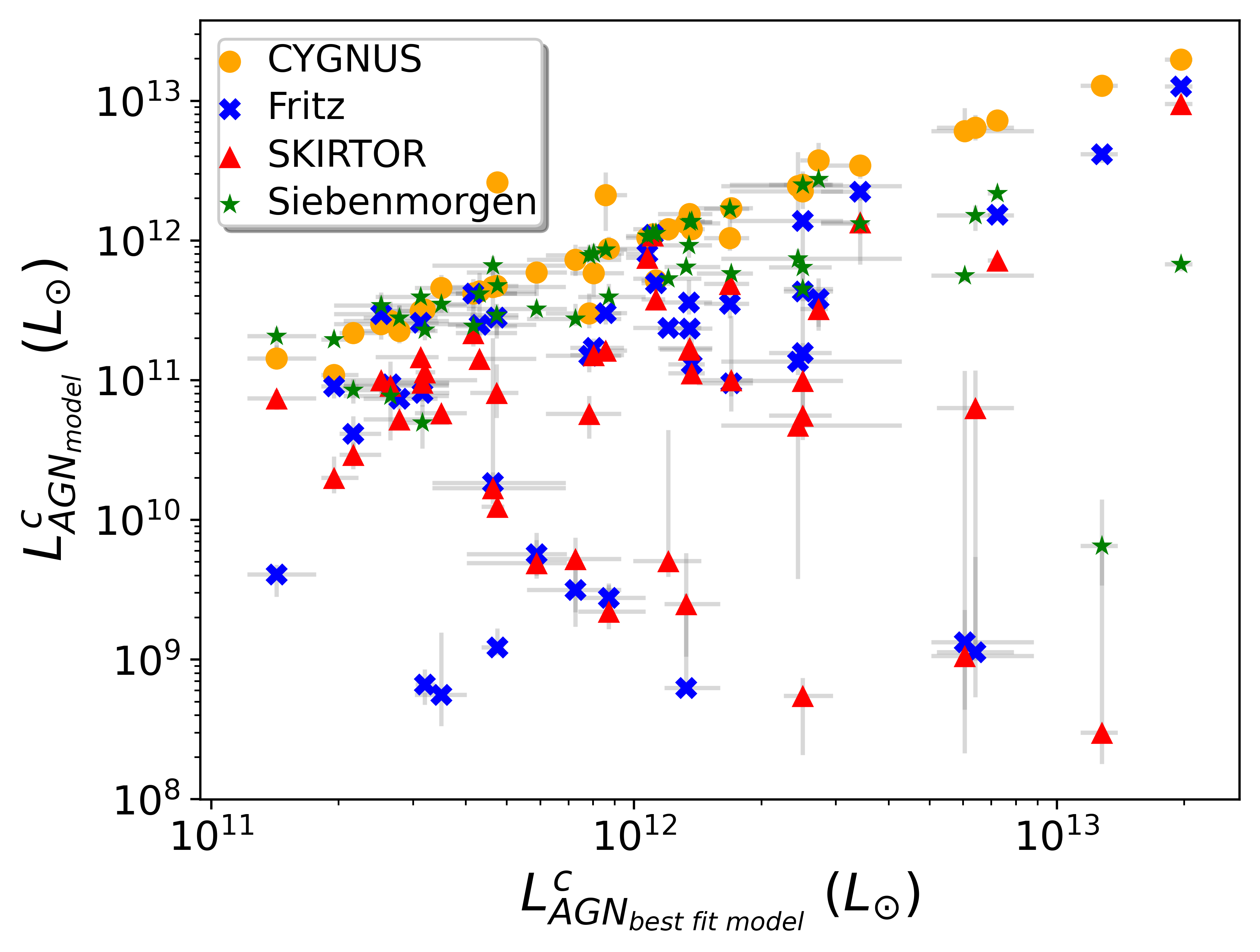}
$$
$$
\includegraphics[width=70mm]{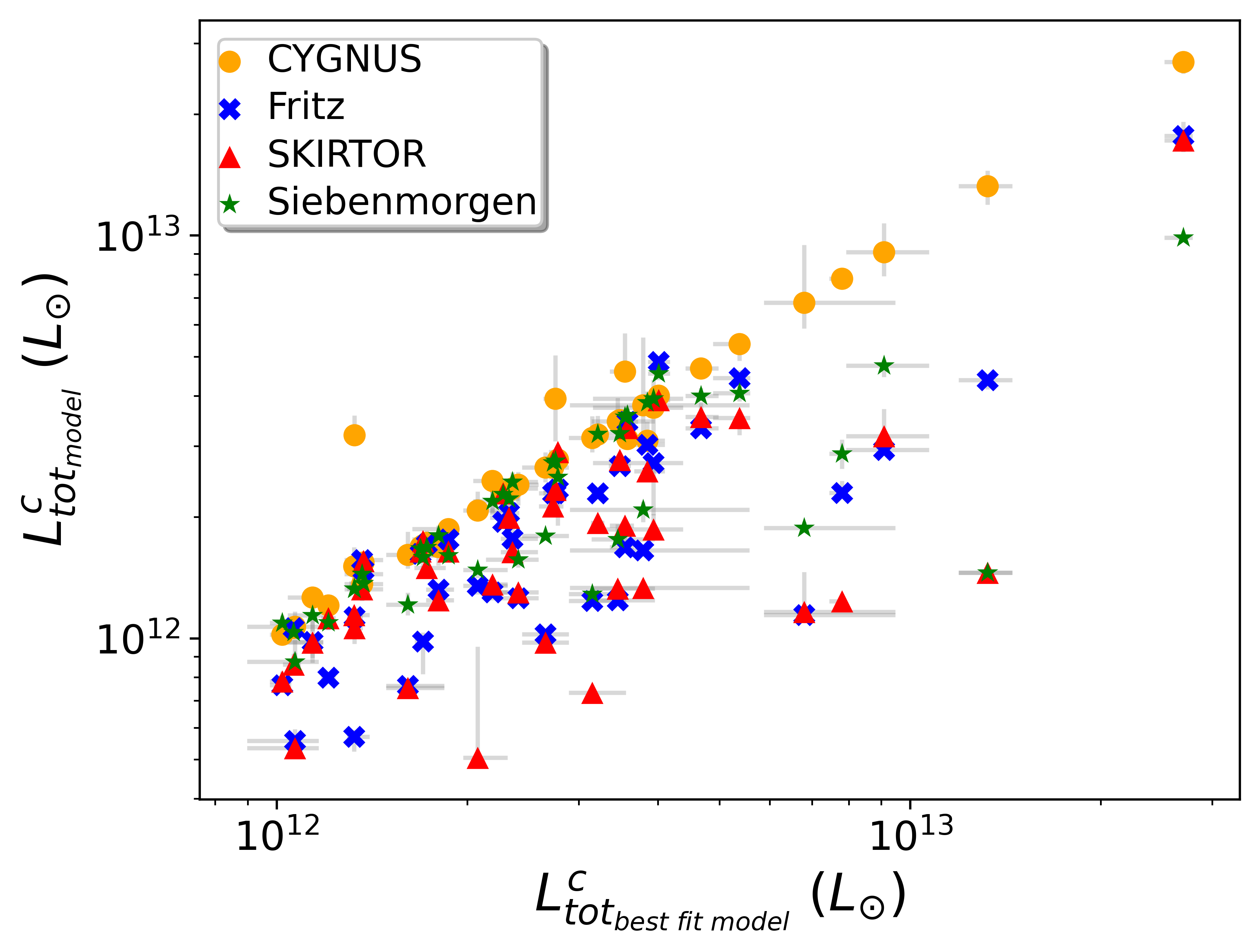} \hspace{30pt}
\includegraphics[width=70mm]{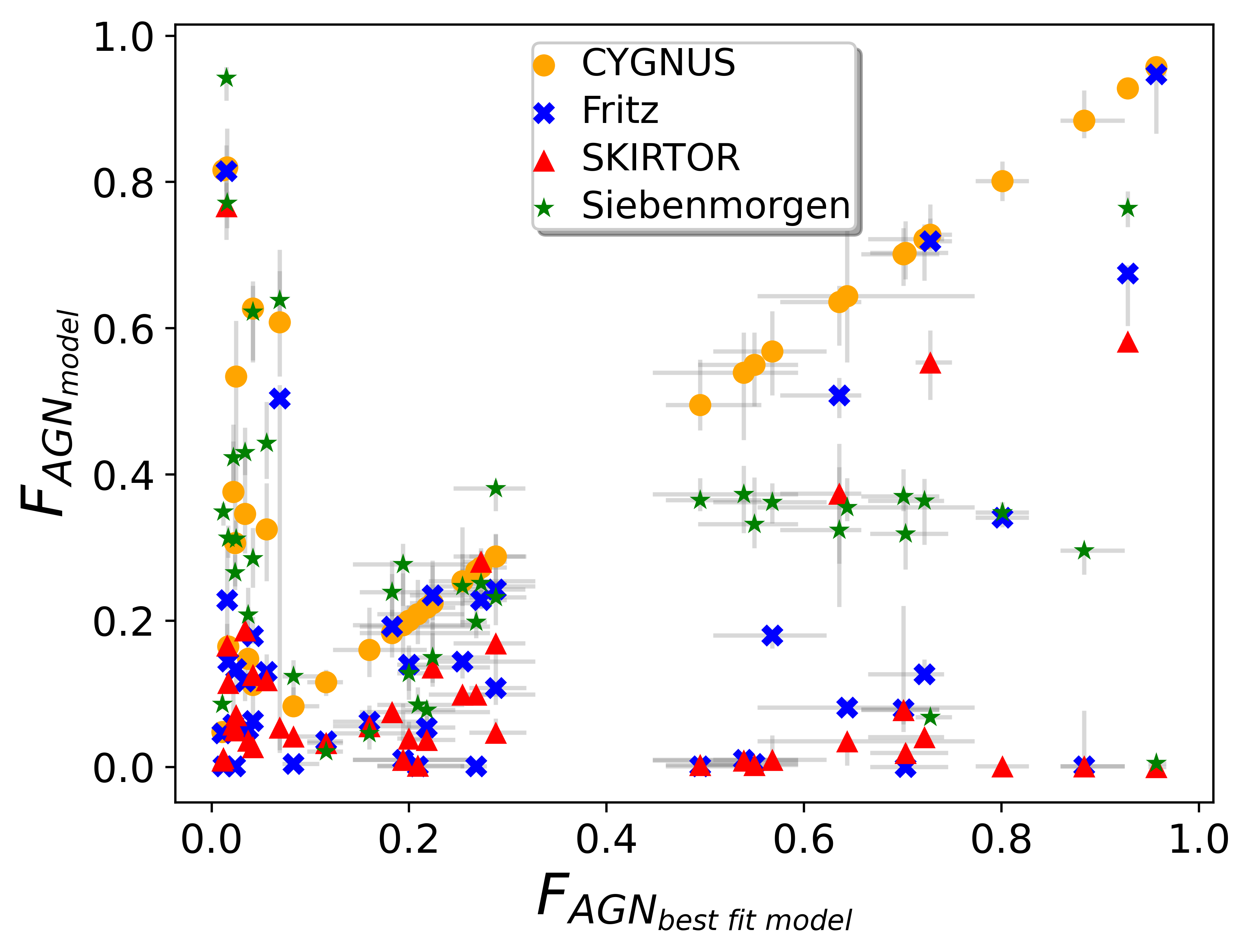}
\vspace{10pt}
\caption{Plot of other physical quantities extracted by the four different combinations of models against the value derived by the best-fitting model: CYGNUS tapered torus model (orange), flared torus model of \citealt{fritz06} (blue), SKIRTOR two-phase flared torus model (red), two-phase fluffy dust torus model of \citealt{sie15} (green).
            {\itshape Top left:} Observed AGN torus luminosity.
            {\itshape Top right:} Anisotropy-corrected AGN torus luminosity.
            {\itshape Bottom left:} Anisotropy-corrected total luminosity.
		  {\itshape Bottom right:} AGN fraction.}
\label{fig:quant2}
\end{figure*}

In Fig. \ref{fig:quant1} we first compare the starburst luminosity predicted by the four different combinations of models. We observe that there is very good agreement between the four estimates and therefore the starburst luminosity can be well constrained for the HERUS sample. We discuss some of the exceptions in Section \ref{sec:individual}. A similarly good agreement between the models, albeit with some more scatter, is found in the comparison we make for the SFR of the starburst. The reason we see more scatter here is because in our method the SFR of a starburst does not simply depend on the starburst luminosity but also on the starburst age, which is a parameter that is self-consistently determined by the fit. We conclude therefore that the starburst luminosity and SFR of local ULIRGs can be robustly estimated irrespective of the AGN torus model that is used in the fitting. We make the corresponding comparison for the luminosities of the spheroidal component. The spheroidal host luminosity is fairly well constrained with some scatter. The comparison of the estimates of the stellar mass is fairly well constrained above $10^{10}L_\odot$. For lower values these quantities are poorly constrained, due to the lack of data.

In Fig. \ref{fig:quant2} we show the observed and corrected AGN torus luminosities, the resulting total corrected luminosities and AGN fractions. Even before applying any anisotropy corrections to the AGN torus luminosities, we see significant variation in the predictions. We observe that we can have galaxies for which the predicted luminosity is $1-2$ orders of magnitude lower than that predicted by the best fit. This is almost exclusively for fits with the \cite{fritz06} and SKIRTOR models, where we find a solution which is starburst-dominated. We note that these fits generally have a very high reduced $\chi^2$. It is very interesting that for the corrected AGN torus luminosities, which may be up to an order of magnitude or more higher than the observed one, we don't seem to have a scatter about the line defined by the best fit, but all models that have a worse fit than CYGNUS, which are usually the \cite{fritz06} or SKIRTOR models, predict a lower corrected AGN torus luminosity. So, using any of these torus models, may systematically underestimate the true luminosity of the AGN by a factor of up to an order of magnitude or more. A similar effect is also seen in the total corrected luminosities. The luminosities predicted by the \cite{fritz06} and SKIRTOR models are systematically lower than the best-fitting CYGNUS model by a factor of an order of magnitude or more. Finally, for best-fitting AGN fractions above 0.1 we see a similar behaviour as seen for the AGN torus and total luminosities, where the \cite{fritz06} and SKIRTOR models show systematically lower AGN fractions compared to the best-fitting CYGNUS model. For best-fitting AGN fractions below 0.1 a more complicated behaviour is observed, which is due to the fact that the AGN fraction is very poorly constrained by the data for any combination of models. In this regime the CYGNUS and \cite{sie15} models usually predict very high AGN fractions, but are not usually the best-fitting models.

\begin{figure}
\centering
\includegraphics[width=1.\linewidth]{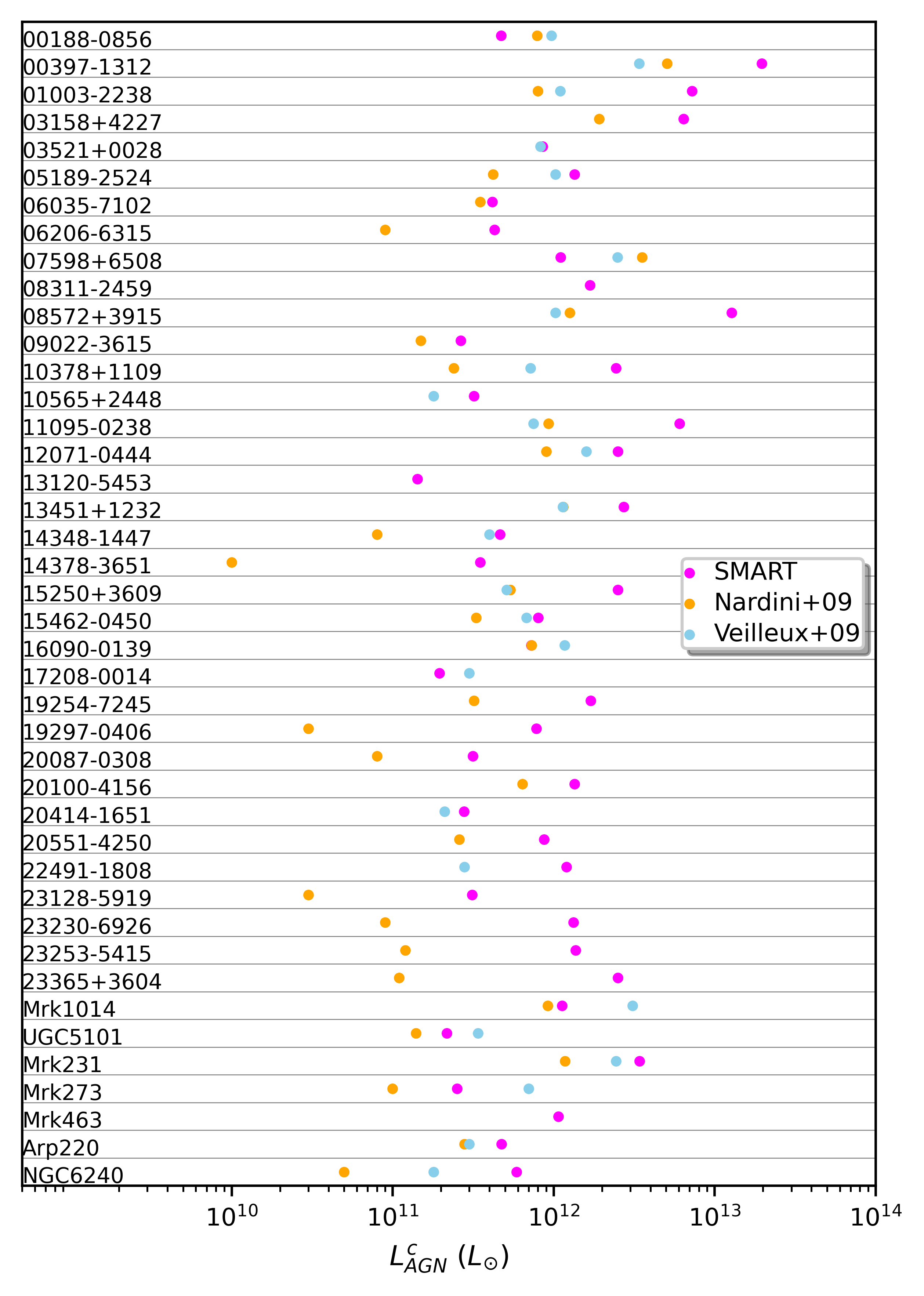}
\caption{Anisotropy-corrected AGN torus luminosities for all objects derived by the best-fitting torus model of SMART, plotted with the AGN torus luminosities extracted by Nardini et al. (2009) and Veilleux et al. (2009)}
\label{fig:lum_AGN}
\end{figure}

In order to assess the reliability of our SED analysis and our estimate of the correction of the AGN luminosity due to the anisotropic emission from the torus, in Fig. \ref{fig:lum_AGN} we compare our results with those of \cite{nard09} and \cite{veilleux09}. \cite{nard09} used Spitzer IRS $5-8~\mu m$ spectra, while the luminosities of \cite{veilleux09} are not based on SED fits, but are averages of five different approaches based on Spitzer IRS $5-35~\mu m$ spectra. In this paper the luminosities of \cite{nard09} have been converted to the cosmology adopted in our study. \cite{veilleux09} assumed the same cosmology as the one adopted in our study. In Section \ref{sec:individual} we examine the results for each object and compare them with those of \cite{nard09} and \cite{veilleux09}.

\begin{figure}
\centering
\includegraphics[width=.95\linewidth]{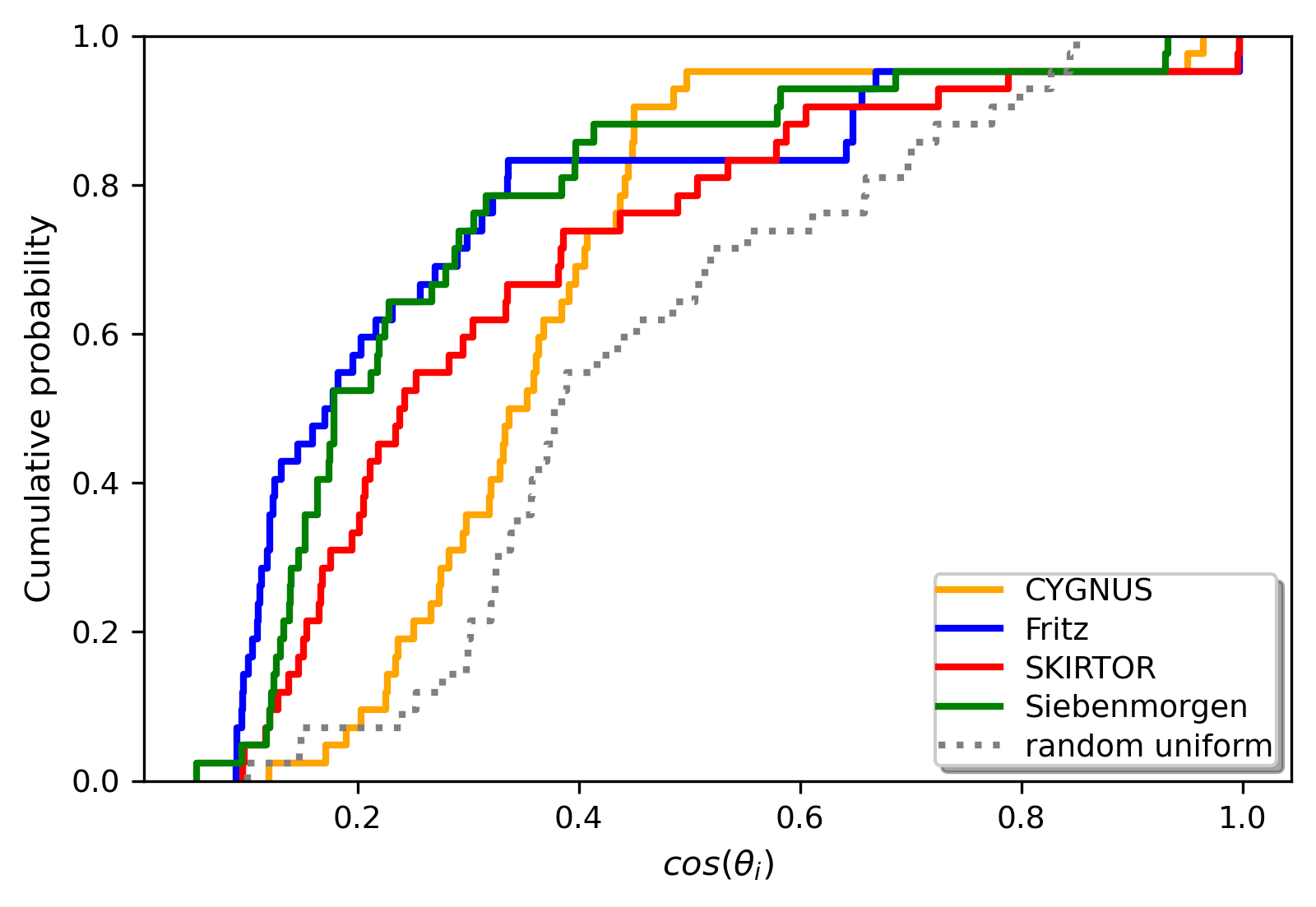}
\caption{Cumulative histograms of the inclinations predicted by the fits with all four combinations of models: CYGNUS tapered torus model (orange), flared torus model of \citealt{fritz06} (blue), SKIRTOR two-phase flared torus model (red), two-phase fluffy dust torus model of \citealt{sie15} (green). We also plot with grey colour the cumulative histogram predicted assuming the inclinations are uniformly distributed at random in $cos (\theta_i)$ in the inclination range $\theta_i=30\degr-85\degr$.}
\label{fig:inclinations}
\end{figure}

\section{Discussion}\label{sec:discussion}

\subsection{Discussion of individual objects}\label{sec:individual}

Before we discuss the wider significance of our results, in this subsection we provide a brief discussion for each of the ULIRGs in our sample. Highlighting first some of the individual characteristics of the ULIRGs constituting the sample, may help us to understand better the properties of the sample as a whole. In the notes below we give information about the optical classification of the system and stage of the merger, as listed in \cite{veilleux02}, \cite{far03} and \cite{far13}. Additionally, we compare the AGN luminosities derived by our best-fitting models with those derived by \cite{nard09} and \cite{veilleux09}, who also quantitatively estimated the AGN luminosities for many ULIRGs presented in this paper.

\vskip 0.25cm
\noindent 
IRAS~00188-0856: This ULIRG is an old merger with a LINER optical spectrum. It is fitted well by all combinations of models, with the starburst dominating the luminosity in all cases. All models predict the presence of an AGN, contributing in the range $4-20$ per cent of the luminosity. \cite{nard09} and \cite{veilleux09} predict similar AGN luminosities more specifically $0.79 \times 10^{12} L_\odot$ and $0.97 \times 10^{12} L_\odot$, respectively. Our best-fitting model, on the contrary, predicts an AGN luminosity of $0.47 \times 10^{12} L_\odot$, as it gives a more starburst-dominated solution.

\vskip 0.25cm
\noindent
IRAS~00397-1312: This ULIRG is an old merger with an H$_{\rm II}$ optical classification. This object is similar to IRAS~08572+3915, which was discussed in more detail in \cite{efs14} and \cite{smart}. It is also one of the objects with deep silicate absorption features and weak PAH emission, which belong to class 3A of \cite{spoon07}. The two smooth models fit this object very well, but the two two-phase models fail to fit the SED, as they cannot produce deep enough silicate absorption features (see also the discussion in \citealt{efs22}). \cite{nard09} predict an AGN luminosity of $5.08 \times 10^{12} L_\odot$, while \cite{veilleux09} suggest a luminosity of $3.40 \times 10^{12} L_\odot$. There is significant difference between these estimates and that of our best-fitting model, which is $19.68 \times 10^{12} L_\odot$. The deep silicate absorption features in this ULIRG and the weak PAH features are due to the fact that it is an AGN-dominated object with its torus viewed almost completely edge-on. This implies an anisotropy correction factor of about 8, according to our best-fitting model. This example may be an indication that in \cite{nard09} and \cite{veilleux09} the anisotropy correction is not taken properly into account.

\vskip 0.25cm
\noindent
IRAS~01003-2238: This ULIRG is an old merger with an H$_{\rm II}$ optical classification. It is not fitted well by any of the models. As discussed in \cite{efs22}, this is probably because we have two AGN in this system, one with its torus viewed face-on, which dominates in the optical/ultraviolet, and one with its torus viewed edge-on, which dominates in the mid-infrared. SMART currently does not provide the option to fit with two AGN, but this option was explored by \cite{efs22} and seems to work very well. There is agreement between the estimates of the AGN luminosity of \cite{nard09} and \cite{veilleux09}, more specifically $0.80 \times 10^{12} L_\odot$ and $1.10 \times 10^{12} L_\odot$, respectively. Our best-fitting model predicts a much higher AGN luminosity of $7.23 \times 10^{12} L_\odot$, but we note that our estimates for this ULIRG are not reliable, as we have not yet developed the functionality to deal with the case of dual AGN.

\vskip 0.25cm
\noindent
IRAS~03158+4227: This object is a pre-merger with a Seyfert 2 optical spectrum. It is another example of a ULIRG that belongs to class 3A of \cite{spoon07}. The CYGNUS model fits the SED very well with a model that combines a highly inclined torus and a starburst. The \cite{fritz06} and SKIRTOR models fail to fit the SED of this object, whereas the \cite{sie15} model gives a more satisfactory fit and predicts a starburst-dominated object with a smaller contribution from an AGN compared to the CYGNUS model. \cite{gowardhan18} studied the molecular outflow of this object in the submillimetre and concluded that AGN alone may not be able to drive the outflow, suggesting the presence of a strong starburst, which is consistent with our best-fitting model. In \cite{nard09} the AGN luminosity is predicted to be $1.92 \times 10^{12} L_\odot$, whereas our best-fitting model predicts a higher AGN luminosity of $6.41 \times 10^{12} L_\odot$, about a factor of three higher. This discrepancy is attributed to the fact that we have a very obscured AGN with a large anisotropy correction and this is what gives the high predicted AGN luminosity.

\vskip 0.25cm
\noindent
IRAS~03521+0028: This object is a pre-merger with a LINER optical spectrum. It is fitted well by all combinations of models, with the starburst dominating the luminosity in all cases. However, a significant contribution from an AGN is predicted, ranging from 7 per cent for the SKIRTOR model to 53 per cent for the CYGNUS model. In \cite{veilleux09} the AGN luminosity is predicted to be $0.83 \times 10^{12} L_\odot$, whereas \cite{nard09} do not predict the presence of an AGN. The prediction of our best-fitting model is very close to that of \cite{veilleux09}, more specifically $0.86 \times 10^{12} L_\odot$.

\vskip 0.25cm
\noindent
IRAS~05189-2524: This object appears to be an advanced merger with a Seyfert 2 optical spectrum. As discussed in \cite{efs22} and \cite{smart}, in this ULIRG there is evidence for polar dust, so we fitted this object with this component enabled. The object is fitted well by all AGN torus models, with the \cite{sie15} and CYGNUS models providing the best fits. According to \cite{nard09}, the estimated AGN luminosity is $0.42 \times 10^{12} L_\odot$, while \cite{veilleux09} suggest a luminosity of $1.03 \times 10^{12} L_\odot$. Our best-fitting model predicts an AGN luminosity similar to \cite{veilleux09}, $1.35 \times 10^{12} L_\odot$.

\vskip 0.25cm
\noindent
IRAS~06035-7102: This ULIRG is a pre-merger with an H$_{\rm II}$ optical classification. It is fitted well by all combinations of models, with the starburst dominating the luminosity in all cases. The contribution of the AGN to the luminosity ranges from 14 to 24 per cent. The AGN luminosity is predicted to be in the range $0.22-0.42 \times 10^{12} L_\odot$. Our results agree with those of \cite{nard09}, who predict an AGN luminosity of $0.35 \times 10^{12} L_\odot$.

\vskip 0.25cm
\noindent
IRAS~06206-6315: This object is a pre-merger with a Seyfert 2 optical spectrum. It is fitted well by all combinations of models. It is a starburst-dominated object with a non-negligible emission from an AGN, which contributes $10-25$ per cent of the luminosity. Our best-fitting models predict an AGN luminosity in the range $0.41-0.43 \times 10^{12} L_\odot$, while \cite{nard09} predict a lower AGN luminosity of $0.09 \times 10^{12} L_\odot$. It is important to note, though, that both methods predict a starburst-dominated object.

\vskip 0.25cm
\noindent
IRAS~07598+6508: This object is an advanced merger with a Seyfert 1 optical spectrum. It is the only ULIRG in the sample which shows the power-law optical continuum that is characteristic of a type 1 AGN. We find, as in \cite{efs22} and \cite{smart}, that an additional polar dust contribution is needed in the near-infrared by all combinations of models. There is very good consistency between the estimates of the AGN luminosity of our models, which are in the range $1.09-1.12 \times 10^{12} L_\odot$. According to \cite{nard09}, the estimated AGN luminosity is $3.54 \times 10^{12} L_\odot$, while \cite{veilleux09} suggest a luminosity of $2.49 \times 10^{12} L_\odot$. A possible explanation for the discrepancy between these estimates and those of our method could be that \cite{nard09} and \cite{veilleux09} do not consider separately the emission from polar dust. In our method, if we add the polar dust luminosity to the AGN luminosity, the estimates will be much closer to those of \cite{nard09} and \cite{veilleux09}.

\vskip 0.25cm
\noindent
IRAS~08311-2459: This object is an early-stage merger with a Seyfert 1 optical spectrum. It is fitted well by all combinations of models. It is predicted to be a starburst-dominated object with a contribution from an AGN, which can range from 12 to 43 per cent.

\vskip 0.25cm
\noindent
IRAS~08572+3915: This object is a pre-merger with a Seyfert 2 optical spectrum. It belongs to class 3A of \cite{spoon07}. \cite{efs14} proposed that the extremely deep silicate absorption feature in this ULIRG and the lack of PAH features are due to the fact that it is an AGN-dominated object with its torus viewed almost completely edge-on. This implies an anisotropy correction factor of about 10, making it possibly the most luminous infrared galaxy in the local ($z<0.1$) Universe. If its torus were viewed from a more face-on inclination, it would clearly appear much more luminous, according to the predictions of the AGN torus models. The anisotropy correction factor attempts to make this correction. We observe that the smooth models fit significantly better the SED compared to the two-phase models. The two-phase models, in particular, cannot produce deep silicate absorption features, like the one observed in IRAS~08572+3915, and predict that the object is starburst-dominated. \cite{nard09} and \cite{veilleux09} predict similar AGN luminosities for this ULIRG, more specifically $1.26 \times 10^{12} L_\odot$ and $1.03 \times 10^{12} L_\odot$, respectively. As in all cases of objects with deep silicate absorption features and weak or no PAH features, our best-fitting model predicts a significantly higher AGN luminosity of $12.79 \times 10^{12} L_\odot$, indicating again that in \cite{nard09} and \cite{veilleux09} the anisotropy correction has not been appropriately considered.

\vskip 0.25cm
\noindent
IRAS~09022-3615: This object has a Seyfert 2 optical spectrum. According to \cite{quin24}, it is a late-stage merger, presenting a tidal tail. It is another example of a ULIRG which appears to be starburst-dominated according to our models, which all fit it very well. Unfortunately, it is an object with very limited optical data, but this does not affect the analysis very much. The contribution from an AGN is predicted to be in the range $5-16$ per cent, with the CYGNUS model predicting the highest value. \cite{nard09} predict an AGN luminosity of $0.15 \times 10^{12} L_\odot$ and this brings their prediction in line with those of our models, which are in the range $0.08-0.27 \times 10^{12} L_\odot$.

\vskip 0.25cm
\noindent
IRAS~10378+1109: This ULIRG is an advanced merger with a LINER optical spectrum. It is an object for which we have good fits with the CYGNUS and \cite{sie15} models but rather poor fits with the \cite{fritz06} and SKIRTOR models. The two models that fit the SED well predict that the AGN may contribute $36-64$ per cent of the luminosity. \cite{nard09} suggest an AGN luminosity of $0.24 \times 10^{12} L_\odot$, while \cite{veilleux09} predict a luminosity of $0.72 \times 10^{12} L_\odot$. Our best-fitting models predict a higher AGN luminosity in the range $0.74-2.44 \times 10^{12} L_\odot$, as they predict a large anisotropy correction, due the obscured AGN in this ULIRG.

\vskip 0.25cm
\noindent
IRAS~10565+2448: This is a late-stage merger with an H$_{\rm II}$/LINER optical spectrum. The CYGNUS, SKIRTOR and \cite{sie15} models predict that the AGN, although weak compared with the starburst, is required for an acceptable SED fit and is contributing $10-27$ per cent of the luminosity. The \cite{fritz06} model, on the contrary, predicts a completely starburst-dominated object. \cite{veilleux09} predict an AGN luminosity of $0.18 \times 10^{12} L_\odot$, while our best-fitting model predicts the value of $0.32 \times 10^{12} L_\odot$. Both studies agree on a starburst-dominated solution.

\vskip 0.25cm
\noindent
IRAS~11095-0238: This is an old merger with a LINER optical spectrum. It is another ULIRG with a deep silicate absorption feature. The near-infrared continuum is not as strong as in the cases of IRAS~08572+3915 and IRAS~00397-1312. The SED is fitted well by the CYGNUS and \cite{sie15} models, predicting a contribution by an obscured AGN in the mid-infrared. By contrast, the \cite{fritz06} and SKIRTOR models predict a completely starburst-dominated object. \cite{nard09} and \cite{veilleux09} predict similar AGN luminosities, more specifically $0.93 \times 10^{12} L_\odot$ and $0.75 \times 10^{12} L_\odot$, respectively. There is significant variation between these estimates and that of our best-fitting model, which is $6.05\times 10^{12} L_\odot$. This discrepancy is observed in all cases of spectra with deep silicate absorption features and can be attributed to the application of the anisotropy correction.

\vskip 0.25cm
\noindent
IRAS~12071-0444: This object appears to be an advanced merger with a Seyfert 2 optical spectrum. We have satisfactory fits with the \cite{sie15} and CYGNUS models but rather poor fits with the \cite{fritz06} and SKIRTOR models. The two models that provide satisfactory fits predict an AGN-dominated object with an AGN fraction in the range $61-64$ per cent. As therefore expected, there is consistency between the predicted AGN luminosity of our best-fitting models, which is in the range $2.25-2.51 \times 10^{12} L_\odot$. In contrast, \cite{nard09} give a lower estimate of $0.90 \times 10^{12} L_\odot$. \cite{veilleux09} align more closely with our method, as they suggest a luminosity of $1.60 \times 10^{12} L_\odot$.

\vskip 0.25cm
\noindent
IRAS~13120-5453: This ULIRG has an H$_{\rm II}$ optical spectrum and appears to be a post-merger single nucleus system. \cite{privon17} observed IRAS~13120-5453 with the Atacama Large Millimeter Array (ALMA) and found it does not have the characteristics of a compact obscured nucleus (CON). The CYGNUS and \cite{sie15} models fit well the SED and predict a starburst-dominated object with a small contribution from an AGN in the range $8-12$ per cent. \cite{nard09} also predict a starburst-dominated object with no sign of an AGN, while our best-fitting model predicts a low AGN luminosity of $0.14 \times 10^{12} L_\odot$.

\vskip 0.25cm
\noindent
IRAS~13451+1232: This object is a pre-merger with a Seyfert 2 optical spectrum. It is the third ULIRG in the sample where we added a contribution from polar dust. The CYGNUS and \cite{sie15} models fit the SED very well. The \cite{fritz06} and SKIRTOR models do not fit very well the $10-50~\mu m$ SED. There is perfect agreement between the estimates of the AGN luminosity by \cite{nard09} and \cite{veilleux09}, which are $1.15 \times 10^{12} L_\odot$ and $1.14 \times 10^{12} L_\odot$, respectively. The predictions of our best-fitting models deviate from those estimates, as they are in the range $2.73-3.73 \times 10^{12} L_\odot$.

\vskip 0.25cm
\noindent
IRAS~14348-1447: This is a pre-merger with a LINER optical spectrum. The CYGNUS and \cite{sie15} provide good fits and predict a starburst-dominated object. Whilst this object does not contain a luminous AGN, the CYGNUS and \cite{sie15} models predict that the AGN contributes 19 and 28 per cent of the luminosity, respectively. \cite{nard09} predict an AGN luminosity of $0.08 \times 10^{12} L_\odot$, whereas \cite{veilleux09} give an estimate of $0.40 \times 10^{12} L_\odot$. The estimate of our best-fitting model is $0.46 \times 10^{12} L_\odot$, therefore perfectly consistent with the estimate of \cite{veilleux09}.

\vskip 0.25cm
\noindent
IRAS~14378-3651: This is a late-stage merger with a LINER/Seyfert 2 optical classification for which we have good fits with the CYGNUS and \cite{sie15} models, which predict an AGN fraction of 31 and 27 per cent, respectively. \cite{nard09} predict a negligible AGN luminosity of $0.01 \times 10^{12} L_\odot$. Our best-fitting models agree on a starburst-dominated solution, but predict an AGN luminosity in the range $0.35-0.46 \times 10^{12} L_\odot$.

\vskip 0.25cm
\noindent
IRAS~15250+3609: This is an advanced merger with a LINER optical classification. It belongs to class 3A of \cite{spoon07}. The CYGNUS, \cite{fritz06} and \cite{sie15} models fit the SED very well, with a deeply obscured AGN dominating in the mid-infrared. The SKIRTOR model predicts a totally starburst-dominated object, but the fit is unsatisfactory. There is very good agreement between the estimates of \cite{nard09} and \cite{veilleux09}, who predict an AGN luminosity of $0.54 \times 10^{12} L_\odot$ and $0.51 \times 10^{12} L_\odot$, respectively. In contrast, our best-fitting model predicts an AGN luminosity of $2.50 \times 10^{12} L_\odot$, due to the large anisotropy correction. This difference is seen in all cases of deeply obscured AGN.

\vskip 0.25cm
\noindent
IRAS~15462-0450: This object appears to be an advanced merger with a Seyfert 2 optical spectrum. The \cite{sie15} model is the best-fitting model of this ULIRG, but all models give satisfactory fits. According to the best-fitting model, the AGN contributes 44 per cent of the luminosity. \cite{nard09} predict an AGN luminosity of $0.33 \times 10^{12} L_\odot$, while \cite{veilleux09} give an estimate of $0.68 \times 10^{12} L_\odot$. A slightly higher value than that predicted by \cite{veilleux09} is indicated by our best-fitting model, more specifically $0.80 \times 10^{12} L_\odot$.

\vskip 0.25cm
\noindent
IRAS~16090-0139: This object is an early-stage merger with a LINER optical spectrum. The CYGNUS model fits better the SED of this ULIRG, but all models provide satisfactory fits. The \cite{fritz06} and SKIRTOR models predict a totally starburst-dominated object, while the CYGNUS and \cite{sie15} models predict an AGN fraction of 21 and 8 per cent, respectively. The predicted AGN luminosity of our best-fitting model perfectly agrees with that of \cite{nard09}, more specifically $0.73 \times 10^{12} L_\odot$ and $0.72 \times 10^{12} L_\odot$, respectively. In contrast, \cite{veilleux09} estimate a higher AGN luminosity of $1.17 \times 10^{12} L_\odot$.

\vskip 0.25cm
\noindent
IRAS~17208-0014: This ULIRG is a late-stage merger with an H$_{\rm II}$ optical classification. All model combinations fit well the SED and predict a starburst-dominated object with a small contribution from an AGN in the range $1-9$ per cent (the highest value given by the \citealt{sie15} model). We therefore find evidence from the SED fitting that this buried nucleus, studied at high resolution with ALMA by \cite{baba22}, is powered mainly by star formation. Our best-fitting models predict an AGN luminosity in the range $0.09-0.20 \times 10^{12} L_\odot$, \cite{nard09} predict a completely starburst-dominated object with no sign of an AGN, while \cite{veilleux09} indicate a low AGN luminosity of $0.30 \times 10^{12} L_\odot$.

\vskip 0.25cm
\noindent
IRAS~19254-7245: This ULIRG is a pre-merger with a Seyfert 2 optical classification. It is the system also known as the Superantennae. As in \cite{efs22}, the CYGNUS model predicts some contribution in the submillimetre from the host galaxy, which may be related to the presence of extended tidal tails. All model combinations predict a contribution from an AGN in the range $8-70$ per cent. The CYGNUS model predicts the strongest contribution of 70 per cent, while the model of \cite{sie15} predicts a contribution of 37 per cent. The \cite{fritz06} and SKIRTOR models predict only a small contribution of 8 per cent, but they do not fit well that part of the spectrum. Our best-fitting models predict an AGN luminosity in the range $0.58-1.70 \times 10^{12} L_\odot$. \cite{nard09} predict a lower AGN luminosity of $0.32 \times 10^{12} L_\odot$.

\vskip 0.25cm
\noindent
IRAS~19297-0406: This ULIRG has an H$_{\rm II}$ optical spectrum and appears to be a merger of four galaxies in the HST/NICMOS image. The CYGNUS and \cite{sie15} models fit well the SED and predict an AGN fraction of 11 and 28 per cent, respectively. \cite{nard09} predict a negligible AGN luminosity of $0.03 \times 10^{12} L_\odot$. Our best-fitting models predict higher values in the range $0.30-0.78 \times 10^{12} L_\odot$.

\vskip 0.25cm
\noindent
IRAS~20087-0308: This is an advanced merger with a LINER optical spectrum. All model combinations fit well the SED of this ULIRG and predict a starburst-dominated object but also the presence of an AGN, contributing in the range $2-12$ per cent of the luminosity. \cite{nard09} also predict that the starburst dominates the luminosity of this ULIRG, with a low AGN luminosity of $0.08 \times 10^{12} L_\odot$. Similarly, our models predict an AGN luminosity in the range $0.05-0.32 \times 10^{12} L_\odot$.

\vskip 0.25cm
\noindent
IRAS~20100-4156: This is a pre-merger with an H$_{\rm II}$ optical spectrum. The CYGNUS torus model fits better the SED of this ULIRG, but all model combinations give a good fit. \cite{gowardhan18} studied the molecular outflow of this object in the submillimetre and suggested the presence of a strong starburst along with the AGN to drive the outflow, which is in agreement with our models. Our best-fitting model predicts an AGN luminosity of $1.35 \times 10^{12} L_\odot$, while \cite{nard09} predict a value approximately a factor of two lower at $0.64 \times 10^{12} L_\odot$. This discrepancy is due to the large anisotropy correction, as in all cases of deeply obscured AGN.

\vskip 0.25cm
\noindent
IRAS~20414-1651: This object appears to be an advanced merger with an H$_{\rm II}$ optical spectrum. All model combinations predict a starburst-dominated object but also the presence of an AGN, contributing in the range $4-21$ per cent of the luminosity. \cite{nard09} predict a completely starburst-dominated object with no sign of an AGN, whereas \cite{veilleux09} predict a low AGN luminosity of $0.21 \times 10^{12} L_\odot$. Our best-fitting models come closer to the estimate of \cite{veilleux09}, as they predict a value in the range $0.22-0.28 \times 10^{12} L_\odot$.

\vskip 0.25cm
\noindent
IRAS~20551-4250: This is a late-stage merger with an H$_{\rm II}$ optical spectrum. The CYGNUS torus model is the best-fitting model of this ULIRG. The \cite{sie15} model provides a satisfactory fit, whereas the fits with the \cite{fritz06} and SKIRTOR models are poor. Both the CYGNUS and \cite{sie15} models predict an AGN-dominated object with an AGN fraction of 55 and 33 per cent, respectively. \cite{sani12}, by analyzing the SED of this object including the Spitzer IRS spectrum, find that the near-infrared spectrum is typical of a galaxy experiencing a very intense starburst, but a highly obscured AGN is identified beyond 5 $\mu m$, which possibly dominates the mid-infrared energy output of the system. They also note that at longer wavelengths star formation is again the main driver of the global spectral shape and features. \cite{yuan10} classified IRAS~20551-4250 as a starburst-AGN composite or H$_{\rm II}$ system, depending on the emission lines used. The presence of a buried AGN, which could explain $20-60$ per cent of the bolometric luminosity, has been suggested, in addition to starburst activity, by \cite{iman10,iman11} and \cite{veilleux13}. The conclusions of all of these studies are consistent with our model for this object. According to \cite{nard09}, the estimated AGN luminosity is $0.26 \times 10^{12} L_\odot$, while our best-fitting model predicts a luminosity of $0.87 \times 10^{12} L_\odot$.

\vskip 0.25cm
\noindent
IRAS~22491-1808: This is a pre-merger with an H$_{\rm II}$ optical spectrum. The CYGNUS and \cite{sie15} are the best-fitting models of this ULIRG. The fit with the SKIRTOR model is very poor.  \cite{lutz99} find that the mid-infrared spectrum is consistent with a starburst. The CYGNUS and \cite{sie15} models predict a significant AGN contribution of 57 and 36 per cent of the luminosity, respectively. Our conclusions and those of \cite{lutz99} may differ because in IRAS~22491-1808 we have a very obscured AGN with a large anisotropy correction and this is what gives the high-predicted AGN contribution. \cite{nard09} agree with \cite{lutz99} on a completely starburst-dominated object. \cite{veilleux09} estimate an AGN luminosity of $0.28 \times 10^{12} L_\odot$. Our best-fitting model predicts a luminosity of $1.20 \times 10^{12} L_\odot$.

\vskip 0.25cm
\noindent
IRAS~23128-5919: This is a late-stage merger with an H$_{\rm II}$/Seyfert 2 optical classification for which we have good fits with the CYGNUS and \cite{sie15} models, which predict an AGN fraction of 29 and 38 per cent, respectively. \cite{nard09} predict a low AGN luminosity of $0.03 \times 10^{12} L_\odot$, whereas our best-fitting model predicts a higher luminosity of $0.31 \times 10^{12} L_\odot$.

\vskip 0.25cm
\noindent
IRAS~23230-6926: This is an advanced merger with a LINER optical spectrum. The CYGNUS is the best-fitting model of this ULIRG and predicts a significant AGN fraction of 50 per cent. The model of \cite{sie15} provides a satisfactory fit and predicts an AGN fraction of 36 per cent. The \cite{fritz06} and SKIRTOR models give very poor fits of the SED. \cite{nard09} predict an AGN luminosity of $0.09 \times 10^{12} L_\odot$, whereas our best-fitting model gives an estimate of $1.33 \times 10^{12} L_\odot$. Our conclusions may differ from those of \cite{nard09} because our method predicts a large anisotropy correction that arises from the presence of an obscured AGN in this ULIRG.

\vskip 0.25cm
\noindent
IRAS~23253-5415: This is an advanced merger with an H$_{\rm II}$ optical spectrum. We have good fits with the \cite{sie15} and CYGNUS models but rather poor fits with the \cite{fritz06} and SKIRTOR models. The two models that fit the SED well predict that the AGN may contribute $38-42$ per cent of the luminosity, with the CYGNUS model predicting the lower value. There is therefore consistency between the estimates of the AGN luminosity of the CYGNUS and \cite{sie15} models, which are in the range $1.20-1.37 \times 10^{12} L_\odot$. As in most cases of obscured AGN with large anisotropy corrections, \cite{nard09} predict a much lower AGN luminosity, $0.15 \times 10^{12} L_\odot$. 

\vskip 0.25cm
\noindent
IRAS~23365+3604: This is a late-stage merger with an H$_{\rm II}$/LINER optical spectrum. The best fit is provided by the CYGNUS model, while the \cite{sie15} model gives a satisfactory fit. The \cite{fritz06} and SKIRTOR models do not fit well the SED of this ULIRG. The best-fitting CYGNUS model predicts a significant AGN fraction of 72 per cent, whereas the model of \cite{sie15} predicts that the AGN contributes 36 per cent of the luminosity. \cite{romero12} found that IRAS~23365+3604 has a composite radio spectrum, which indicates activity from an AGN but also a population of supernovae. \cite{nard09} predict a low AGN luminosity of $0.11 \times 10^{12} L_\odot$. The large anisotropy correction predicted by our best-fitting model is a result of the very obscured AGN in IRAS~23365+3604 and this gives the high predicted AGN luminosity of $2.50 \times 10^{12} L_\odot$.

\vskip 0.25cm
\noindent
Mrk~1014: This is a Seyfert 1 galaxy, considered to be in a post-merger. It is the other object where \cite{efs22} concluded that there is evidence for a dual AGN. We note, as well, that it is difficult to fit with a single torus model with any of the model combinations. For this object there is significant variation between the estimates of the AGN luminosity of \cite{nard09} and \cite{veilleux09}, more specifically $0.92 \times 10^{12} L_\odot$ and $3.10 \times 10^{12} L_\odot$, respectively. Our best-fitting model predicts the value of $1.13 \times 10^{12} L_\odot$, but it is noted that we do not have satisfactory fits for this object, since we currently cannot fit with a dual AGN.

\vskip 0.25cm
\noindent
UGC~5101: The optical spectrum of this object is a combination of a LINER and a Seyfert 2. The morphology has been interpreted as a late-stage merger between two large spiral galaxies \citep{sanders88}. This ULIRG is not fitted very well by any of the model combinations, but all of them predict it contains both a starburst and an AGN, which is in line with the estimates of \cite{far03}. Near-infrared spectroscopy by \cite{iman01} shows that this object is likely to contain a heavily obscured buried AGN. Our models predict an AGN luminosity in the range $0.03-0.22 \times 10^{12} L_\odot$. These estimates are fairly consistent with those of \cite{nard09} and \cite{veilleux09}, which are $0.14 \times 10^{12} L_\odot$ and $0.34 \times 10^{12} L_\odot$, respectively.

\vskip 0.25cm
\noindent
Mrk~231: This object is an advanced merger with a Seyfert 1 optical spectrum, one of very few in this sample. All models provide good fits, except from the SKIRTOR model. The infrared luminosity is thought to be powered by both starburst and AGN activity \citep{cutri84,condon91} and this is consistent with our SED fitting results. Our results also confirm the studies of \cite{far03} and \cite{lons03}. Our best-fitting models predict an AGN luminosity in the range $1.32-3.43 \times 10^{12} L_\odot$, \cite{nard09} indicate the value of $1.18 \times 10^{12} L_\odot$, while \cite{veilleux09} suggest a luminosity of $2.45 \times 10^{12} L_\odot$.

\vskip 0.25cm
\noindent
Mrk~273: This ULIRG is an advanced merger with a Seyfert 2 optical spectrum. All models fit this object very well, with a solution which is starburst-dominated with a small contribution from an obscured AGN in the mid-infrared in the range $8-24$ per cent. Our best-fitting models predict an AGN luminosity in the range $0.25-0.34 \times 10^{12} L_\odot$. \cite{nard09} predict a negligible AGN luminosity of $0.10 \times 10^{12} L_\odot$, while \cite{veilleux09} predict a higher AGN luminosity of $0.70 \times 10^{12} L_\odot$.

\vskip 0.25cm
\noindent
Mrk~463: This ULIRG is a pre-merger with a Seyfert 2 optical spectrum. The fits with the CYGNUS, \cite{fritz06} and SKIRTOR models are similar, predicting an object which is AGN-dominated in the mid-infrared and starburst-dominated in the far-infrared and submillimetre. The \cite{sie15} model, by contrast, predicts a totally AGN-dominated object with a small contribution from the starburst in the submillimetre. This is the only ULIRG in the sample where the \cite{sie15} model shows this extreme behaviour. 

\vskip 0.25cm
\noindent
Arp~220: This galaxy is a pre-merger with a LINER/Seyfert 2 optical spectrum. It is not fitted well by any of the models. The models that come closer are the \cite{sie15} and CYGNUS models. Both models predict a contribution from an AGN in the mid-infrared in the range $35-82$ per cent. The \cite{fritz06} and SKIRTOR models predict an object completely dominated by a starburst. Our best-fitting model predicts an AGN luminosity of $0.47 \times 10^{12} L_\odot$. The AGN luminosities reported by \cite{nard09} and \cite{veilleux09} fall slightly below, more specifically $0.28 \times 10^{12} L_\odot$ and $0.30 \times 10^{12} L_\odot$, respectively.

\vskip 0.25cm
\noindent
NGC~6240: This is a late-stage merger with a LINER optical spectrum. CYGNUS is the best-fitting model of this object and predicts an AGN fraction of 54 per cent. The model of \cite{sie15} gives a satisfactory fit and predicts that the AGN contributes 37 per cent of the luminosity. The SKIRTOR and \cite{fritz06} models give starburst-dominated solutions with a very small contribution from an AGN, but the fits are poorer. We note, however, that \cite{kollats20} found evidence for three AGN in NGC~6240. \cite{nard09} predict a negligible AGN luminosity of $0.05 \times 10^{12} L_\odot$, whereas \cite{veilleux09} predict a higher luminosity of $0.18 \times 10^{12} L_\odot$. Our best-fitting models predict a higher AGN luminosity in the range $0.32-0.59 \times 10^{12} L_\odot$. This difference between our approach and those of \cite{nard09} and \cite{veilleux09} arises once more from the large anisotropy correction, due to the presence of an obscured AGN in NGC~6240.

\subsection{Wider significance of our results}

Our results suggest that the SFR of a ULIRG can be robustly estimated with SED fitting, independently of the assumed AGN torus model. The exception is the model of \cite{sie15} which, because of its assumption of fluffy grains, has a higher emissivity in the far-infrared and therefore sometimes can predict a lower SFR. In this sample we have a single case where we see this effect, in particular Mrk~463, where the \cite{sie15} model predicts a totally AGN-dominated object. This may be due to the fact that this object has no traces of PAH features in its Spitzer spectrum. However, there are other ULIRGS with weak or no PAH features like IRAS~05189-2524 and Mrk~231, where the fit with the \cite{sie15} model does not give a completely AGN-dominated solution. Our results therefore support the idea that the far-infrared and submillimetre emission of ULIRGs is nearly always dominated by starburst activity. If we consider these objects as analogs for submillimere galaxies, we can conclude from our analysis that we still need to explain the tension between the observational discovery of high redshift submillimetre galaxies with SFRs of thousands of solar masses per year and galaxy formation models that predict that such systems should not exist (e.g. \citealt{lacey16}). The usual interpretation for this tension is that the IMF of starbursts is top-heavy. Our analysis cannot shed light on this issue, as our results are not sensitive to the IMF.

Although the SFR of the starburst in the HERUS ULIRGs is predicted to be insensitive to the assumed torus model, we find that the AGN torus luminosities are, in most cases, underestimated by an order of magnitude or more, if an AGN torus model other than the smooth CYGNUS tapered disc is assumed. \cite{pap25} carried out a similar analysis for 200 galaxies at $z=2$ in the ELAIS-N1 field, which was one of the fields studied by HELP. This sample is selected at 250 $\mu m$, ensuring the presence of a starburst. Although the galaxies in the sample of \cite{pap25} do not include mid-infrared spectroscopy, the authors find that there is similarly very large uncertainty in the AGN torus luminosity and AGN fraction.

In Fig. \ref{fig:inclinations} we plot cumulative histograms of the inclinations predicted by the fits with all four combinations of models. We also plot with grey colour the cumulative histogram predicted assuming the inclinations are uniformly distributed at random in $cos(\theta_i)$ in the inclination range $\theta_i=30\degr-85\degr$. This would be the expected distribution assuming all ULIRGs have a similar AGN torus, which is oriented at random. One could argue that more face-on AGN would be expected, because they are more luminous in the 30-60 $\mu m$ range and therefore are more likely to be included in a flux-limited sample. However, in most ULIRGS in our sample we find from the fits that the 60 $\mu m$ emission is dominated by the starburst, which is emitting essentially isotropically.

We observe that all models predict higher inclinations than what is expected from a uniform distribution. The inclination of the torus is very strongly constrained by the depth of the silicate absorption feature. A deeper absorption feature gives a higher inclination. There are a number of possible reasons for this: First of all, the torus models that are currently available in the literature, although successful in many respects, still may not capture the true geometry of the torus and this leads to overestimated inclinations. Another possible reason is that the dust model used for the radiative transfer models of the torus is not appropriate for this environment. All models, except the \cite{sie15} model, assume a dust mixture appropriate for the general interstellar medium, but the dust mixture in the AGN environment may be somewhat different with preference for smaller or larger grains. \cite{gonz23} explored torus models with a number of different dust models and found a preference for larger grains. A third reason may be that we see here evidence that some of the silicate absorption in ULIRGs is not related to the torus itself but to absorption either by circumnuclear star formation regions or by the host galaxy. This would also skew the distribution of predicted torus inclinations towards higher values. A possible method to resolve this issue is to compare the predicted sizes of the AGN tori with observations in the (sub)millimetre with ALMA or with the Northern Extended Millimeter Array (NOEMA). The size of the torus depends very much on the model parameters but also on the intrinsic AGN luminosity, which in turn depends on the inclination and the associated anisotropy correction. We plan to pursue such studies in future work.

Our finding that the smooth tapered disc CYGNUS model fits better the SEDs of ULIRGs may provide hints about the formation of AGN tori. Traditionally it has been difficult to explain theoretically how tori can maintain their geometrical thickness and various mechanisms have been proposed \citep{krol88}. The advantage of the tapered disc model is that its geometrical thickness, especially in the outer parts of the torus, is much smaller. One may therefore conclude that, even though flared discs may exist in AGN, they last for only a limited time and then transition to something like the tapered disc geometry, which is more long-lived. This may explain why we find most of the ULIRGs are better fitted with tapered discs.

\section{Conclusions}\label{sec:conclusion}

We explored further the physical properties of the HERUS sample of 42 local ULIRGs at $z<0.27$ with the recently developed SED fitting code SMART. SMART fits SEDs exclusively with radiative transfer models and currently allows fitting with four different AGN torus models, namely the CYGNUS model of \cite{efstathiou95}, which assumes a smooth tapered disc geometry, the smooth flared disc model of \cite{fritz06}, the two-phase flared disc model SKIRTOR of \cite{stal16} and the two-phase model of \cite{sie15}, which assumes fluffy grains and an anisotropic sphere geometry. Additionally, a starburst and a spheroidal host galaxy model, as well as polar dust, can be included in the fitting. For three of the ULIRGs in the HERUS sample (IRAS~05189-2524, IRAS~07598+6508 and IRAS~13451+1232) we added polar dust as an additional component (see also \citealt{smart}), as this provided a better fit.

Our main conclusions can be summarized as follows:

\begin{enumerate}

\item First of all, we find that the starburst luminosity and the SFR averaged over the age of the starburst, which is self-consistently determined from the fit, are robustly estimated for all combinations of models. This may be partly due to the fact that we use a single starburst model, whereas we explore four different AGN torus models. However, it is important to note that we do not expect to have too much variation in starburst models, as they peak at far-infrared wavelengths, where the emission is more or less isotropic. We plan to incorporate more starburst models in SMART in future work.

\vskip 0.5cm

\item The estimates of stellar mass are also found to be independent of the assumed radiative transfer model. This is also as expected, since the AGN contributes very little in the optical and near-infrared, which mainly determine the estimates of stellar mass. Again, we note that we only use a single model for the spheroidal component. However, we consider very unlikely that the uniformity in the estimate of the stellar mass is due to this.

\vskip 0.5cm

\item We find that the smooth AGN torus model of \cite{efstathiou95} provides overall better fits compared to the other AGN torus models and is therefore the obscurer geometry that best approximates the torus in the galaxies in the HERUS sample. This conclusion was also reached by \cite{efs22}.

\vskip 0.5cm

\item An important result of our study is that we find significant differences in the anisotropy-corrected AGN and total luminosities, as well as the AGN fraction. The CYGNUS model usually predicts the highest luminosities and AGN fraction, whereas the predictions from the other AGN torus models are up to a factor of $1-2$ orders of magnitude lower.  

\vskip 0.5cm

\item In almost half of the galaxies our estimates of the AGN luminosity are higher than the values derived by \cite{nard09}. In most cases, we attribute this systematic discrepancy to the fact that we take into account the anisotropy correction of the AGN luminosity. The comparison with the AGN luminosities reported by \cite{veilleux09} is more interesting, as we observe more consistency between our estimates and those of \cite{veilleux09}, but again there are cases where we predict much higher AGN luminosities.

\end{enumerate}

\section*{Acknowledgements}

We would like to thank an anonymous referee for useful comments and suggestions. CV and AE acknowledge support from the projects CYGNUS (contract number 4000126896) and CYGNUS+ (contract number 4000139319) funded by the European Space Agency.

\section*{Data Availability}

The data underlying this article are available in the article, in public databases like Cornell Atlas of Spitzer/Infrared Spectrograph Sources (CASSIS) or are publicly available in the literature.




\appendix

\section{Plots and other SED fitting results discussed in the paper}

In this Appendix we give the plots of all of the SED fits with the four different model combinations for the HERUS sample. In addition, we provide tables with extracted physical quantities of the fits with the four model combinations for the HERUS sample. 

\begin{figure*}
	\begin{center}
      {\includegraphics[width=51mm]{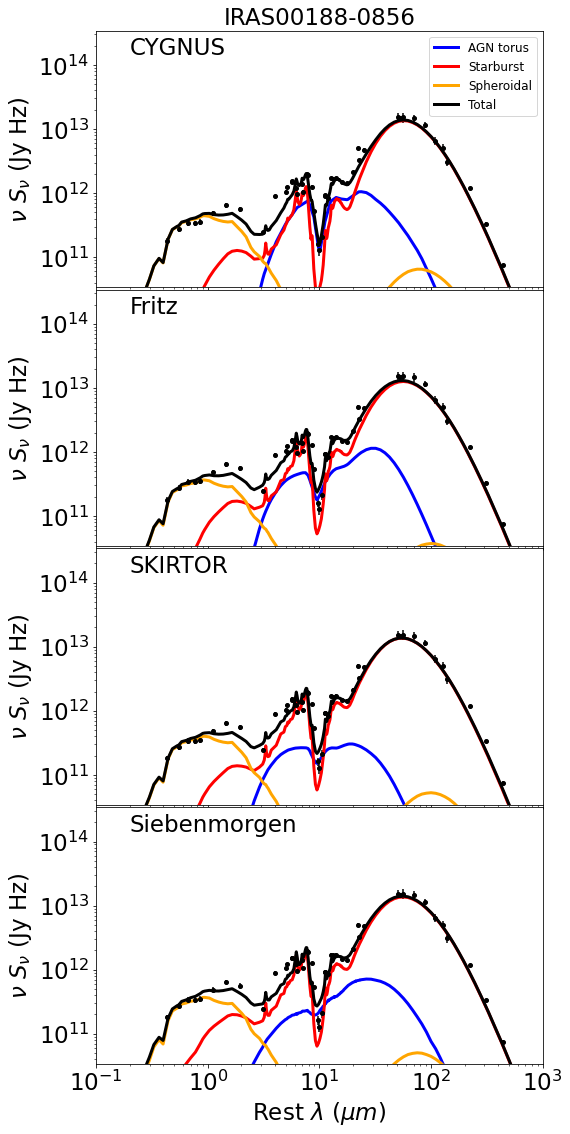}}    \hspace{20pt}
      {\includegraphics[width=51mm]{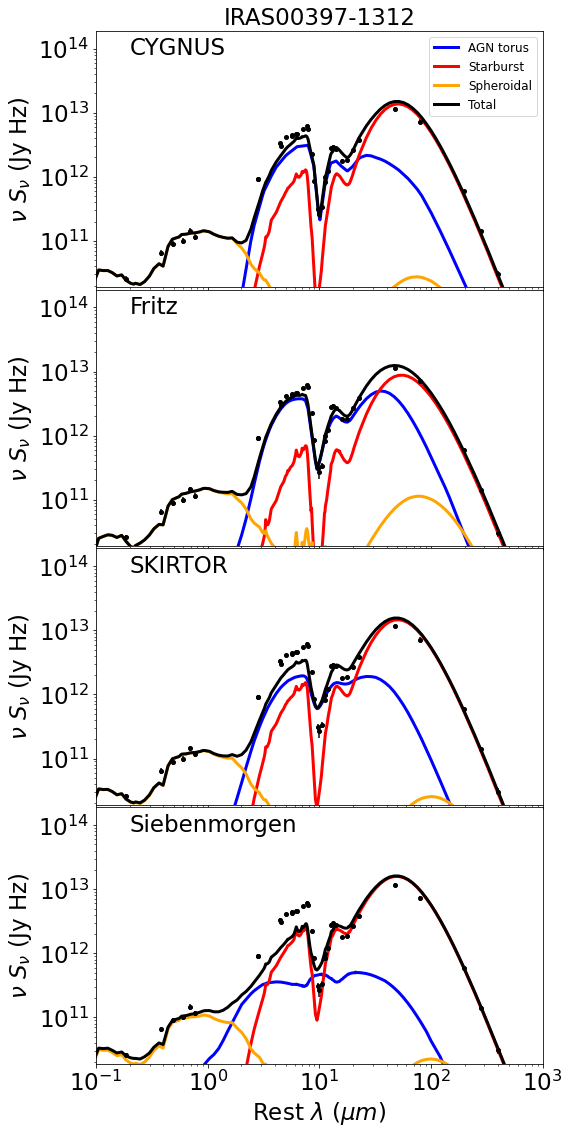}}    \hspace{20pt}
      {\includegraphics[width=51mm]{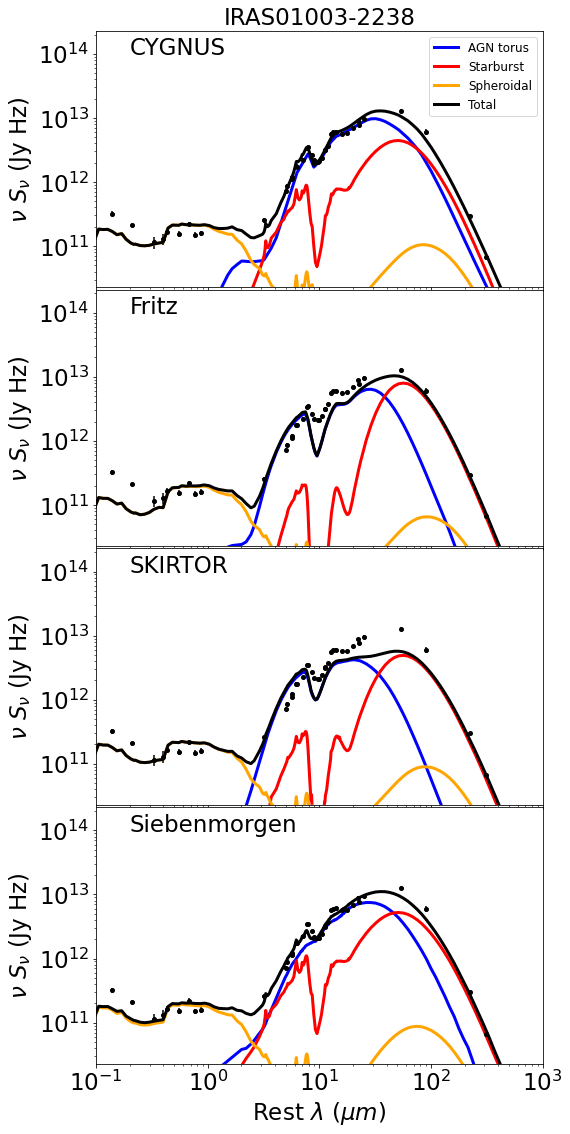}} 
      $$
      $$
      $$
      $$
      {\includegraphics[width=51mm]{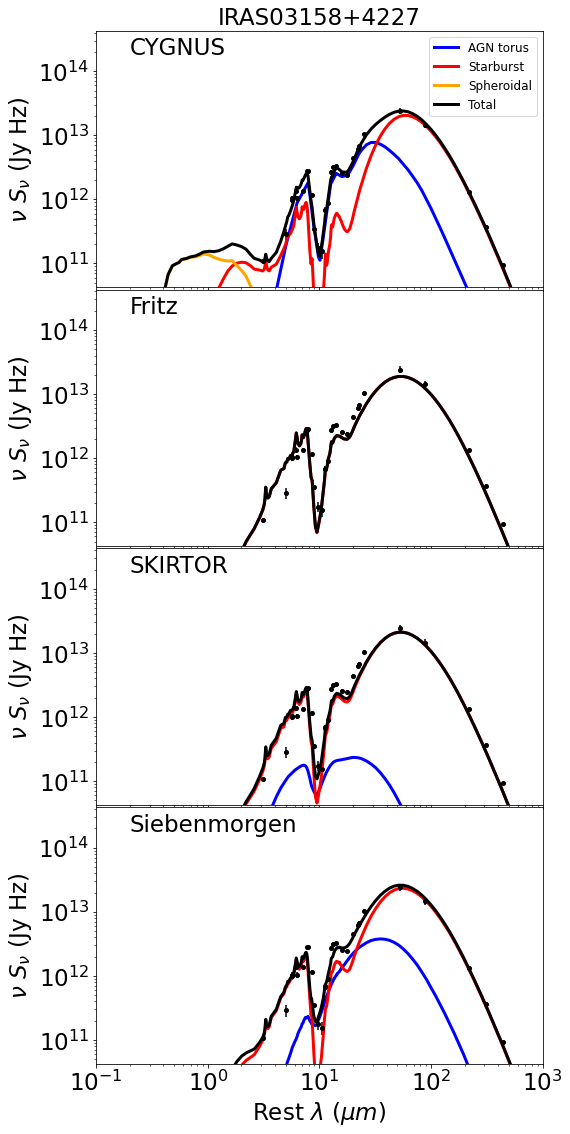}} \hspace{20pt}
      {\includegraphics[width=51mm]{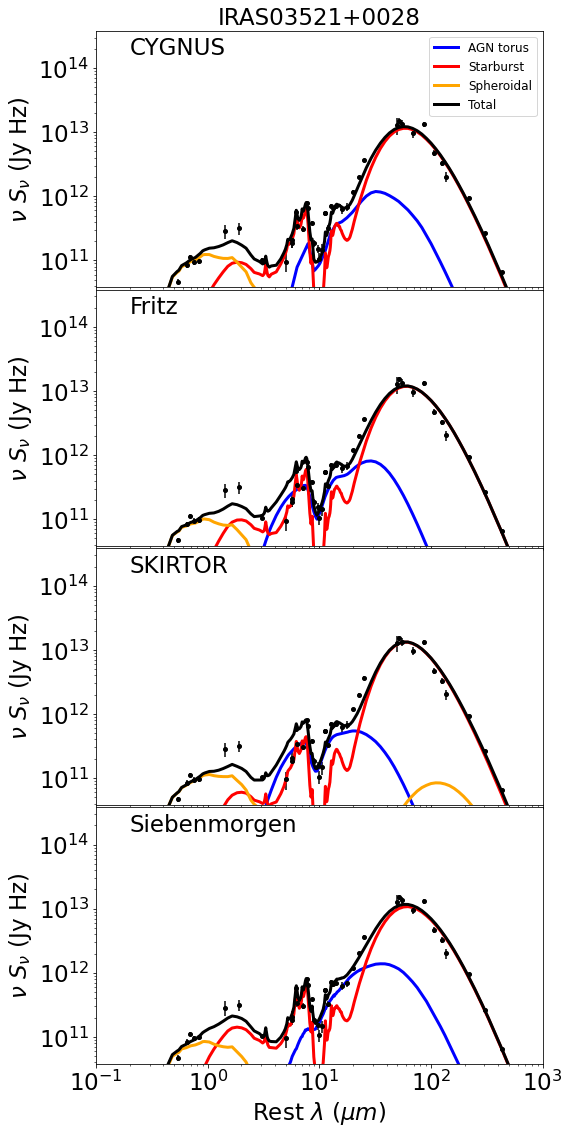}} \hspace{20pt}     
      {\includegraphics[width=51mm]{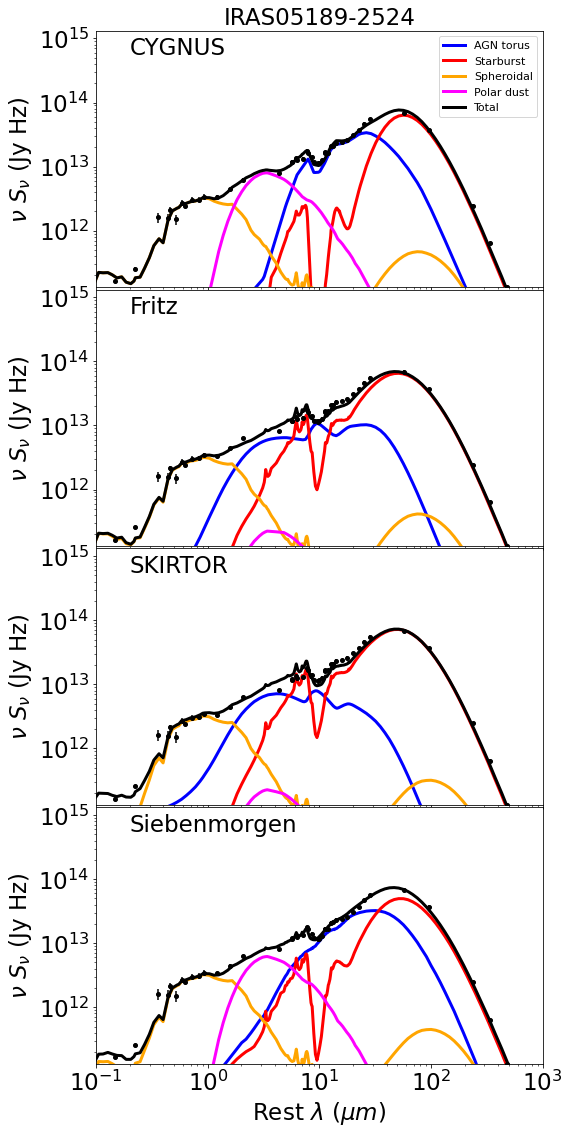}} 
      \\
\vspace{10pt}
\caption{Comparison SED fit plots of the first six objects from the list of the HERUS sample: AGN torus (blue), starburst (red), spheroidal host (orange), polar dust (magenta), total (black). In each comparison plot the top row shows fits with the CYGNUS tapered torus model, the second row shows fits with the CYGNUS tapered torus model replaced by the Fritz et al. (2006) flared torus model, the third row replaces the CYGNUS tapered torus model with the SKIRTOR two-phase flared torus model, while the bottom row replaces the CYGNUS tapered torus model with the Siebenmorgen et al. (2015) two-phase fluffy dust torus model.}
     \label{fig:HERUS-resultsA}
\end{center}
\end{figure*}

\begin{figure*}
	\begin{center}
      {\includegraphics[width=52mm]{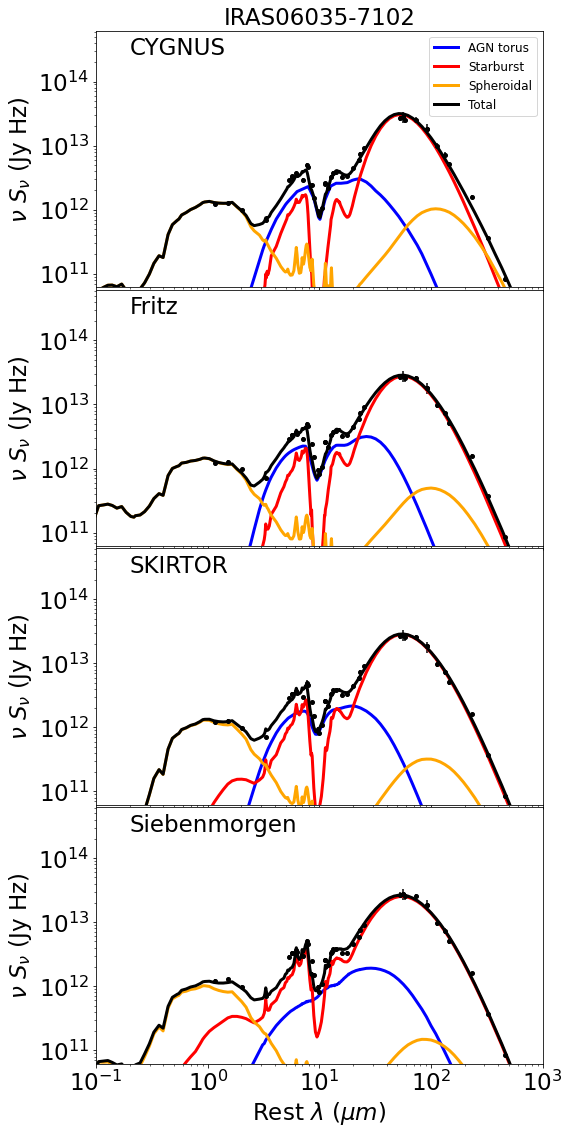}} \hspace{20pt}
      {\includegraphics[width=51mm]{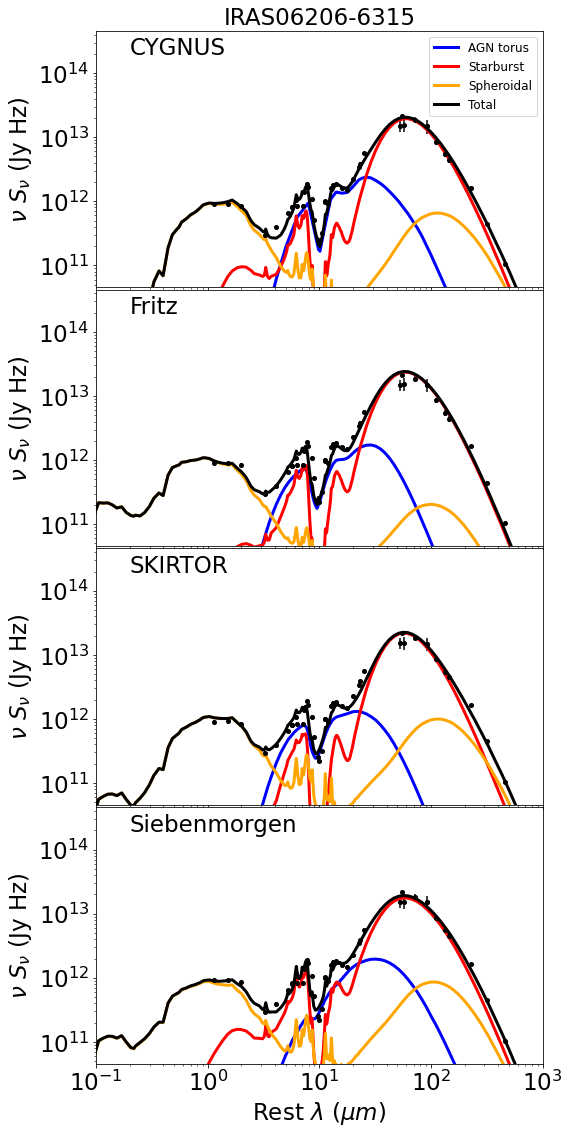}} \hspace{20pt}
      {\includegraphics[width=51mm]{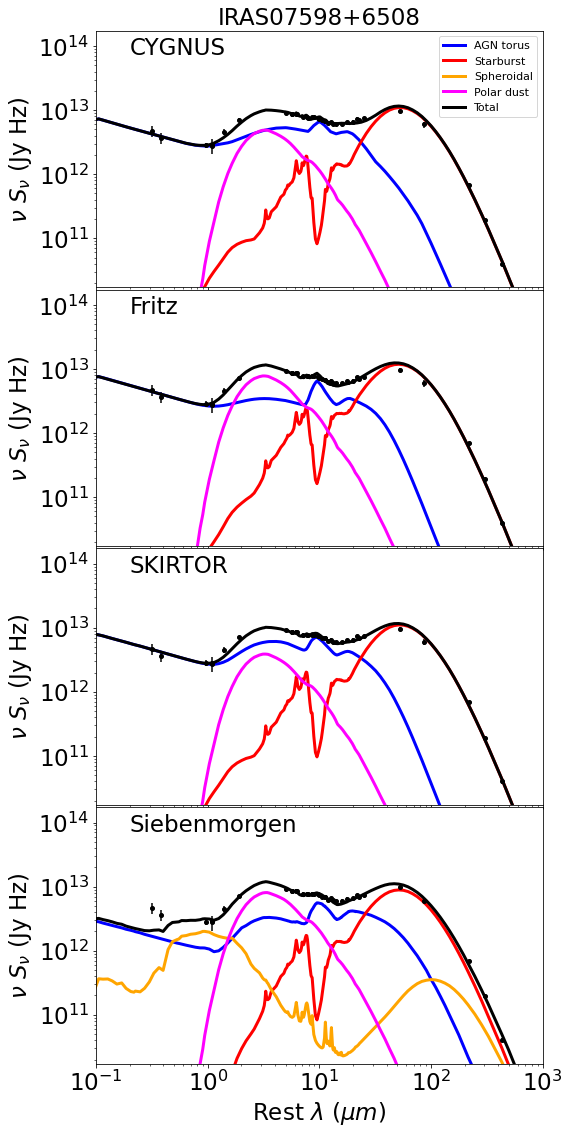}}
      $$
      $$
      $$
      $$      
      {\includegraphics[width=51mm]{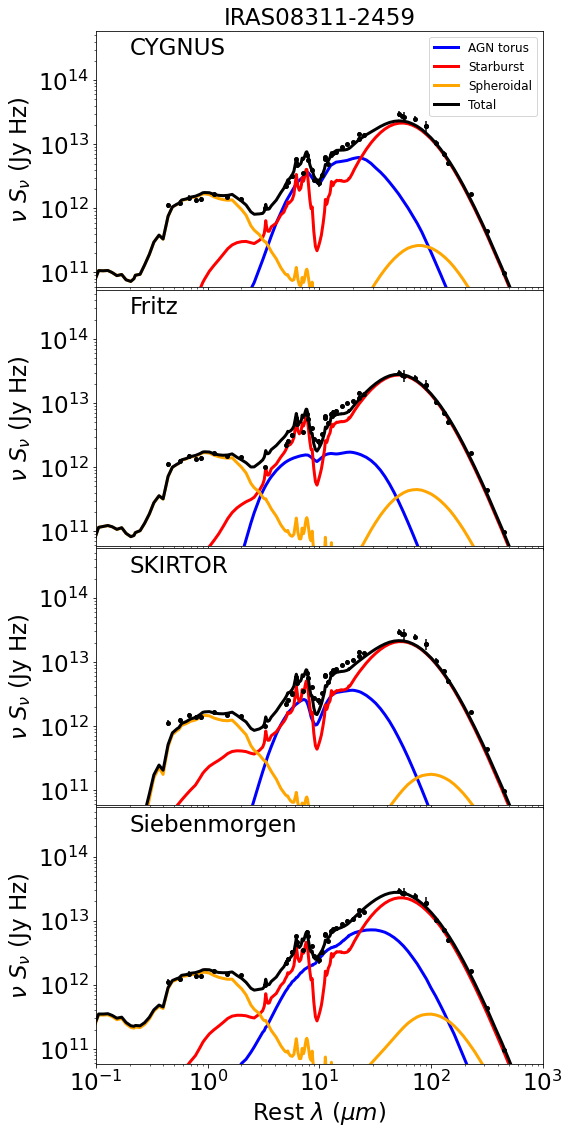}} \hspace{20pt}
      {\includegraphics[width=51mm]{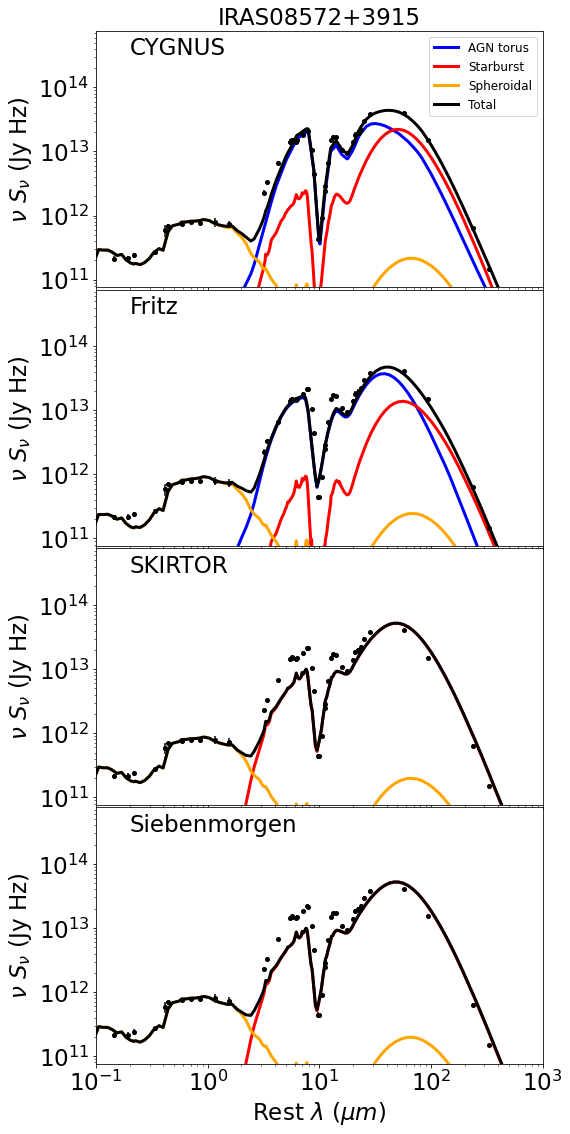}} \hspace{20pt}
      {\includegraphics[width=51mm]{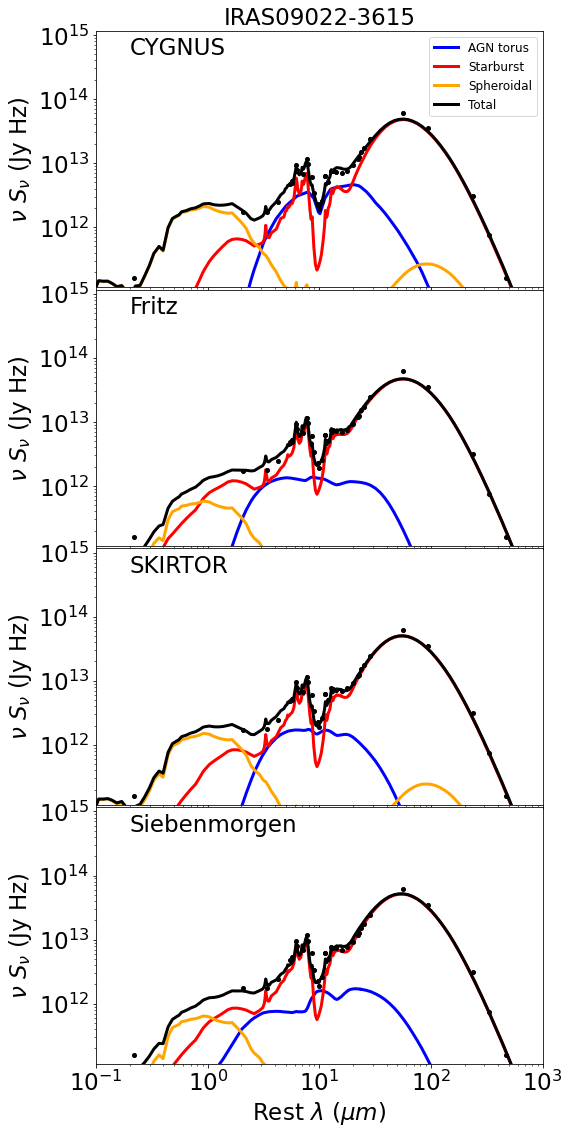}} 
      \\
\vspace{10pt}
\caption{Comparison SED fit plots of the second six objects from the list of the HERUS sample: AGN torus (blue), starburst (red), spheroidal host (orange), polar dust (magenta), total (black). In each comparison plot the top row shows fits with the CYGNUS tapered torus model, the second row shows fits with the CYGNUS tapered torus model replaced by the Fritz et al. (2006) flared torus model, the third row replaces the CYGNUS tapered torus model with the SKIRTOR two-phase flared torus model, while the bottom row replaces the CYGNUS tapered torus model with the Siebenmorgen et al. (2015) two-phase fluffy dust torus model.}
     \label{fig:HERUS-resultsB}
\end{center}
\end{figure*}

\begin{figure*}
	\begin{center}
      {\includegraphics[width=51mm]{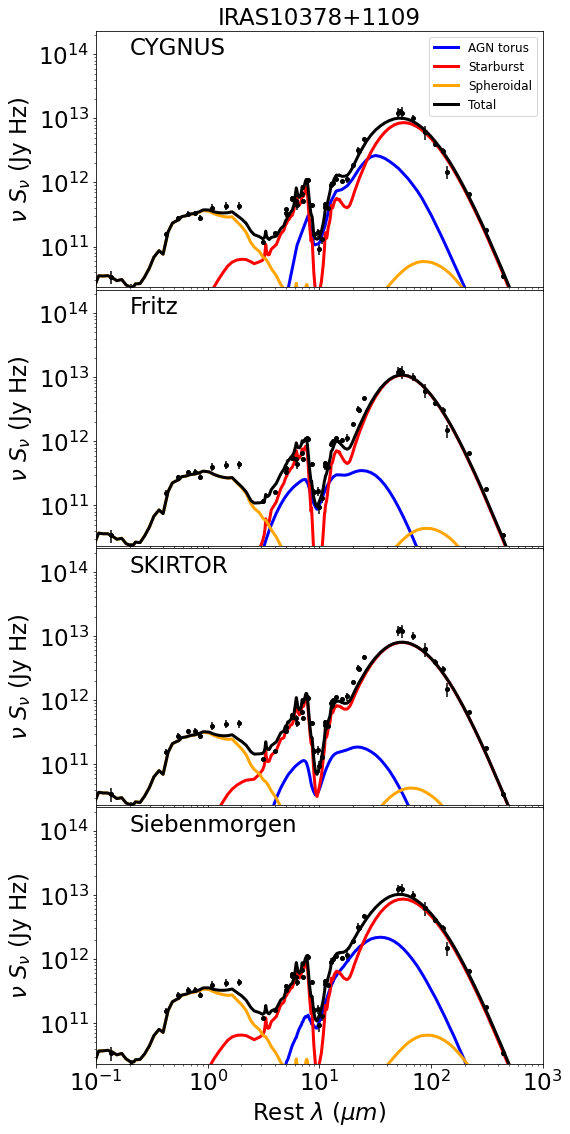}} \hspace{20pt}
      {\includegraphics[width=51mm]{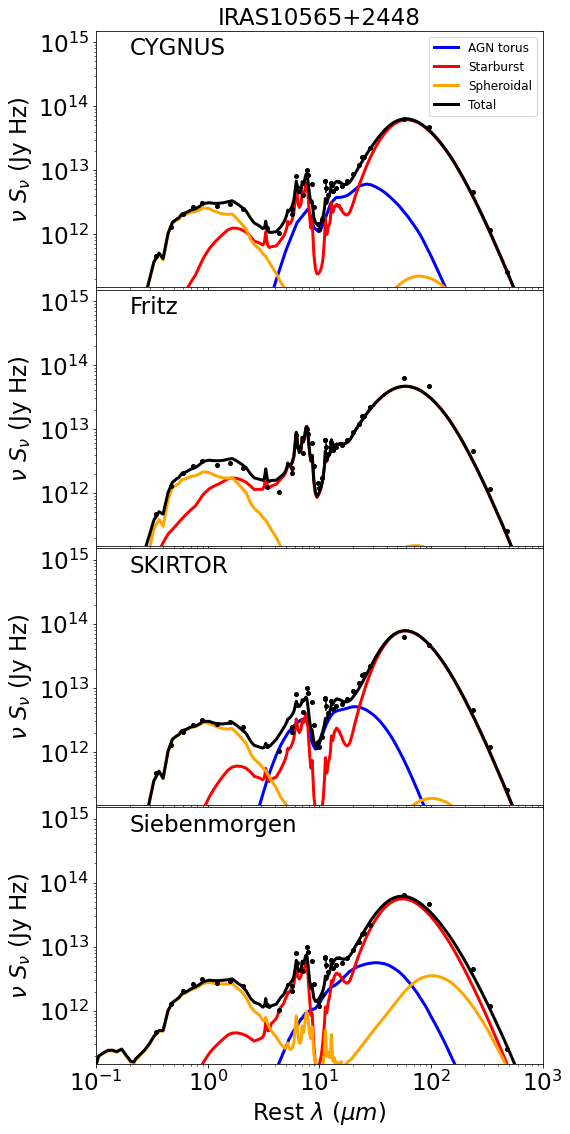}} \hspace{20pt}
      {\includegraphics[width=51mm]{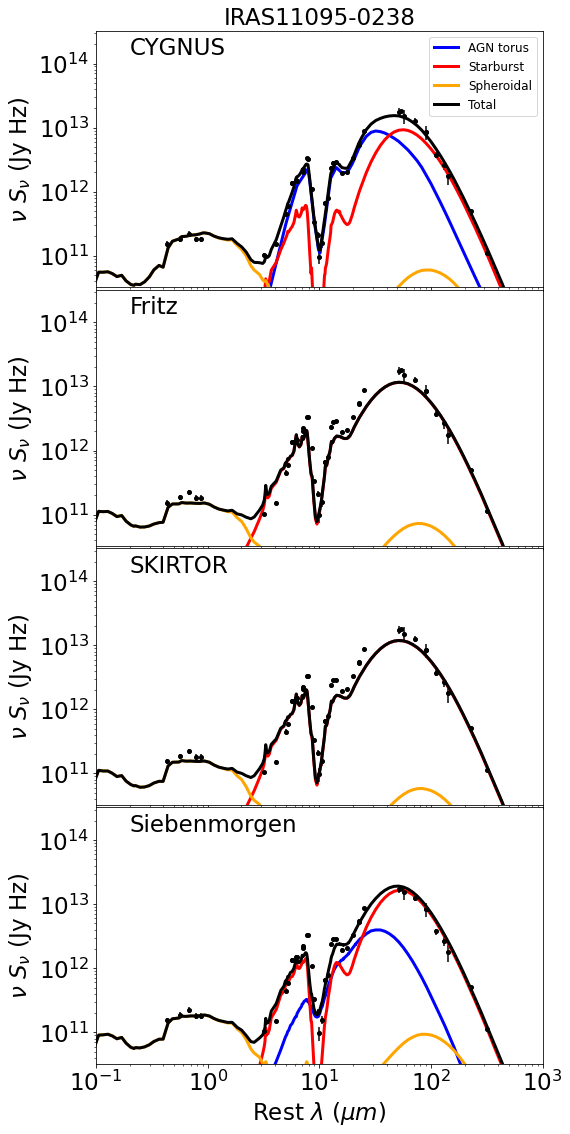}} 
      $$
      $$
      $$
      $$      
      {\includegraphics[width=51mm]{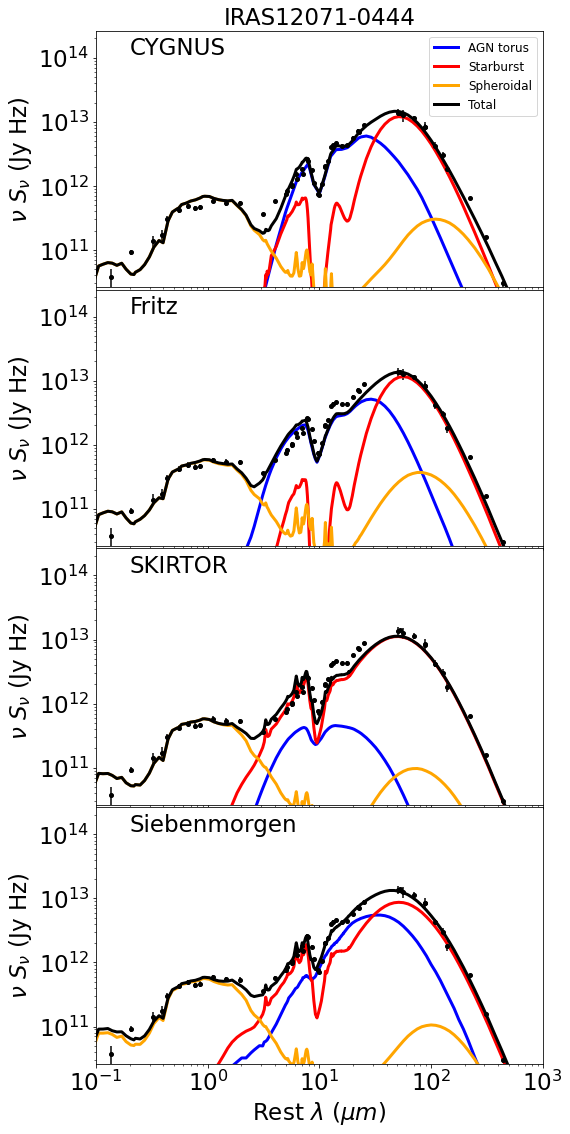}} \hspace{20pt}
      {\includegraphics[width=51mm]{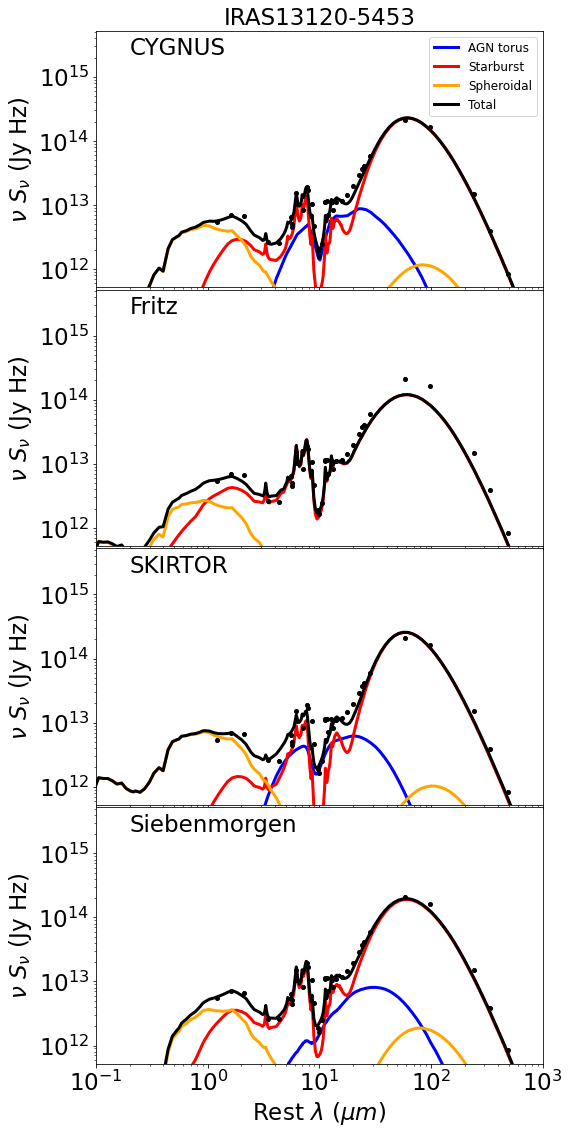}} \hspace{20pt}
      {\includegraphics[width=51mm]{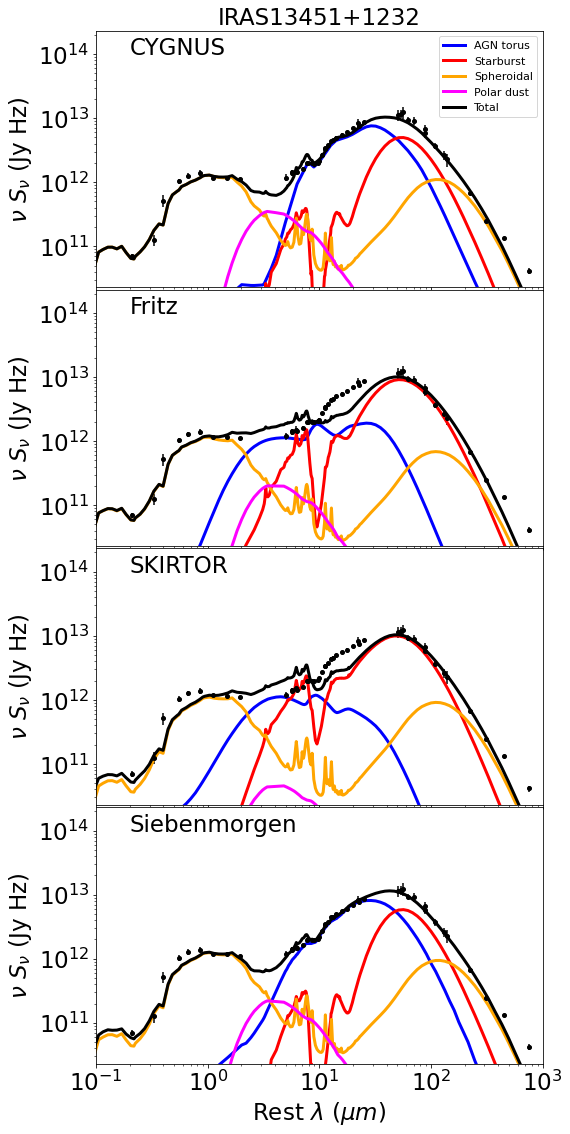}} 
      \\
\vspace{10pt}
\caption{Comparison SED fit plots of the third six objects from the list of the HERUS sample: AGN torus (blue), starburst (red), spheroidal host (orange), polar dust (magenta), total (black). In each comparison plot the top row shows fits with the CYGNUS tapered torus model, the second row shows fits with the CYGNUS tapered torus model replaced by the Fritz et al. (2006) flared torus model, the third row replaces the CYGNUS tapered torus model with the SKIRTOR two-phase flared torus model, while the bottom row replaces the CYGNUS tapered torus model with the Siebenmorgen et al. (2015) two-phase fluffy dust torus model.}
     \label{fig:HERUS-resultsC}
\end{center}
\end{figure*}

\begin{figure*}
	\begin{center}
      {\includegraphics[width=51mm]{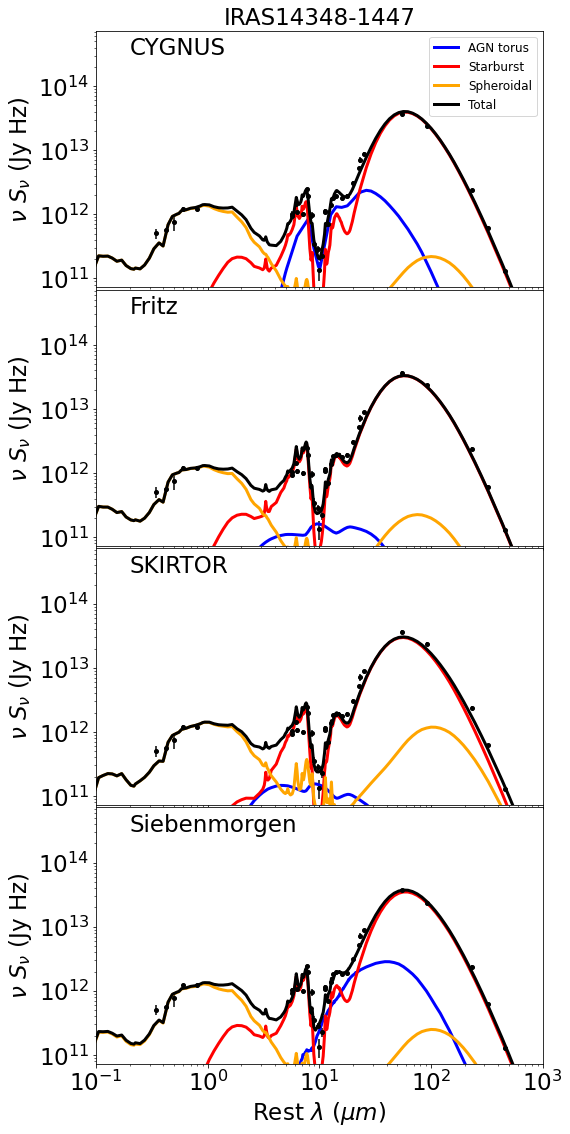}} \hspace{20pt}
      {\includegraphics[width=51mm]{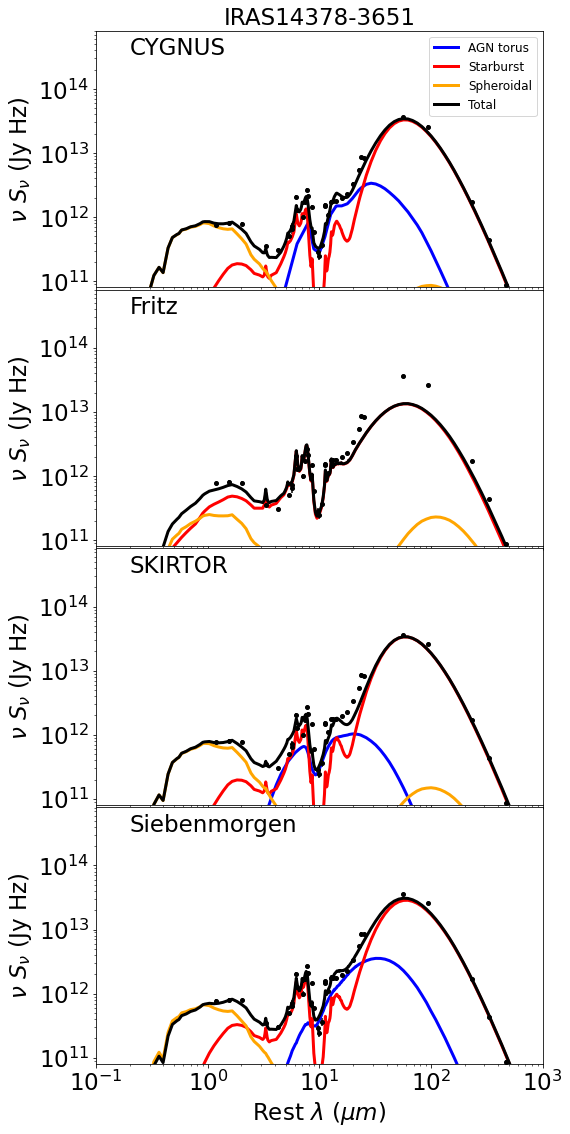}} \hspace{20pt}    
      {\includegraphics[width=51mm]{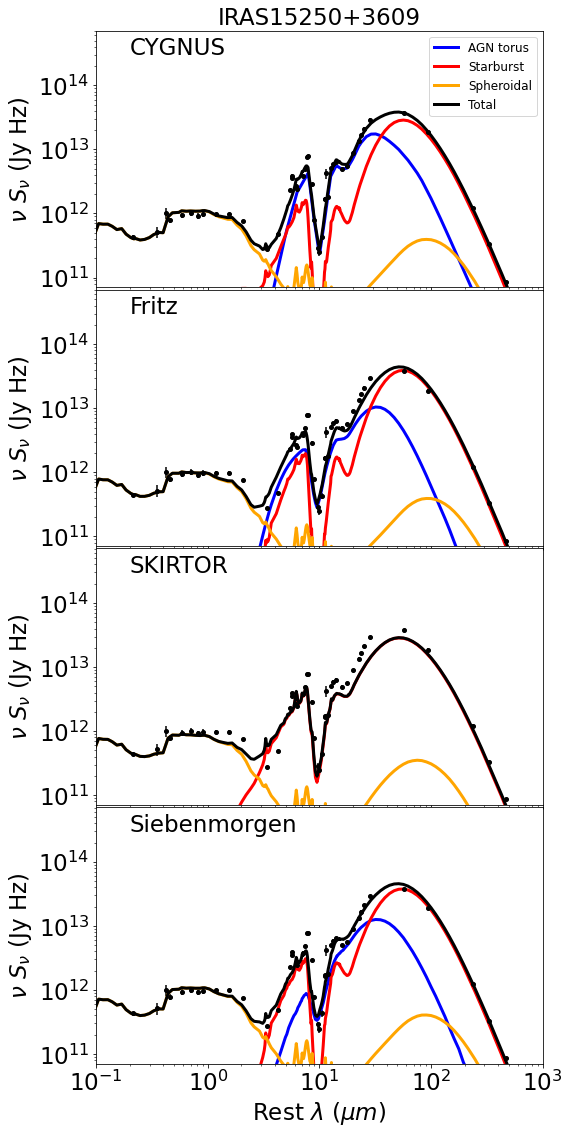}} 
      $$
      $$
      $$
      $$      
      {\includegraphics[width=51mm]{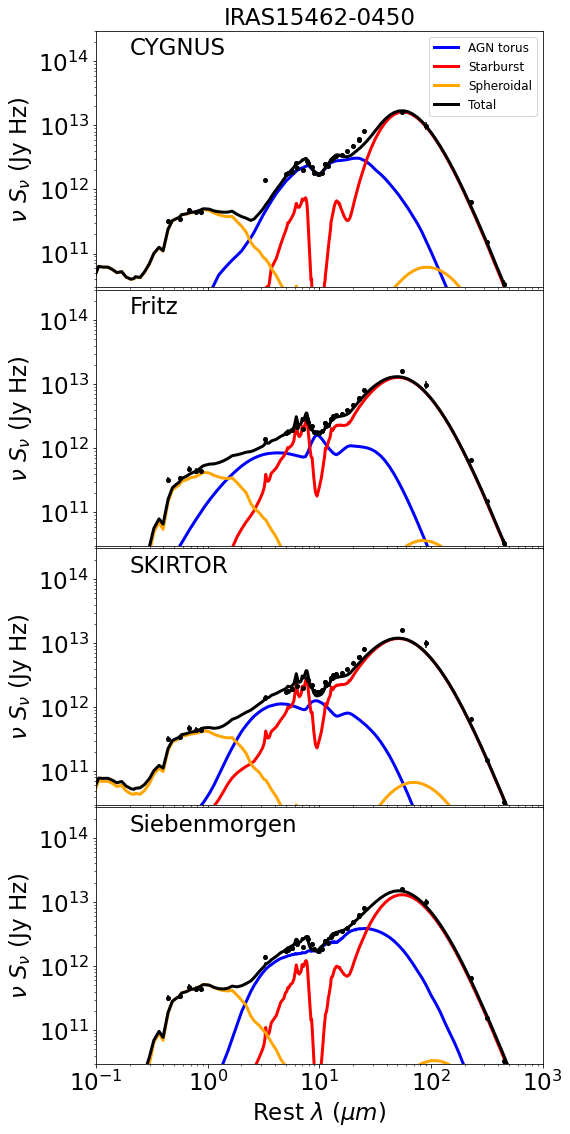}}  \hspace{20pt}   
      {\includegraphics[width=51mm]{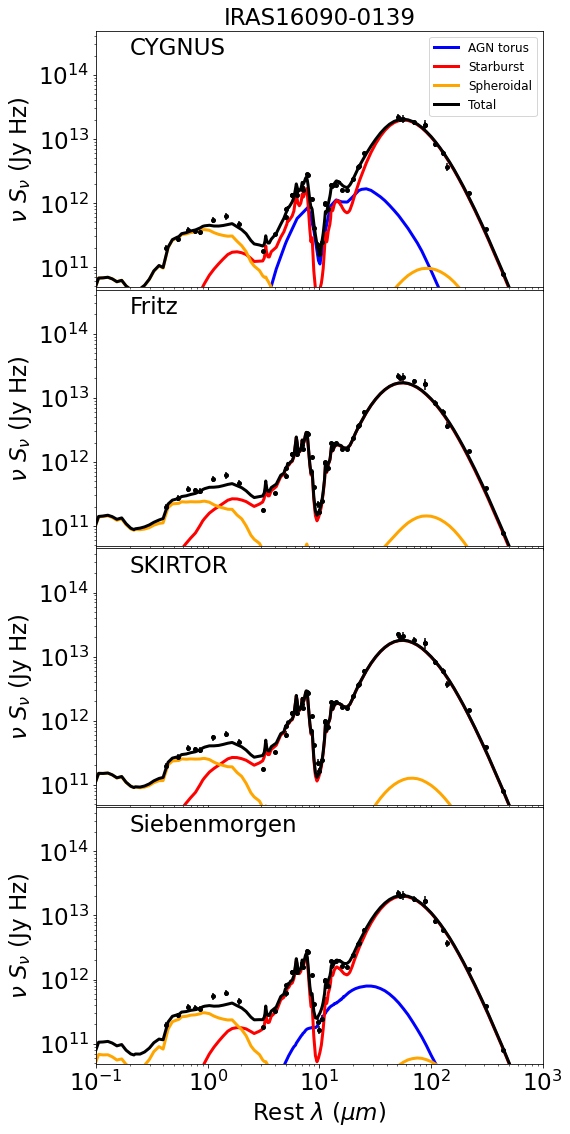}} \hspace{20pt}
      {\includegraphics[width=51mm]{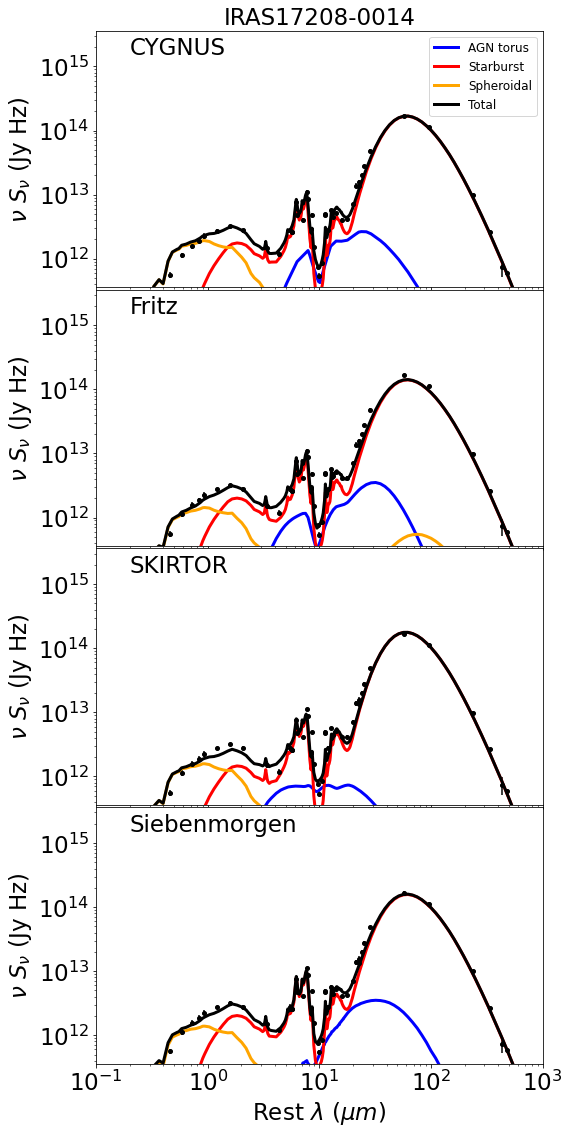}} 
      \\
\vspace{10pt}
\caption{Comparison SED fit plots of the fourth six objects from the list of the HERUS sample: AGN torus (blue), starburst (red), spheroidal host (orange), total (black). In each comparison plot the top row shows fits with the CYGNUS tapered torus model, the second row shows fits with the CYGNUS tapered torus model replaced by the Fritz et al. (2006) flared torus model, the third row replaces the CYGNUS tapered torus model with the SKIRTOR two-phase flared torus model, while the bottom row replaces the CYGNUS tapered torus model with the Siebenmorgen et al. (2015) two-phase fluffy dust torus model.}
     \label{fig:HERUS-resultsD}
\end{center}
\end{figure*}

\begin{figure*}
	\begin{center}
      {\includegraphics[width=51mm]{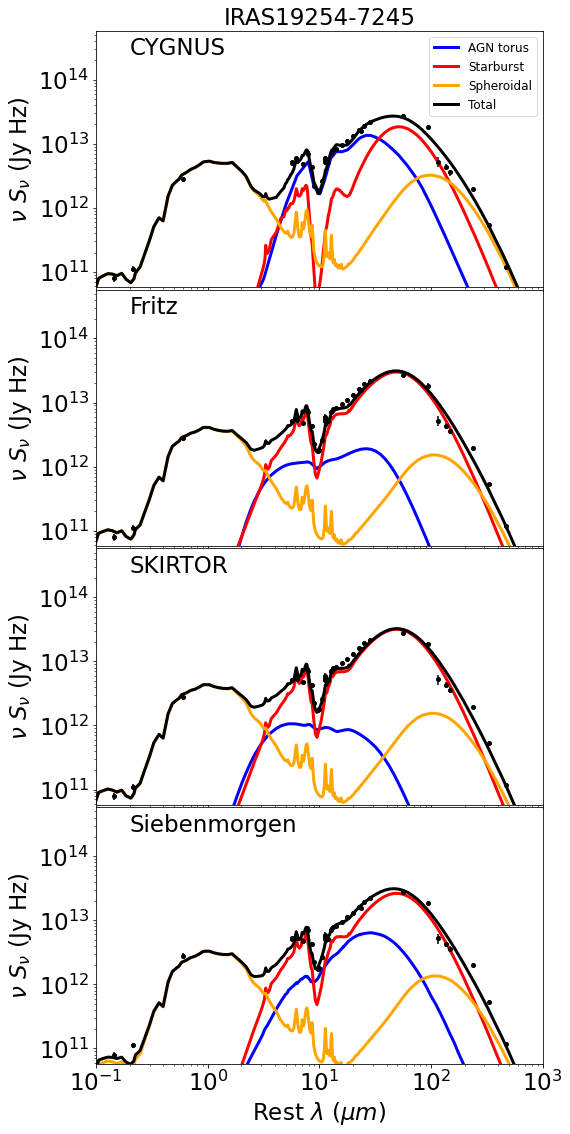}} \hspace{20pt}
      {\includegraphics[width=51mm]{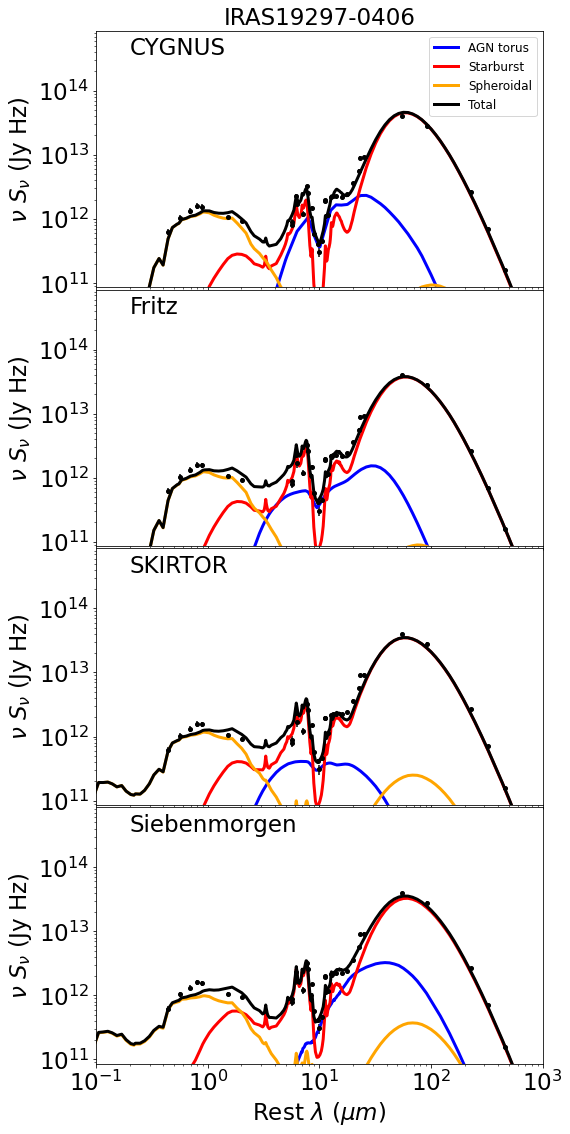}}  \hspace{20pt}   
      {\includegraphics[width=51mm]{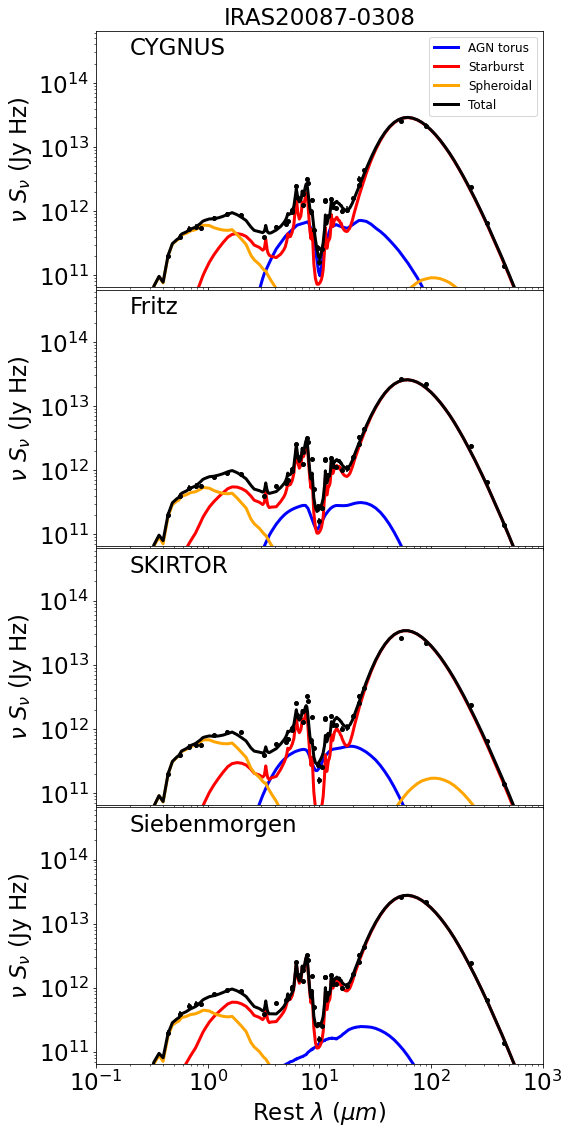}}
      $$
      $$
      $$
      $$      
      {\includegraphics[width=51mm]{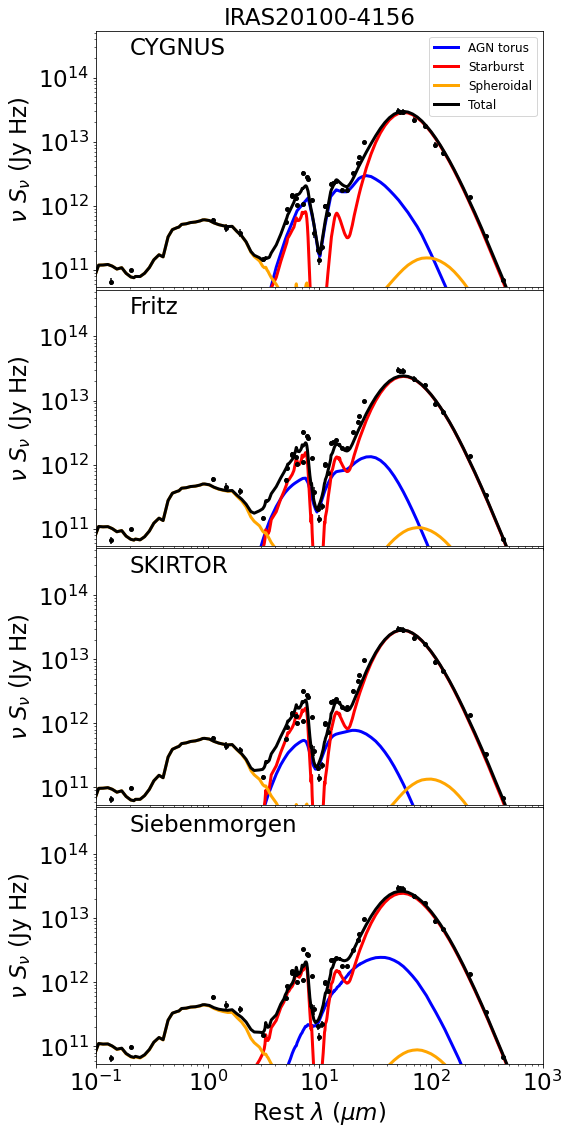}}   \hspace{20pt}   
      {\includegraphics[width=51mm]{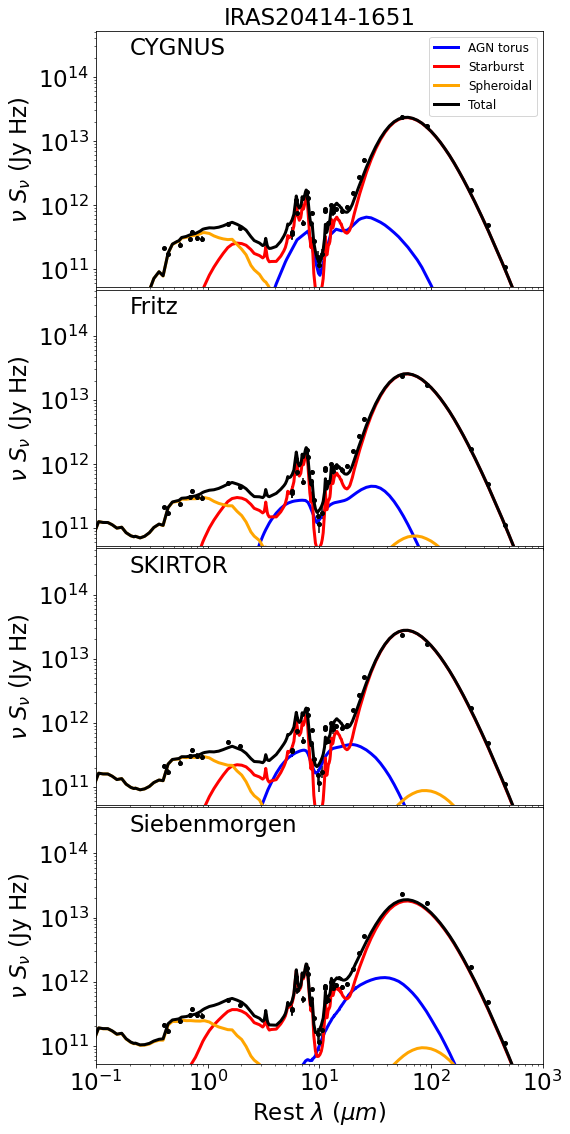}} \hspace{20pt}
      {\includegraphics[width=51mm]{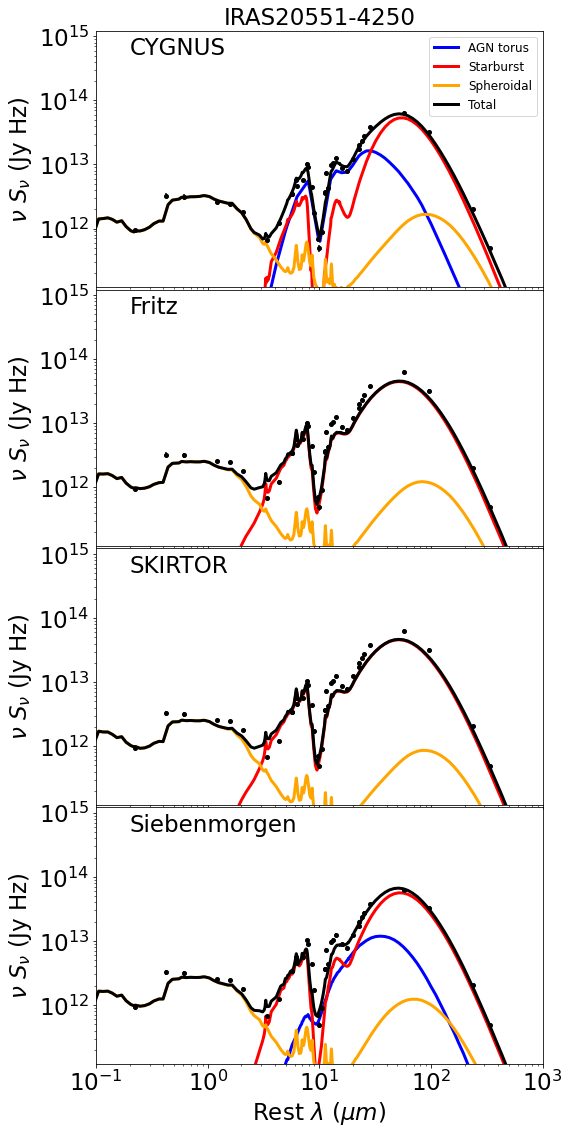}} 
      \\
\vspace{10pt}
\caption{Comparison SED fit plots of the fifth six objects from the list of the HERUS sample: AGN torus (blue), starburst (red), spheroidal host (orange), total (black). In each comparison plot the top row shows fits with the CYGNUS tapered torus model, the second row shows fits with the CYGNUS tapered torus model replaced by the Fritz et al. (2006) flared torus model, the third row replaces the CYGNUS tapered torus model with the SKIRTOR two-phase flared torus model, while the bottom row replaces the CYGNUS tapered torus model with the Siebenmorgen et al. (2015) two-phase fluffy dust torus model.}
     \label{fig:HERUS-resultsE}
\end{center}
\end{figure*}

\begin{figure*}
	\begin{center}
      {\includegraphics[width=51mm]{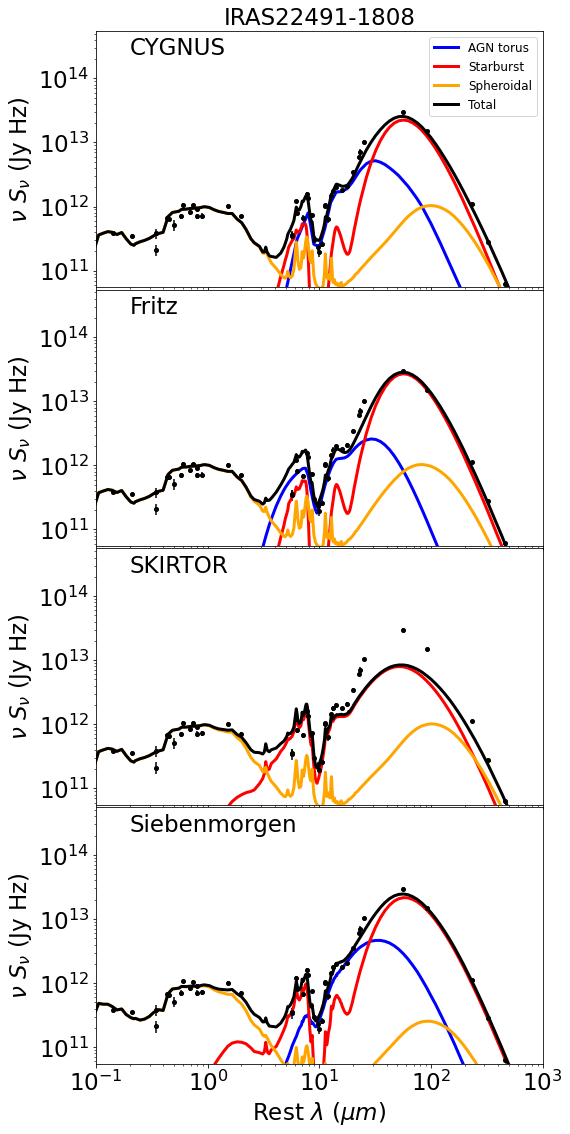}} \hspace{20pt}
      {\includegraphics[width=51mm]{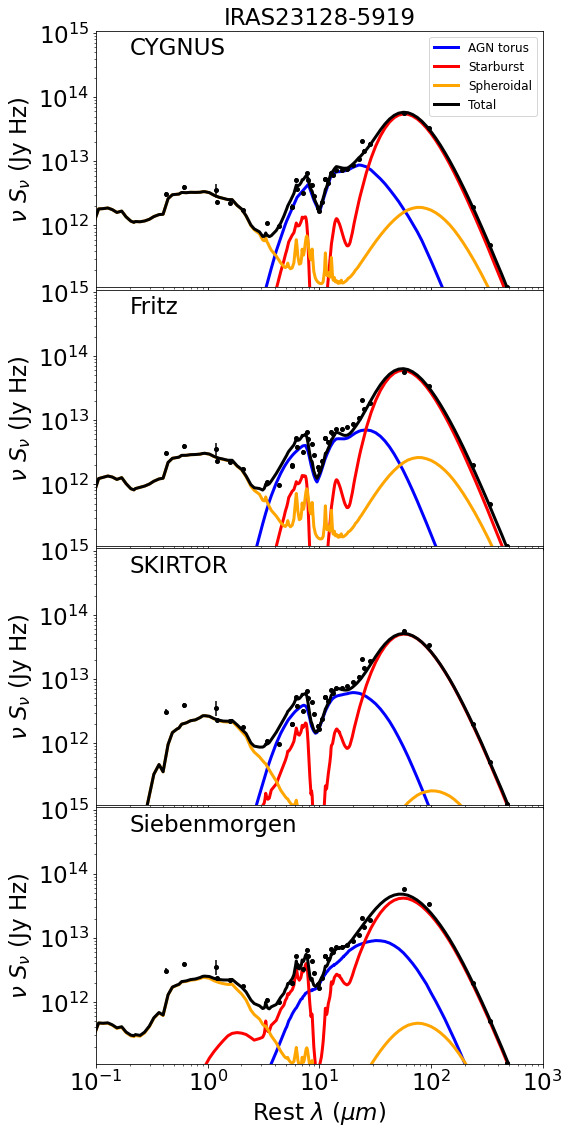}}  \hspace{20pt} 
      {\includegraphics[width=51mm]{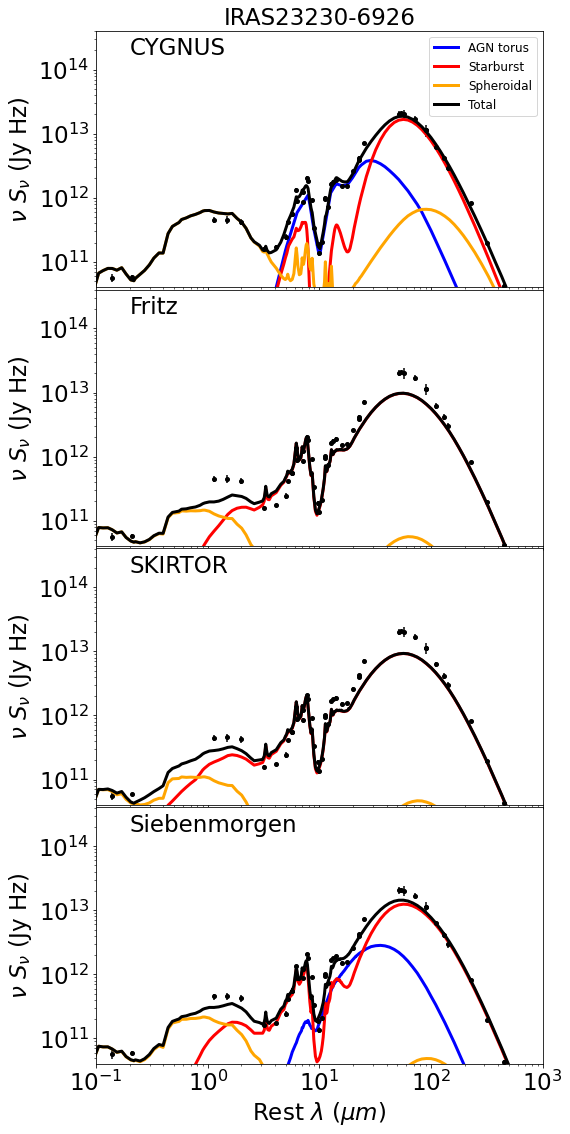}} 
      $$
      $$
      $$
      $$      
      {\includegraphics[width=51mm]{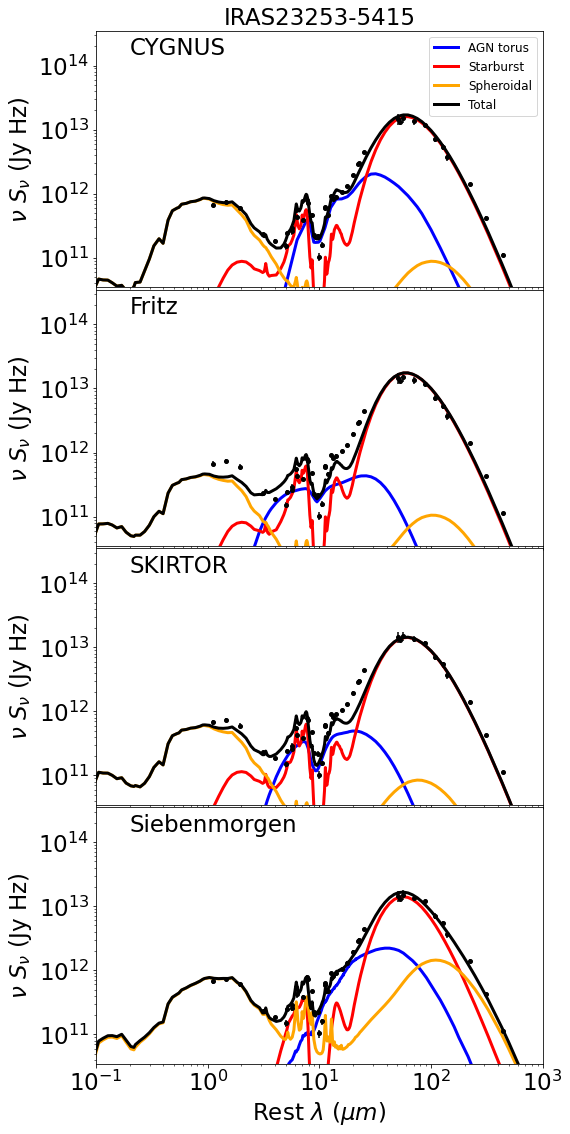}}  \hspace{20pt}     
      {\includegraphics[width=51mm]{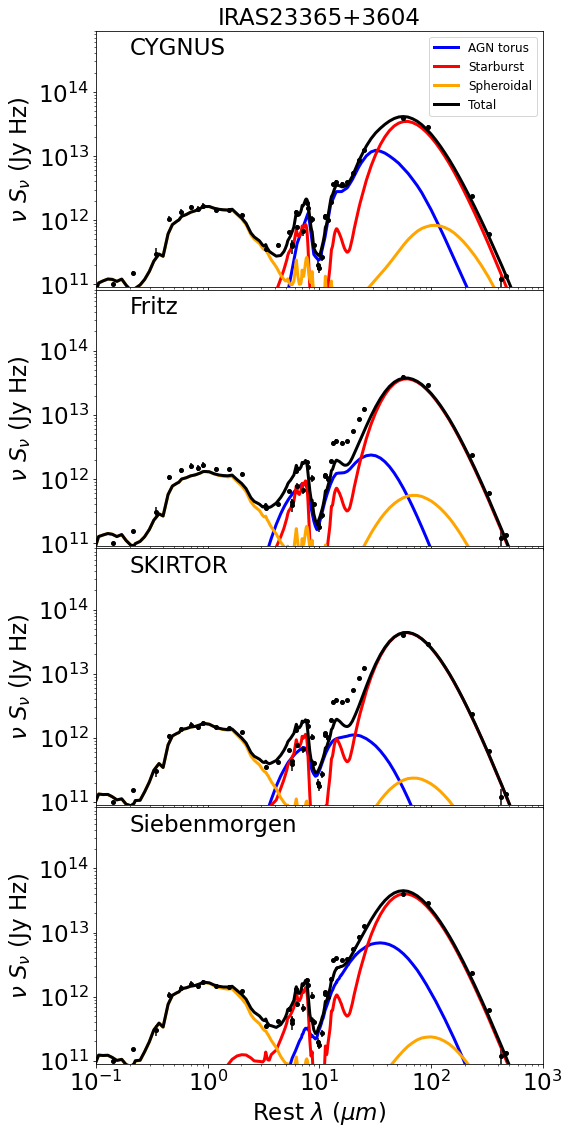}} \hspace{20pt}
      {\includegraphics[width=51mm]{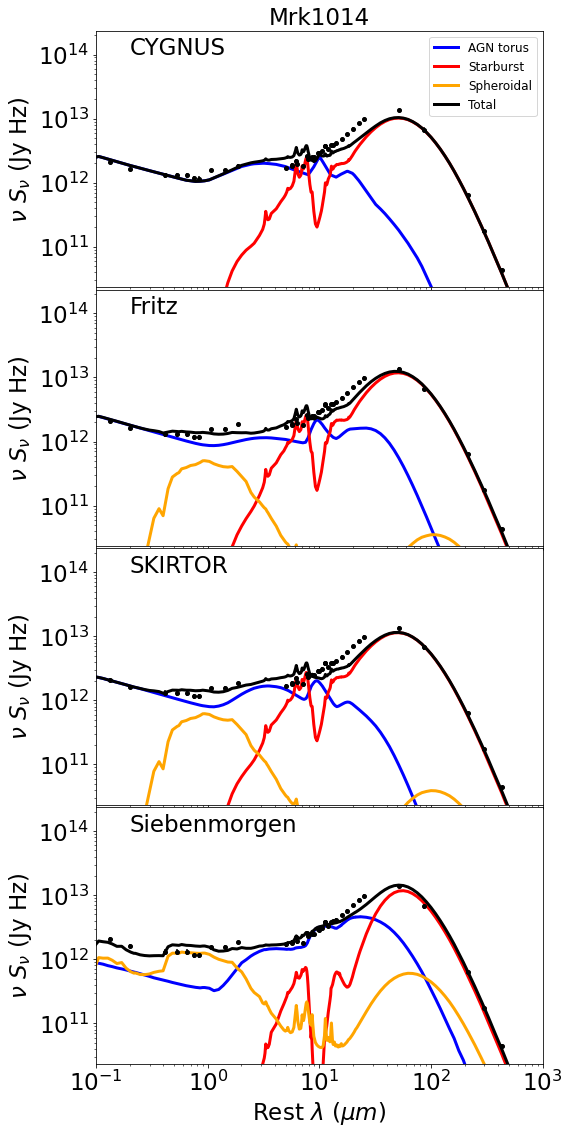}}
      \\
\vspace{10pt}
\caption{Comparison SED fit plots of the sixth six objects from the list of the HERUS sample: AGN torus (blue), starburst (red), spheroidal host (orange), total (black). In each comparison plot the top row shows fits with the CYGNUS tapered torus model, the second row shows fits with the CYGNUS tapered torus model replaced by the Fritz et al. (2006) flared torus model, the third row replaces the CYGNUS tapered torus model with the SKIRTOR two-phase flared torus model, while the bottom row replaces the CYGNUS tapered torus model with the Siebenmorgen et al. (2015) two-phase fluffy dust torus model.}
     \label{fig:HERUS-resultsF}
\end{center}
\end{figure*}

\begin{figure*}
	\begin{center}
      {\includegraphics[width=51mm]{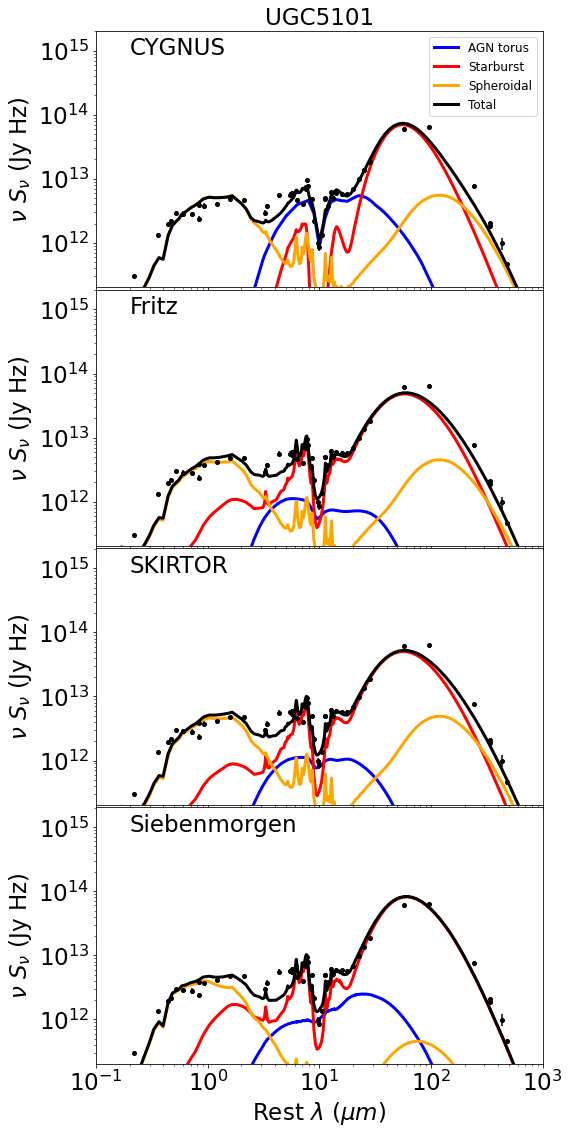}} \hspace{20pt} 
      {\includegraphics[width=51mm]{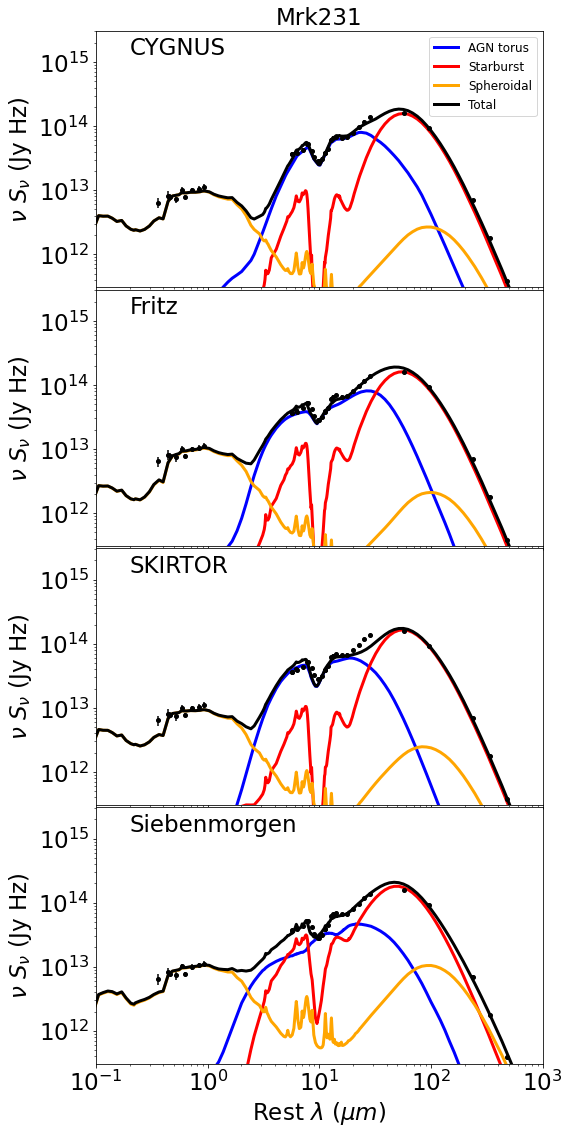}} \hspace{20pt} 
      {\includegraphics[width=51mm]{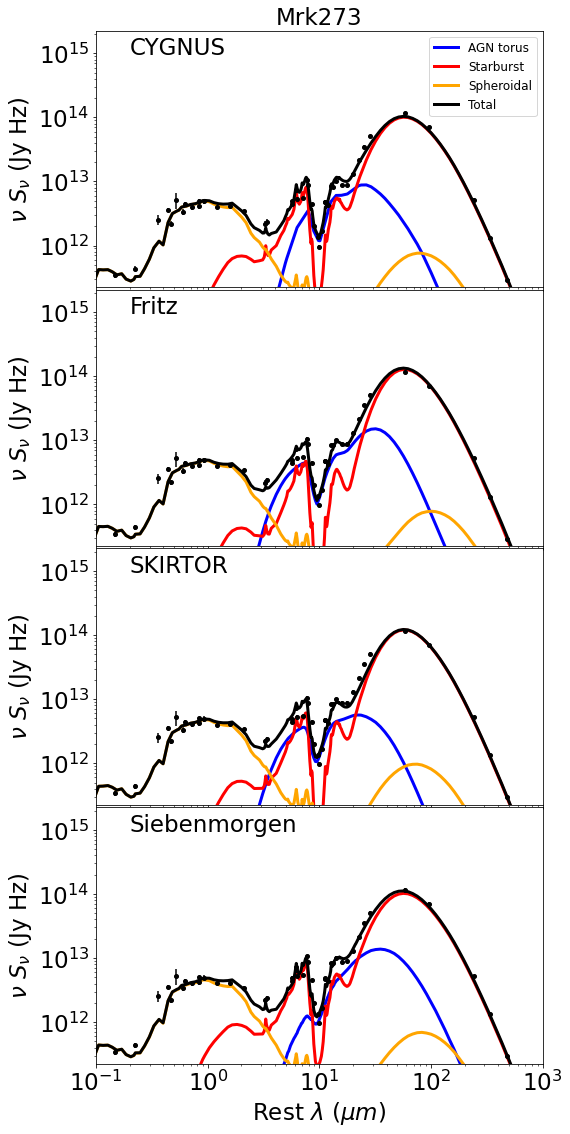}} 
      $$
      $$
      $$
      $$      
      {\includegraphics[width=51mm]{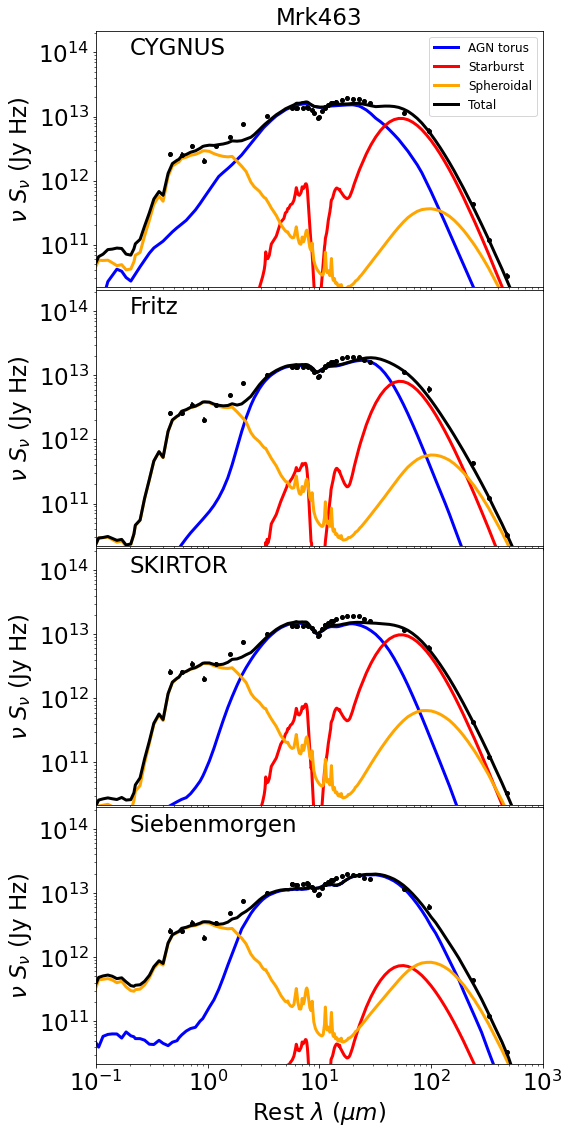}} \hspace{20pt}   
      {\includegraphics[width=51mm]{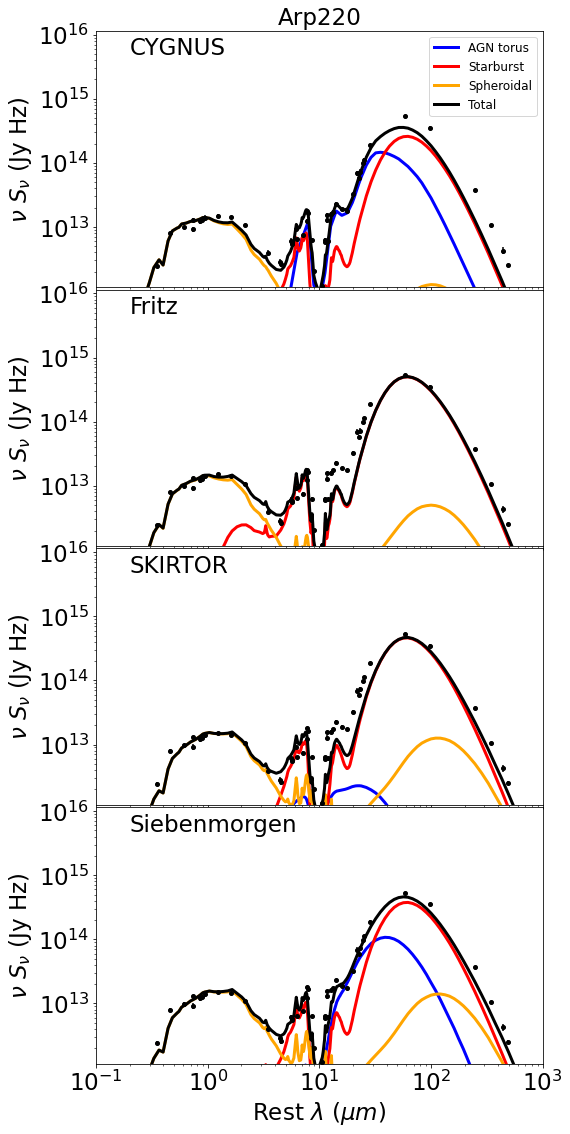}} \hspace{20pt} 
      {\includegraphics[width=51mm]{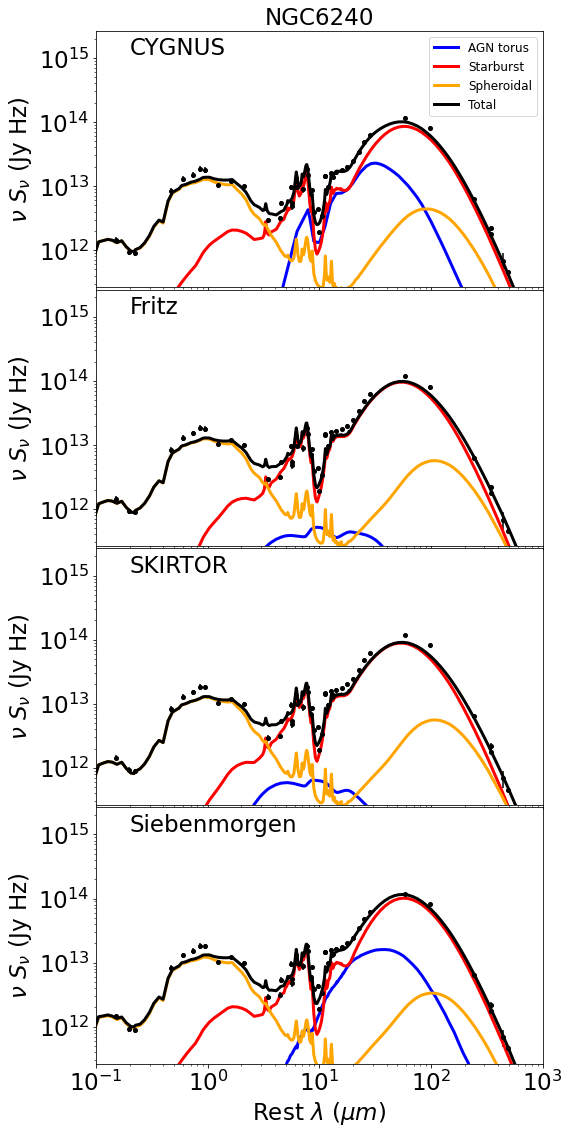}} 
      \\
\vspace{10pt}
\caption{Comparison SED fit plots of the last six objects from the list of the HERUS sample: AGN torus (blue), starburst (red), spheroidal host (orange), total (black). In each comparison plot the top row shows fits with the CYGNUS tapered torus model, the second row shows fits with the CYGNUS tapered torus model replaced by the Fritz et al. (2006) flared torus model, the third row replaces the CYGNUS tapered torus model with the SKIRTOR two-phase flared torus model, while the bottom row replaces the CYGNUS tapered torus model with the Siebenmorgen et al. (2015) two-phase fluffy dust torus model.}
     \label{fig:HERUS-resultsG}
\end{center}
\end{figure*}

\begin{table*}
	\centering
\caption{Selected extracted physical quantities for the galaxies in the HERUS sample, using the CYGNUS tapered torus model. For the AGN and total luminosities the anisotropy-corrected luminosities are given. All of the luminosities are $1-1000~\mu m$ luminosities.}
	\label{tab:extractedA_CYGNUS}
 \resizebox{\textwidth}{!}{
\begin{tabular}{ lcccccccc }
 \hline 
Name & $z$ & $L_{AGN}^c$ & $L_{SB}$ & $L_{sph}$ & $L_{p}$ & $L_{tot}^c$ & $\dot{M}^{age}_{*}$ & $\dot{M}_{sph}$ \\ 
     &     & $10^{12} L_\odot$ & $10^{11} L_\odot$ & $10^{10} L_\odot$ & $10^{10} L_\odot$ & $10^{12} L_\odot$ & $M_\odot yr^{-1}$ & $M_\odot yr^{-1}$ \\  
 \hline IRAS00188-0856 & 0.12840 &    $0.47^{0.06}_{-0.06}$  &    $17.63^{0.88}_{-0.48}$  &    $3.85^{9.03}_{-0.48}$  &    -  &    $2.33^{0.09}_{-0.15}$  &    $592.60^{7.10}_{-142.20}$  &    $0.38^{2.57}_{-0.14}$ \\  IRAS00397-1312 & 0.26170 &    $19.68^{1.23}_{-1.69}$  &    $70.25^{3.10}_{-2.90}$  &    $5.36^{0.81}_{-0.28}$  &    -  &    $26.98^{0.92}_{-1.78}$  &    $1791.00^{140.90}_{-122.30}$  &    $6.19^{0.60}_{-0.82}$ \\  IRAS01003-2238 & 0.11780 &    $7.23^{0.10}_{-0.36}$  &    $5.49^{0.21}_{-0.44}$  &    $2.30^{0.14}_{-0.18}$  &    -  &    $7.81^{0.08}_{-0.35}$  &    $134.80^{8.31}_{-10.78}$  &    $2.94^{0.05}_{-0.28}$ \\  IRAS03158+4227 & 0.13440 &    $6.41^{1.50}_{-1.21}$  &    $26.35^{2.91}_{-1.61}$  &    $1.17^{1.23}_{-0.56}$  &    -  &    $9.09^{1.62}_{-1.16}$  &    $917.30^{39.90}_{-62.42}$  &    $0.15^{1.45}_{-0.15}$ \\  IRAS03521+0028 & 0.15190 &    $2.10^{0.97}_{-0.94}$  &    $19.51^{0.79}_{-2.43}$  &    $1.71^{1.37}_{-0.51}$  &    -  &    $3.94^{1.10}_{-0.85}$  &    $667.30^{32.45}_{-95.50}$  &    $0.00^{0.00}_{-0.00}$ \\  IRAS05189-2524 & 0.04256 &    $1.54^{0.10}_{-0.40}$  &    $7.66^{0.46}_{-0.47}$  &    $2.86^{0.16}_{-0.13}$  &    $11.06^{0.98}_{-1.28}$  &    $2.46^{0.06}_{-0.39}$  &    $190.60^{8.60}_{-10.85}$  &    $0.29^{0.03}_{-0.04}$ \\
\hline 
 \end{tabular}}
\end{table*}

\begin{table*}
	\centering
\caption{Other extracted physical quantities, along with the fitted torus inclination $\theta_i$, for the galaxies in the HERUS sample, using the CYGNUS tapered torus model}
	\label{tab:extractedB_CYGNUS}
 \resizebox{\textwidth}{!}{\begin{tabular}{ lcccccccc }
 \hline 
Name & $\dot{M}_{tot}$ & ${M}^{*}_{sph}$ & ${M}^{*}_{SB}$ & ${M}^{*}_{tot}$ & $F_{AGN}$ & $A$ & $\theta_i$ \\ 
     & $M_\odot yr^{-1}$ & $10^{10} M_\odot$ & $10^{9} M_\odot$ & $10^{10} M_\odot$ & $   $ & $   $ & $\degr$ \\  
 \hline IRAS00188-0856 &    $592.80^{7.29}_{-139.50}$  &    $14.54^{8.02}_{-0.23}$  &    $17.25^{0.03}_{-5.91}$  &    $16.27^{7.42}_{-0.31}$  &    $0.20^{0.02}_{-0.01}$  &    $2.54^{0.31}_{-0.14}$  &    $76.46^{4.46}_{-0.37}$ \\  IRAS00397-1312 &    $1797.00^{141.50}_{-122.00}$  &    $9.60^{2.77}_{-1.60}$  &    $13.54^{1.47}_{-2.54}$  &    $10.95^{2.87}_{-1.50}$  &    $0.73^{0.02}_{-0.02}$  &    $8.30^{0.20}_{-0.31}$  &    $83.12^{1.11}_{-0.87}$ \\  IRAS01003-2238 &    $137.60^{8.47}_{-10.55}$  &    $2.21^{0.06}_{-0.22}$  &    $1.70^{0.09}_{-0.21}$  &    $2.35^{0.09}_{-0.13}$  &    $0.93^{0.00}_{-0.01}$  &    $5.61^{0.06}_{-0.09}$  &    $69.32^{0.27}_{-0.20}$ \\  IRAS03158+4227 &    $917.30^{40.01}_{-62.44}$  &    $3.54^{2.48}_{-1.89}$  &    $24.91^{1.76}_{-2.13}$  &    $5.99^{2.55}_{-1.71}$  &    $0.70^{0.04}_{-0.04}$  &    $6.40^{0.75}_{-1.18}$  &    $76.88^{0.62}_{-0.60}$ \\  IRAS03521+0028 &    $667.30^{32.45}_{-95.50}$  &    $7.58^{3.44}_{-1.57}$  &    $19.90^{1.01}_{-2.48}$  &    $9.53^{3.59}_{-1.47}$  &    $0.53^{0.08}_{-0.16}$  &    $7.67^{0.87}_{-2.46}$  &    $70.52^{1.82}_{-1.40}$ \\  IRAS05189-2524 &    $190.90^{8.61}_{-10.87}$  &    $10.24^{0.59}_{-0.44}$  &    $3.73^{0.83}_{-0.59}$  &    $10.70^{0.53}_{-0.45}$  &    $0.63^{0.03}_{-0.07}$  &    $2.90^{0.08}_{-0.55}$  &    $63.61^{1.75}_{-0.60}$ \\
\hline 
 \end{tabular}}
\end{table*}

\begin{table*}
	\centering
\caption{Selected extracted physical quantities for the galaxies in the HERUS sample, using the flared torus model of Fritz et al. (2006). For the AGN and total luminosities the anisotropy-corrected luminosities are given. All of the luminosities are $1-1000~\mu m$ luminosities.}
	\label{tab:extractedA_Fritz}
 \resizebox{\textwidth}{!}{\begin{tabular}{ lcccccccc }
 \hline 
Name & $z$ & $L_{AGN}^c$ & $L_{SB}$ & $L_{sph}$ & $L_{p}$ & $L_{tot}^c$ & $\dot{M}^{age}_{*}$ & $\dot{M}_{sph}$ \\ 
     &     & $10^{12} L_\odot$ & $10^{11} L_\odot$ & $10^{10} L_\odot$ & $10^{10} L_\odot$ & $10^{12} L_\odot$ & $M_\odot yr^{-1}$ & $M_\odot yr^{-1}$ \\  
 \hline IRAS00188-0856 & 0.12840 &    $0.28^{0.07}_{-0.07}$  &    $17.52^{0.53}_{-0.56}$  &    $2.74^{0.18}_{-0.14}$  &    -  &    $2.05^{0.07}_{-0.03}$  &    $560.80^{93.10}_{-55.20}$  &    $0.28^{0.07}_{-0.07}$ \\  IRAS00397-1312 & 0.26170 &    $12.62^{1.10}_{-0.97}$  &    $46.20^{3.64}_{-7.54}$  &    $12.04^{1.65}_{-4.95}$  &    -  &    $17.67^{0.38}_{-1.18}$  &    $1087.00^{108.00}_{-128.00}$  &    $10.40^{0.83}_{-3.72}$ \\  IRAS01003-2238 & 0.11780 &    $1.53^{0.05}_{-0.05}$  &    $7.18^{2.40}_{-0.19}$  &    $1.80^{0.23}_{-0.06}$  &    -  &    $2.29^{0.16}_{-0.03}$  &    $190.80^{57.30}_{-7.09}$  &    $2.33^{0.01}_{-0.11}$ \\  IRAS03158+4227 & 0.13440 &    $0.00^{0.00}_{-0.00}$  &    $29.33^{0.88}_{-0.81}$  &    $0.01^{0.01}_{-0.01}$  &    -  &    $2.94^{0.09}_{-0.08}$  &    $909.90^{23.37}_{-31.70}$  &    $0.00^{0.00}_{-0.00}$ \\  IRAS03521+0028 & 0.15190 &    $0.30^{0.02}_{-0.05}$  &    $20.08^{1.41}_{-0.70}$  &    $1.00^{0.09}_{-0.07}$  &    -  &    $2.29^{0.14}_{-0.04}$  &    $781.50^{17.17}_{-34.89}$  &    $0.00^{0.00}_{-0.00}$ \\  IRAS05189-2524 & 0.04256 &    $0.23^{0.03}_{-0.02}$  &    $10.35^{0.42}_{-0.41}$  &    $2.66^{0.18}_{-0.14}$  &    $0.34^{0.16}_{-0.17}$  &    $1.30^{0.04}_{-0.02}$  &    $244.20^{8.97}_{-11.19}$  &    $0.25^{0.02}_{-0.02}$ \\
\hline 

 \end{tabular}}
\end{table*}

\begin{table*}
	\centering
\caption{Other extracted physical quantities, along with the fitted torus inclination $\theta_i$, for the galaxies in the HERUS sample, using the flared torus model of Fritz et al. (2006)}
	\label{tab:extractedB_Fritz}
 \resizebox{\textwidth}{!}{\begin{tabular}{ lcccccccc }
 \hline 
Name & $\dot{M}_{tot}$ & ${M}^{*}_{sph}$ & ${M}^{*}_{SB}$ & ${M}^{*}_{tot}$ & $F_{AGN}$ & $A$ & $\theta_i$ \\ 
     & $M_\odot yr^{-1}$ & $10^{10} M_\odot$ & $10^{9} M_\odot$ & $10^{10} M_\odot$ & $   $ & $   $ & $\degr$ \\  
 \hline IRAS00188-0856 &    $561.20^{93.01}_{-55.28}$  &    $11.43^{0.88}_{-1.01}$  &    $15.85^{1.10}_{-0.89}$  &    $13.04^{0.77}_{-0.98}$  &    $0.14^{0.03}_{-0.04}$  &    $1.70^{0.15}_{-0.08}$  &    $81.61^{2.02}_{-1.51}$ \\  IRAS00397-1312 &    $1098.00^{107.50}_{-128.00}$  &    $9.25^{1.28}_{-0.71}$  &    $17.04^{1.26}_{-1.06}$  &    $10.96^{1.42}_{-0.83}$  &    $0.72^{0.05}_{-0.01}$  &    $3.16^{0.13}_{-0.10}$  &    $78.70^{2.76}_{-0.83}$ \\  IRAS01003-2238 &    $193.10^{57.40}_{-6.97}$  &    $1.92^{0.08}_{-0.04}$  &    $2.05^{0.20}_{-0.22}$  &    $2.15^{0.06}_{-0.08}$  &    $0.68^{0.01}_{-0.07}$  &    $1.97^{0.04}_{-0.01}$  &    $83.73^{0.56}_{-0.72}$ \\  IRAS03158+4227 &    $909.90^{23.37}_{-31.70}$  &    $0.04^{0.08}_{-0.03}$  &    $16.71^{0.45}_{-0.64}$  &    $1.70^{0.11}_{-0.07}$  &    $0.00^{0.00}_{-0.00}$  &    $1.52^{0.77}_{-0.58}$  &    $76.62^{23.41}_{-7.04}$ \\  IRAS03521+0028 &    $781.50^{17.17}_{-34.89}$  &    $5.45^{0.38}_{-0.43}$  &    $21.56^{1.24}_{-0.18}$  &    $7.61^{0.55}_{-0.45}$  &    $0.13^{0.00}_{-0.02}$  &    $1.87^{0.02}_{-0.08}$  &    $83.66^{0.89}_{-0.70}$ \\  IRAS05189-2524 &    $244.50^{8.98}_{-11.20}$  &    $9.92^{0.69}_{-0.63}$  &    $4.66^{0.48}_{-0.95}$  &    $10.44^{0.64}_{-0.66}$  &    $0.18^{0.03}_{-0.01}$  &    $0.90^{0.03}_{-0.09}$  &    $50.07^{5.37}_{-4.62}$ \\
\hline 
 \end{tabular}}
\end{table*}

\begin{table*}
	\centering
\caption{Selected extracted physical quantities for the galaxies in the HERUS sample, using the SKIRTOR two-phase flared torus model. For the AGN and total luminosities the anisotropy-corrected luminosities are given. All of the luminosities are $1-1000~\mu m$ luminosities.}
	\label{tab:extractedA_SKIRTOR}
 \resizebox{\textwidth}{!}{\begin{tabular}{ lcccccccc }
 \hline 
Name & $z$ & $L_{AGN}^c$ & $L_{SB}$ & $L_{sph}$ & $L_{p}$ & $L_{tot}^c$ & $\dot{M}^{age}_{*}$ & $\dot{M}_{sph}$ \\ 
     &     & $10^{12} L_\odot$ & $10^{11} L_\odot$ & $10^{10} L_\odot$ & $10^{10} L_\odot$ & $10^{12} L_\odot$ & $M_\odot yr^{-1}$ & $M_\odot yr^{-1}$ \\  
 \hline IRAS00188-0856 & 0.12840 &    $0.08^{0.05}_{-0.03}$  &    $18.90^{0.90}_{-1.11}$  &    $3.44^{5.28}_{-0.61}$  &    -  &    $1.99^{0.17}_{-0.10}$  &    $494.90^{22.54}_{-51.04}$  &    $0.45^{1.33}_{-0.11}$ \\  IRAS00397-1312 & 0.26170 &    $9.46^{2.06}_{-1.37}$  &    $76.89^{2.64}_{-3.00}$  &    $5.14^{0.32}_{-0.47}$  &    -  &    $17.21^{1.92}_{-1.09}$  &    $2175.00^{148.60}_{-205.00}$  &    $5.85^{0.51}_{-0.60}$ \\  IRAS01003-2238 & 0.11780 &    $0.72^{0.00}_{-0.00}$  &    $4.94^{0.04}_{-0.04}$  &    $2.17^{0.08}_{-0.08}$  &    -  &    $1.24^{0.01}_{-0.01}$  &    $128.70^{3.79}_{-3.79}$  &    $2.93^{0.12}_{-0.12}$ \\  IRAS03158+4227 & 0.13440 &    $0.06^{0.05}_{-0.06}$  &    $31.10^{4.96}_{-1.66}$  &    $0.01^{0.03}_{-0.01}$  &    -  &    $3.18^{0.53}_{-0.23}$  &    $842.00^{106.80}_{-62.89}$  &    $0.00^{0.01}_{-0.00}$ \\  IRAS03521+0028 & 0.15190 &    $0.16^{0.02}_{-0.02}$  &    $21.48^{0.46}_{-0.61}$  &    $2.92^{0.77}_{-0.33}$  &    -  &    $2.34^{0.04}_{-0.05}$  &    $628.90^{63.16}_{-51.88}$  &    $0.75^{0.53}_{-0.40}$ \\  IRAS05189-2524 & 0.04256 &    $0.17^{0.02}_{-0.01}$  &    $11.61^{0.24}_{-0.54}$  &    $2.58^{0.13}_{-0.10}$  &    $0.39^{0.21}_{-0.28}$  &    $1.36^{0.02}_{-0.05}$  &    $272.90^{8.35}_{-9.49}$  &    $0.07^{0.03}_{-0.02}$ \\
\hline 
 \end{tabular}}
\end{table*}

\begin{table*}
	\centering
\caption{Other extracted physical quantities, along with the fitted torus inclination $\theta_i$, for the galaxies in the HERUS sample, using the SKIRTOR two-phase flared torus model}
	\label{tab:extractedB_SKIRTOR}
 \resizebox{\textwidth}{!}{\begin{tabular}{ lcccccccc }
 \hline 
Name & $\dot{M}_{tot}$ & ${M}^{*}_{sph}$ & ${M}^{*}_{SB}$ & ${M}^{*}_{tot}$ & $F_{AGN}$ & $A$ & $\theta_i$ \\ 
     & $M_\odot yr^{-1}$ & $10^{10} M_\odot$ & $10^{9} M_\odot$ & $10^{10} M_\odot$ & $   $ & $   $ & $\degr$ \\  
 \hline IRAS00188-0856 &    $495.30^{22.46}_{-49.61}$  &    $13.27^{4.57}_{-2.19}$  &    $14.16^{1.18}_{-0.98}$  &    $14.69^{4.57}_{-2.03}$  &    $0.04^{0.02}_{-0.01}$  &    $1.22^{0.32}_{-0.09}$  &    $67.57^{4.98}_{-5.18}$ \\  IRAS00397-1312 &    $2181.00^{149.20}_{-204.10}$  &    $8.93^{1.68}_{-1.39}$  &    $11.99^{0.43}_{-0.39}$  &    $10.18^{1.60}_{-1.48}$  &    $0.55^{0.04}_{-0.05}$  &    $5.46^{0.38}_{-0.80}$  &    $79.88^{0.96}_{-2.07}$ \\  IRAS01003-2238 &    $131.60^{3.68}_{-3.68}$  &    $2.22^{0.06}_{-0.06}$  &    $2.72^{0.35}_{-0.35}$  &    $2.49^{0.03}_{-0.03}$  &    $0.58^{0.00}_{-0.00}$  &    $1.31^{0.00}_{-0.00}$  &    $84.48^{0.09}_{-0.09}$ \\  IRAS03158+4227 &    $842.00^{106.80}_{-62.90}$  &    $0.04^{0.17}_{-0.04}$  &    $16.75^{3.30}_{-2.16}$  &    $1.73^{0.49}_{-0.26}$  &    $0.02^{0.01}_{-0.02}$  &    $1.46^{0.20}_{-0.14}$  &    $81.31^{9.87}_{-2.89}$ \\  IRAS03521+0028 &    $629.80^{62.35}_{-51.85}$  &    $5.66^{0.97}_{-1.18}$  &    $19.02^{1.72}_{-1.73}$  &    $7.59^{0.97}_{-1.33}$  &    $0.07^{0.01}_{-0.01}$  &    $1.34^{0.08}_{-0.07}$  &    $77.82^{6.44}_{-2.22}$ \\  IRAS05189-2524 &    $272.90^{8.31}_{-9.48}$  &    $11.14^{0.55}_{-0.64}$  &    $4.37^{0.76}_{-0.42}$  &    $11.58^{0.59}_{-0.69}$  &    $0.13^{0.01}_{-0.00}$  &    $0.95^{0.07}_{-0.08}$  &    $37.94^{5.18}_{-6.35}$ \\
\hline 
 \end{tabular}}
\end{table*}

\begin{table*}
	\centering
\caption{Selected extracted physical quantities for the galaxies in the HERUS sample, using the two-phase fluffy dust torus model of Siebenmorgen et al. (2015). For the AGN and total luminosities the anisotropy-corrected luminosities are given. All of the luminosities are $1-1000~\mu m$ luminosities.}
	\label{tab:extractedA_Sie}
 \resizebox{\textwidth}{!}{\begin{tabular}{ lcccccccc }
 \hline 
Name & $z$ & $L_{AGN}^c$ & $L_{SB}$ & $L_{sph}$ & $L_{p}$ & $L_{tot}^c$ & $\dot{M}^{age}_{*}$ & $\dot{M}_{sph}$ \\ 
     &     & $10^{12} L_\odot$ & $10^{11} L_\odot$ & $10^{10} L_\odot$ & $10^{10} L_\odot$ & $10^{12} L_\odot$ & $M_\odot yr^{-1}$ & $M_\odot yr^{-1}$ \\  
 \hline IRAS00188-0856 & 0.12840 &    $0.29^{0.03}_{-0.09}$  &    $18.94^{0.75}_{-0.49}$  &    $2.96^{1.40}_{-0.37}$  &    -  &    $2.22^{0.09}_{-0.11}$  &    $590.40^{34.68}_{-61.30}$  &    $0.31^{0.41}_{-0.13}$ \\  IRAS00397-1312 & 0.26170 &    $0.67^{0.07}_{-0.07}$  &    $91.27^{1.49}_{-1.54}$  &    $4.37^{0.29}_{-0.21}$  &    -  &    $9.88^{0.14}_{-0.18}$  &    $2659.00^{58.65}_{-127.50}$  &    $5.25^{0.29}_{-0.46}$ \\  IRAS01003-2238 & 0.11780 &    $2.17^{0.24}_{-0.21}$  &    $6.50^{0.52}_{-0.24}$  &    $1.95^{0.14}_{-0.10}$  &    -  &    $2.87^{0.24}_{-0.24}$  &    $160.50^{10.09}_{-7.78}$  &    $2.57^{0.15}_{-0.12}$ \\  IRAS03158+4227 & 0.13440 &    $1.51^{0.27}_{-0.34}$  &    $32.49^{1.26}_{-2.02}$  &    $0.01^{0.01}_{-0.01}$  &    -  &    $4.75^{0.17}_{-0.30}$  &    $963.30^{56.61}_{-65.37}$  &    $0.00^{0.00}_{-0.00}$ \\  IRAS03521+0028 & 0.15190 &    $0.86^{0.10}_{-0.05}$  &    $18.68^{0.71}_{-0.23}$  &    $1.14^{0.90}_{-0.22}$  &    -  &    $2.76^{0.09}_{-0.12}$  &    $803.30^{28.74}_{-38.91}$  &    $0.35^{0.43}_{-0.34}$ \\  IRAS05189-2524 & 0.04256 &    $1.35^{0.18}_{-0.21}$  &    $7.17^{0.95}_{-0.80}$  &    $2.89^{0.32}_{-0.20}$  &    $8.53^{1.12}_{-0.91}$  &    $2.19^{0.12}_{-0.15}$  &    $181.30^{19.76}_{-16.49}$  &    $0.17^{0.05}_{-0.06}$ \\
\hline 
 \end{tabular}}
\end{table*}

\begin{table*}
	\centering
\caption{Other extracted physical quantities, along with the fitted torus inclination $\theta_i$, for the galaxies in the HERUS sample, using the two-phase fluffy dust torus model of Siebenmorgen et al. (2015)}
	\label{tab:extractedB_Sie}
 \resizebox{\textwidth}{!}{\begin{tabular}{ lcccccccc }
 \hline 
Name & $\dot{M}_{tot}$ & ${M}^{*}_{sph}$ & ${M}^{*}_{SB}$ & ${M}^{*}_{tot}$ & $F_{AGN}$ & $A$ & $\theta_i$ \\ 
     & $M_\odot yr^{-1}$ & $10^{10} M_\odot$ & $10^{9} M_\odot$ & $10^{10} M_\odot$ & $   $ & $   $ & $\degr$ \\  
 \hline IRAS00188-0856 &    $590.80^{34.54}_{-61.27}$  &    $12.07^{0.56}_{-1.69}$  &    $17.89^{0.62}_{-2.95}$  &    $13.78^{0.63}_{-1.72}$  &    $0.13^{0.02}_{-0.04}$  &    $2.44^{0.09}_{-0.32}$  &    $72.23^{4.21}_{-1.73}$ \\  IRAS00397-1312 &    $2663.00^{59.35}_{-126.60}$  &    $6.39^{1.10}_{-0.40}$  &    $13.78^{0.86}_{-0.39}$  &    $7.90^{1.06}_{-0.58}$  &    $0.07^{0.01}_{-0.01}$  &    $1.16^{0.05}_{-0.04}$  &    $54.38^{0.68}_{-1.16}$ \\  IRAS01003-2238 &    $163.20^{9.89}_{-7.88}$  &    $1.94^{0.11}_{-0.10}$  &    $3.06^{0.60}_{-0.32}$  &    $2.23^{0.13}_{-0.07}$  &    $0.76^{0.02}_{-0.03}$  &    $2.21^{0.16}_{-0.11}$  &    $79.71^{0.92}_{-1.57}$ \\  IRAS03158+4227 &    $963.30^{56.61}_{-65.37}$  &    $0.04^{0.05}_{-0.04}$  &    $21.38^{2.14}_{-1.78}$  &    $2.17^{0.19}_{-0.17}$  &    $0.32^{0.06}_{-0.05}$  &    $2.55^{0.19}_{-0.25}$  &    $82.74^{2.18}_{-1.26}$ \\  IRAS03521+0028 &    $804.10^{28.03}_{-39.10}$  &    $4.04^{0.24}_{-0.83}$  &    $24.24^{0.97}_{-0.81}$  &    $6.52^{0.30}_{-0.88}$  &    $0.31^{0.02}_{-0.02}$  &    $2.82^{0.31}_{-0.14}$  &    $79.94^{1.50}_{-1.61}$ \\  IRAS05189-2524 &    $181.40^{19.82}_{-16.45}$  &    $10.89^{1.02}_{-0.35}$  &    $4.19^{0.50}_{-0.72}$  &    $11.34^{0.94}_{-0.39}$  &    $0.62^{0.04}_{-0.07}$  &    $2.36^{0.12}_{-0.15}$  &    $71.55^{1.10}_{-0.90}$ \\
\hline 
 \end{tabular}}
\end{table*}


\bsp	
\label{lastpage}

\begin{thebibliography}{}

\bibitem[Aalto et al.(2015)]{aalto15} 
Aalto S., Mart{\'\i}n S., Costagliola F. et al., 2015, \aap, 584, A42

\bibitem[Antonucci(1993)]{anton93} 
Antonucci R.~R.~J., 1993, \araa, 31, 473

\bibitem[Babba et al.(2022)]{baba22}
Baba S., Imanishi M., Izumi T. et al., 2022, \apj, 928, 184

\bibitem[Barger et al.(1998)]{barg98} 
Barger A.~J., Cowie L.~L., Sanders D.~B. et al., 1998, \nat, 394, 248

\bibitem[Boquien et al.(2019)]{boq19} 
Boquien M., Burgarella D., Roehlly Y. et al., 2019, \aap, 622, A103

\bibitem[Bridge et al.(2013)]{bridge13} 
Bridge C.~R., Blain A., Borys C.~J.~K. et al., 2013, \apj, 769, 91

\bibitem[Bruzual \& Charlot(1993)]{bru93} 
Bruzual  G. \& Charlot S., 1993, \apj, 405, 538 

\bibitem[Bruzual \& Charlot(2003)]{bru03} 
Bruzual G. \& Charlot S., 2003, \mnras, 344, 1000 

\bibitem[Canalizo \& Stockton(2001)]{canal01} 
Canalizo G. \& Stockton A., 2001, \apj, 555, 719

\bibitem[Carnall et al.(2018)]{carnall18} 
Carnall A.~C.. McLure R.~J., Dunlop J.~S. \& Dav{\'e} R., 2018, \mnras, 480, 4379

\bibitem[Casey et al.(2014)]{casey14}
Casey C.~M., Narayanan D. \& Cooray A., 2014, \physrep, 541, 45

\bibitem[Chevallard \& Charlot(2016)]{cheval16} 
Chevallard J. \& Charlot S., 2016, \mnras, 462, 1415

\bibitem[Clements et al.(2018)]{cle18} 
Clements D.~L., Pearson C., Farrah D. et al., 2018, \mnras, 475, 2097

\bibitem[Condon et al.(1991)]{condon91}
Condon J.~J., Huang Z.~P., Yin Q.~F. \& Thuan T.X., 1991, \apj, 378, 65

\bibitem[Cutri et al.(1984)]{cutri84}
Cutri R.~M., Rieke G.~H. \& Lebofsky M.~J., 1984, \apj, 287, 566

\bibitem[da Cunha et al.(2008)]{dac08} 
da Cunha E., Charlot S. \& Elbaz D., 2008, \mnras, 388, 1595

\bibitem[Dasyra et al.(2006)]{dasyra06} 
Dasyra K.~M., Tacconi L.~J., Davies R.~I. et al., 2006, \apj, 651, 835 

\bibitem[Efstathiou \& Rowan-Robinson(1995)]{efstathiou95} 
Efstathiou A. \&  Rowan-Robinson M., 1995, \mnras, 273, 649

\bibitem[Efstathiou et al.(2000)]{efstathiou00} 
Efstathiou A., Rowan-Robinson M. \& Siebenmorgen R., 2000, \mnras, 313, 734

\bibitem[Efstathiou \& Rowan-Robinson(2003)]{efs03} 
Efstathiou A. \& Rowan-Robinson M., 2003, \mnras, 343, 322. 

\bibitem[Efstathiou(2006)]{efstathiou06} 
Efstathiou A., 2006, \mnras, 371, L70

\bibitem[Efstathiou \& Siebenmorgen(2009)]{efstathiou09} 
Efstathiou A. \& Siebenmorgen R., 2009, \aap, 502, 541

\bibitem[Efstathiou et al.(2014)]{efs14} 
Efstathiou A., Pearson C., Farrah D. et al., 2014, \mnras, 437, L16

\bibitem[Efstathiou et al.(2021)]{efs21} 
Efstathiou A., {Ma{\l}ek} K., {Burgarella} D. et al., 2021, \mnras, 503, L11

\bibitem[Efstathiou et al.(2022)]{efs22} 
Efstathiou A., Farrah D., Afonso J. et al., 2022, \mnras, 512, 5183

\bibitem[Eisenhardt et al.(2012)]{eisen12}
Eisenhardt P.~R.~M., Wu J., Tsai C.~W. et al. 2012, \apj, 755, 173

\bibitem[Farrah et al.(2002)]{far02} 
Farrah D., Serjeant S., Efstathiou A., et al., 2002, \mnras, 335, 1163

\bibitem[Farrah et al.(2003)]{far03} 
Farrah D., Afonso J., Efstathiou A. et al., 2003, \mnras, 343, 585

\bibitem[Farrah et al.(2013)]{far13} 
Farrah D., Lebouteiller V., Spoon H.~W.~W. et al., 2013, \apj, 776, 38 

\bibitem[Farrah et al.(2022)]{far22} 
Farrah D., Efstathiou A., Afonso J. et al., 2022, \mnras, 513, 4770

\bibitem[Fischer et al.(2010)]{fisch10} 
Fischer J., Sturm E., Gonz{\'a}lez-Alfonso E. et al., 2010, \aap, 518, L41

\bibitem[Foreman-Mackey et al.(2013)]{foreman13} 
Foreman-Mackey D., Hogg D.~W., Lang D. \& Goodman J., 2013, \pasp, 125, 306

\bibitem[Fritz et al.(2006)]{fritz06} 
Fritz J., Franceschini A. \& Hatziminaoglou E., 2006, \mnras, 366, 767

\bibitem[Genzel et al.(1998)]{genzel98} 
Genzel R., Lutz D., Sturm E. et al., 1998, \apj, 498, 579

\bibitem[Genzel et al.(2001)]{genzel01} 
Genzel R., Tacconi L.~J, Rigopoulou D. et al., 2001, \apj, 563, 527

\bibitem[Gonz{\'a}lez-Alfonso et al.(2013)]{gonz13} 
Gonz{\'a}lez-Alfonso E., Fischer J., Bruderer S. et al., 2013, \aap, 550, A25

\bibitem[Gonz{\'a}lez-Mart{\'\i}n et al.(2023)]{gonz23}
Gonz{\'a}lez-Mart{\'\i}n O., Ramos Almeida C., Fritz J. et al., 2023, \aap, 676, A73

\bibitem[Gowardhan et al.(2018)]{gowardhan18}
Gowardhan A., Spoon H., Riechers D.~A. et al., 2018, \apj, 859, 35

\bibitem[Hailey-Dunsheath et al.(2012)]{hail12} 
Hailey-Dunsheath, S., Sturm E., Fischer J. et al., 2012, \apj, 755, 57

\bibitem[Hastings(1970)]{hastings70}
Hastings W.~K., 1970, Biometrika, 57, 97

\bibitem[Hou et al.(2011)]{hou11} 
Hou L.~G., Han J.~L., Kong M.~Z. \& Wu Xue-Bing, 2011, \apj, 732, 72

\bibitem[Houck et al.(1985)]{houck85} 
Houck J.~R., Schneider D.~P., Danielson G.~E. et al., 1985, \apj, 290, L5

\bibitem[Houck et al.(2004)]{houck04} 
Houck J.~R., Roellig T.~L., van Cleve J. et al., 2004, \apjs, 154, 18

\bibitem[Hughes et al.(1998)]{hugh98}
Hughes D.~H., Serjeant S., Dunlop J. et al., 1998, \nat, 394, 241

\bibitem[Imanishi et al.(2001)]{iman01}
Imanishi M., Dudley C.~C. \& Maloney P.~R., 2001, \apj, 558, L93

\bibitem[Imanishi et al.(2010)]{iman10}
Imanishi M., Nakagawa T., Shirahata M. et al., 2010, \apj, 721, 1233

\bibitem[Imanishi et al.(2011)]{iman11}
Imanishi M., Imase K., Oi N. \& Ichikawa K., 2011, \aj, 141, 156
 
\bibitem[Imanishi et al.(2019)]{iman19} 
Imanishi M., Nakanishi K. \& Izumi T., 2019, \apjs, 241, 19

\bibitem[Johnson et al.(2021)]{johnson21} 
Johnson B. D., Leja J., Conroy C. \& Speagle J. S., 2021, \apjs, 254, 22

\bibitem[Johnson et al.(2013)]{john13} 
Johnson S.~P., Wilson G.~W., Tang Y. \& Scott K.~S., 2013, \mnras, 436, 2535

\bibitem[Kollatschny et al.(2020)]{kollats20} 
Kollatschny W., Weilbacher P.~M., Ochmann M.~W. et al., 2020, \aap, 633, A79

\bibitem[Krolik \& Begelman(1988)]{krol88} 
Krolik J.~H. \& Begelman M.~C., 1988, \apj, 329, 702

\bibitem[Lacey et al.(2016)]{lacey16} 
Lacey C.~G., Baugh C.~M., Frenk C.~S. et al., 2016, \mnras, 462, 3854

\bibitem[Lonsdale et al.(2003)]{lons03}
Lonsdale Carol J., Lonsdale Colin J., Smith H. ~E. \& Diamond P., 2003, \apj, 592, 804

\bibitem[Lutz et al.(1999)]{lutz99}
Lutz D., Veilleux S. \& Genzel R., 1999, \apj, 517, L13

\bibitem[Metropolis et al.(1953)]{metropolis53} 
Metropolis N., Rosenbluth A.~W., Rosenbluth M.~N. et al., 1953, \jcp, 21, 1087

\bibitem[Nardini et al.(2009)]{nard09} 
Nardini E., Risaliti G., Salvati M. et al. 2009, \mnras, 399, 1373 

\bibitem[Noll et al.(2009)]{noll09} 
Noll S., Burgarella D., Giovannoli E. et al., 2009, \aap, 507, 1793

\bibitem[Papadopoulos et al.(2025)]{pap25} 
Papadopoulos M., Papadopoulou Lesta V., Michos I. et al. 2025, \mnras, 536, 2433

\bibitem[Pearson et al.(2016)]{pea16} 
Pearson C., Rigopoulou D., Hurley P. et al. 2016, \apjs, 227, 9

\bibitem[Privon et al.(2017)]{privon17}
Privon G.~C., Aalto S. \& Falstad N. et al., 2017, \apj, 835, 213

\bibitem[Quinatoa et al.(2024)]{quin24}
Quinatoa D., Yang C., Ibar E. et al., 2024, \mnras, 527, 6321

\bibitem[Ramos Almeida \& Ricci(2017)]{ramos17} 
Ramos Almeida C. \& Ricci C, 2017, Nature Astronomy, 1, 679

\bibitem[Rigopoulou et al.(1999)]{rigop99}
Rigopoulou D., Spoon H.~W.~W., Genzel R. et al., 1999, \aj, 118, 2625 

\bibitem[Rodr{\'\i}guez Zaur{\'\i}n et al.(2010)]{rodr10} 
Rodr{\'\i}guez Zaur{\'\i}n J., Tadhunter C.~N. \& Gonz{\'a}lez Delgado R.~M., 2010, \mnras, 403, 1317

\bibitem[Romero-Ca{\~n}izales et al.(2012)]{romero12} 
Romero-Ca{\~n}izales C., P{\'e}rez-Torres M. {\'A}ngel \& Alberdi A., 2012, \mnras, 422, 510

\bibitem[Rowan-Robinson et al.(1993)]{rowan93}
Rowan-Robinson M., Efstathiou A., Lawrence A. et al., 1993, \mnras, 261, 513

\bibitem[Rowan-Robinson(2000)]{rowan00}
Rowan-Robinson M., 2000, \mnras, 316, 885
 
\bibitem[Rowan-Robinson et al.(2018)]{rowan18}
Rowan-Robinson M., Wang L., Farrah D. et al., 2018, \aap, 619, A169

\bibitem[Sani \& Nardini(2012)]{sani12}
Sani E. \& Nardini E., 2012, Advances in Astronomy, 2012, 783451

\bibitem[Sanders et al.(1988)]{sanders88} 
Sanders D.~B., Soifer B.~T., Elias J.~H. et al., 1988, \apj, 325, 74  

\bibitem[Saunders et al.(2000)]{sau00} 
Saunders W., Sutherland W.~J., Maddox S.~J. et al., 2000, \mnras, 317, 55

\bibitem[Shirley et al.(2019)]{shi19} 
Shirley R., Roehlly Y., Hurley P.~D. et al., 2019, \mnras, 490, 634

\bibitem[Shirley et al.(2021)]{shi21} 
Shirley R., Duncan K., Campos Varillas M.~C. et al., 2021, \mnras, 507, 129

\bibitem[Siebenmorgen et al.(2015)]{sie15} 
Siebenmorgen R., Heymann F. \& Efstathiou A., 2015, \aap, 583, A120 

\bibitem[Soifer et al.(1986)]{soif86} 
Soifer B.~T. Sanders D.~B., Neugebauer G. et al., 1986, \apj, 303, L41

\bibitem[Spoon et al.(2007)]{spoon07} 
Spoon H.~W.~W., Marshall J.~A., Houck J.~R. et al., 2007, \apjl, 654, L49

\bibitem[Stalevski et al.(2016)]{stal16} 
Stalevski M., Ricci C., Ueda Y. et al., 2016, \mnras, 458, 2288

\bibitem[Sturm et al.(2011)]{sturm11} 
Sturm E., Gonz{\'a}lez-Alfonso E., Veilleux S. et al., 2011, \apjl, 733, L16

\bibitem[Surace et al.(1998)]{surace98} 
Surace J.~A., Sanders D.~B., Vacca W.~D. et al., 1998, \apj, 492, 116

\bibitem[Tacconi et al.(2002)]{tacc02} 
Tacconi L.~J., Genzel R., Lutz D. et al., 2002, \apj, 580, 73 

\bibitem[Varnava \& Efstathiou(2024a)]{smart} 
Varnava C. \& Efstathiou A., 2024a, \mnras, 531, 2304

\bibitem[Varnava \& Efstathiou(2024b)]{ascl} 
Varnava C. \& Efstathiou A., 2024b, Astrophysics Source Code Library, record ascl:2406.003

\bibitem[Varnava et al.(2024)]{varn24} 
Varnava C., Efstathiou A. \& Farrah D., 2024, \mnras, 534, 2585

\bibitem[Veilleux et al.(2002)]{veilleux02}
Veilleux S., Kim D.~C. \& Sanders D.~B., 2002, \apjs, 143, 315

\bibitem[Veilleux et al.(2009)]{veilleux09}
Veilleux S., Rupke D.~S.~N., Kim D.~C. et al., 2009, \apjs , 182, 628 

\bibitem[Veilleux et al.(2013)]{veilleux13}
Veilleux S., Melendez M., Sturm E. et al., 2013, \apj, 776, 27

\bibitem[Verma et al.(2002)]{ver02} 
Verma A., Rowan-Robinson M., McMahon R. et al., 2002, \mnras, 335, 574. 

\bibitem[Yuan et al.(2010)]{yuan10} 
Yuan T.~T., Kewley L.~J. \& Sanders D.~B., 2010, \apj, 709, 884


\end{thebibliography}
\end{document}